\documentclass[onecolumn,showpacs,aps,prd]{revtex4}
\usepackage{graphicx,subfig,rotate,color}
\usepackage{amsmath,amssymb,latexsym}
\def\arccosh{{\rm arccosh}}

\def\ETA{\zeta}
\usepackage{graphicx}
\usepackage{dcolumn}
\usepackage{amsmath}
\usepackage{epsfig}
\RequirePackage{xspace}

\begin{document}

\def\bp{{b^\prime}}
\def\tp{{t^\prime}}
\def\acpt{\alpha}
\def\op{{\cal O}}
\def\lsim{\mathrel{\lower4pt\hbox{$\sim$}}\hskip-12pt\raise1.6pt\hbox{$<$}\;}
\def\Dd{\psi}
\def\pp{\lambda}
\def\ket{\rangle}
\def\BAR{\bar}
\def\xba{\bar}
\def\fa{{\cal A}}
\def\fm{{\cal M}}
\def\fl{{\cal L}}
\def\ufs{\Upsilon(5S)}
\def\gsim{\mathrel{\lower4pt\hbox{$\sim$}}
\hskip-10pt\raise1.6pt\hbox{$>$}\;}
\def\ufour{\Upsilon(4S)}
\def\xcp{X_{CP}}
\def\ynotcp{Y}
\vspace*{-.5in}
\def\ETAp{\ETA^\prime}
\def\bfb{{\bf B}}
\def\fd{r_D}
\def\fb{r_B}
\def\ed{\ETA_D}
\def\eb{\ETA_B}
\def\hatA{\hat A}
\def\hatfd{{\hat r}_D}
\def\hated{{\hat\ETA}_D}
\def\D{{\bf D}}
\def\pcc{(+ charge conjugate)}

\def\uglu{\hskip 0pt plus 1fil
minus 1fil} \def\uglux{\hskip 0pt plus .75fil minus .75fil}

\def\slashed#1{\setbox200=\hbox{$ #1 $}
    \hbox{\box200 \hskip -\wd200 \hbox to \wd200 {\uglu $/$ \uglux}}}

\def\slpar{\slashed\partial}
\def\sla{\slashed a}
\def\slb{\slashed b}
\def\slc{\slashed c}
\def\sld{\slashed d}
\def\sle{\slashed e}
\def\slf{\slashed f}
\def\slg{\slashed g}
\def\slh{\slashed h}
\def\sli{\slashed i}
\def\slj{\slashed j}
\def\slk{\slashed k}
\def\sll{\slashed l}
\def\slm{\slashed m}
\def\sln{\slashed n}
\def\slo{\slashed o}
\def\slp{\slashed p}
\def\slq{\slashed q}
\def\slr{\slashed r}
\def\sls{\slashed s}
\def\slt{\slashed t}
\def\slu{\slashed u}
\def\slv{\slashed v}
\def\slw{\slashed w}
\def\slx{\slashed x}
\def\sly{\slashed y}
\def\slz{\slashed z}
\def\slE{\slashed E}

%
%
%
%

\title{
\vskip 10mm
\large\bf
Detecting Fourth Generation Quarks at Hadron Colliders 
}

\author{David Atwood}
\affiliation{Dept. of Physics and Astronomy, Iowa State University, 
Ames,
IA 50011}
\author{Sudhir Kumar Gupta}
\affiliation{Dept. of Physics and Astronomy, Iowa State University, 
Ames,
IA 50011}
\author{Amarjit Soni}
\affiliation{ Theory Group, Brookhaven National Laboratory, Upton, NY
11973}

\begin{abstract}

Although there is no compelling evidence, at present, against the 
Standard Model (SM), in the past few years, a number of 2-3 sigma 
tensions have appeared which could be alleviated simply by adding 
another generation of fermions. Furthermore, a fourth generation could 
help resolve the issue of baryogenesis and the understanding of the 
hierarchy problem.

In this paper, we consider the phenomenology of the fourth generation 
heavy quarks which would be pair produced at the LHC. We show that if 
such a quark with a mass in the phenomenologically interesting range of 
400~GeV--600~GeV decays to a light quark and a W-boson, it will produce 
a signal in a number of channels which can be seen above the background 
from the three generation Standard Model processes. In particular, such 
quarks could be seen in channels where multiple jets are present with 
large missing momentum and either a single hard lepton, an opposite sign 
hard lepton pair or a same sign lepton pair.

In the same sign dilepton channel there is little background and so an 
excess of such pairs at large invariant mass will indicate the presence 
of heavy down type quarks. More generally, in our study, the main tool 
we use to determine the mass of the heavy quark in each of the channels 
we consider is to use the kinematics of the decay of such quarks to 
resolve the momenta of the unobserved neutrinos. We show how this can be 
carried out, even in cases where the kinematics is under-determined by 
use of the approximation, which holds quite well, that the two heavy 
quarks are nearly at rest in the center of mass frame.

Since it is very likely that at least the lightest heavy quark decays in 
the mode we consider, this means that it should be observed at the LHC. 
Indeed, it is expected that the mass splitting between the quarks is 
less than $m_W$ so that if the Cabbibo-Kobayshi-Maskawa
(CKM) matrix element between the fourth and lower 
generations are not too small, both members of the fourth generation 
quark doublet will decay in this way. If this is so, the combined signal 
of these two quarks will make the signal for the fourth generation 
somewhat more prominent.
 
\end{abstract}


\maketitle

\section{Introduction}\label{introduction}

The Standard Model with three generations (SM3) has been very successful 
in explaining all experimental results to date, in particular CP 
violation in the K- and B-meson systems is well understood, to an 
accuracy of about 15-20\%, in terms of the CKM matrix of that 
theory~\cite{cref,kmref}. Recently, however, some ``possible evidence" 
for deviations from SM3 in B decays~\cite{Lunghi:2007ak,Lunghi:2008aa, 
Lunghi:2009sm,Lenz:2006hd,Bona:2008jn,Bona:2009cj} is claimed. Although 
these effects can be explained by several physics beyond the Standard 
Model~\cite{APS1,APS2,AJB081,AJB082,MN08,lang,paridi} scenarios, the 
simplest viable explanation seems to be an extension of the Standard 
Model to 4 generations (SM4) 
~\cite{Hung:2007ak, Frampton:1999xi, Soni:2010xh, AS08,AJB10, GH10, Eberhardt:2010bm} where the mass of the 
new heavy quarks is in the range 400-600~GeV.

If further studies in the B system continue to show deviations from the 
SM3 predictions, it may be difficult to resolve which extension of SM3 
is responsible. The most direct way to determine the nature of the new 
physics which may be involved is to produce it ``on shell''. Indeed, 
hadron colliders, particularly the LHC, are ideally suited for this 
task. Since the LHC generates a significant rate of parton interactions 
up to energies of $\sim$1~TeV it may be able to produce direct evidence 
of 
the new physics although this is not guaranteed in all cases. For 
instance, if the explanation lies in warped space 
ideas~\cite{APS1,AJB082,MN08} then it appears that the relevant 
particles have to be at least approximately 3~TeV~\cite{AMS03} rendering 
their 
detection at LHC rather difficult~\cite{ADPS07,shri1,shri2}.

If the new physics is an additional sequential fourth generation, there 
would be two new heavy quarks, a heavy charge +2/3 quark ($\tp$) and 
charge -1/3 quark ($\bp$). These quarks will be produced at the LHC 
predominantly by gluon-gluon fusions and should be produced at the LHC 
with appreciable rates~\cite{MGM08}. For example at 10~TeV center of 
mass energy cross-section for pair producing 500~GeV quarks is large, 
$\approx$ 1 pb, rising to around 4 pb at 14~TeV and LHC experiments may 
well be able to study up to about 1~TeV~\cite{MGM08,Whiteson10}, which 
is well above the perturbative 
bound~\cite{chanowitz_furman_hinchliffe,cfl}. We also note that 
experiments have already been searching for the heavier quarks and 
provided (95\% CL) bounds: $m_\tp \gsim 311$~GeV; $m_\bp \gsim 
338$~GeV~\cite{Murat_ICHEP10,Whiteson10,Conway_BF10,CDF09, 
Aaltonen:2011vr, Abazov:2011vy, Chatrchyan:2011em}. These bounds are a 
little higher than the one quoted in Ref.~\cite{Nakamura:2010zzi} of 
$m_\tp \gsim 256$ GeV; $m_\bp \gsim 128$ GeV at $95\%$ CL.

For the analysis to follow an important characteristic of the fourth 
generation quark doublet is that the mass splitting between the $\bp$- 
and $\tp$-quarks is constrained by electroweek precision tests to be 
small, likely less than $m_W$ \cite{splitting}.

In addition to resolving the phenomenological hints of physics beyond 
SM3 that are alluded to in the above discussion, if a fourth generation 
is present, it may be helpful in explaining the long standing hierarchy 
problem within the SM. With very massive quarks in the new generation it 
has been proposed that electroweak symmetry breaking may well become a 
dynamical feature of the model ~\cite{ 
Holdom:2009rf,Holdom:1986rn,King:1990he, 
Hill:1990ge,Carpenter:1989ij,Hung:2009hy, Hung:2010xh,Burdman:2007sx}.

Another issue which is naturally addressed by the presence of a fourth 
generation is the origin of baryogenesis in the early universe. The tiny 
amount of CP violation allowed in the context of the CKM matrix of SM3 
is much too small to supply the CP violation necessary for baryogenesis 
in the early universe. However, if a fourth generation is present, then 
there are two more additional phases in the CKM matrix. The effects of 
these new phases is significantly enhanced~\cite{Hou:2008xd} by the 
larger masses of the new generation and so the natural CP violation of 
SM4 is perhaps large enough~\cite{Ham:2004xh, Fok:2008yg, 
Kikukawa:2009mu} to satisfy that Sakharov~\cite{Sakharov:1967dj} 
condition for baryogenesis. It has been pointed out \cite{Fok:2008yg}, 
however, that in the Standard Model with a fourth generation there might 
not be a first order phase transition hence it may still be the case 
that additional physics is required to satisfy that condition for 
baryogenesis. The presence of the two additional phases in the SM4 
mixing matrix also can have many interesting phenomenological 
implications especially in observables that in SM3 are predicted to 
yield null results
~\cite{Soni:2010xh,AS08,AJB10,GH10,Gershon:2006mt,Eilam:2009hz}.

%
%

Motivated by these considerations, in this paper we will consider the 
strategies
for detecting fourth generation quarks at the hadron 
colliders. In Section~\ref{decayrates} we present the expressions for 
the decay distributions of the decay of a fourth generation quarks 
considering, in particular the energy spectrum of the lepton that is 
produced by the decay of such a quark.

In Section~\ref{samples} we discuss the various event samples which are 
most likely to be useful in obtaining signals of heavy quarks. In 
particular, we will consider signals consisting of multiple hard jets 
with missing momentum in combination with either one hard lepton, an 
opposite sign dilepton pair or a same sign dilepton pair. We also set 
out a set of basic cuts which are helpful in enhancing the signal with 
respect to the SM3 background. For each of the three kinds of event 
samples, we then discuss how the kinematics can be used to determine the 
mass of the heavy quark. In general a significant signal versus the 
background will be seen in a histogram of the reconstructed mass. In 
general we highlight two important challenges in reconstruction of the 
mass. First of all, in the presence of a large number of jets, there 
will be a potentially large combinatorial background; secondly in some 
cases there are not enough kinematic constraints to reconstruct the 
neutrino momenta which make up the missing momentum. In the case of the 
same sign dilepton pairs the SM3 background is much smaller than the 
signal which arises in $\bp$-pair production thus an excess of high 
invariant mass same sign pairs is a clear signal for new physics in 
general and could be produced by $\bp$-quarks.

In Section~\ref{backgrounds} we discuss in detail the Standard Model 
(i.e. SM3) backgrounds which contribute to the event samples while in 
Section~\ref{summary} we present our conclusions.

\section{Decay Rates of Heavy Quarks}\label{decayrates}

Let us now consider the main decay modes for the $\tp$- and $\bp$- 
quarks. A similar discussion is found in~\cite{Das:2010fh, 
Arhrib:2006pm} and in \cite{Holdom:2007ap,Chao:2011th} the threshold 
effects at the interface between two body and three body decays are 
discussed.

If the $\tp$ is more massive than the $\bp$ and the splitting is greater 
than $m_W$ then the main decay mode of the $\tp$ will be $\tp\to\bp W$. 
If the splitting is less than $m_W$ then it can decay through the three 
body modes $\tp\to\bp W^*$ where $W^*$ is a virtual $W$ which either 
decays leptonically or hadronically. The heavy top can also decay 
through the two body mode to lighter quarks, $\tp\to b W^+$ etc.. As we 
shall show below, unless $V_{\tp i}\leq 10^{-3}$ the two body mode will 
generally dominates over the three body mode. In this scenario the 
lighter $\bp$ should undergo a two body decay to u- or c-quarks with the 
relative branching ratios depending on the values of $V_{i\bp}$.

If the $\tp$ is lighter than the $\bp$; the $\tp$ will decay via a 2 
body mode to generation 1-3 quarks; in analogy with the $\bp$ case 
above, the relative branching ratios depend on the values of $V_{\tp 
i}$. If the splitting is greater than $m_W$, the $\bp$ will decay via 
$\bp\to \tp W^-$. If the splitting is less than $m_W$ the dominant decay 
mode might either be the three body mode $\bp\to\tp W^*$ or the two body 
mode to generation 1-3 quarks, $\bp\to q W^-$ depending on the value of 
$V_{i \bp}$. Again, the two body mode will dominate unless $V_{i 
\bp}\leq 10^{-3}$.

Six scenarios for the complete decay chains of a fourth generation quark 
are thus possible depending on whether the relative mass of these quarks 
and the CKM matrix coupling to the lighter generations:

\begin{enumerate}

\item\label{scA} $m_{\tp}>m_{\bp}+m_W$ in which case $\tp\to\bp W^+$ and 
$\bp\to q W^-$ where $q=u,c,t$ (with branching ratio depending on the 
CKM elements).

\item\label{scB} If $m_{\bp}+m_W>m_\tp>m_\bp$ and $V_{\tp i}$ is 
relatively large, the dominant decay of $\tp$ is $\tp\to q W^+$ 
with a small branching ratio to the three body decay $\tp\to \bp W^*$. 
Again $\bp\to q W^-$ where $q=u,c,t$.

\item\label{scC} If $m_{\bp}+m_W>m_\tp>m_\bp$ but $V_{\tp i}$ is small 
enough, the dominant decay of $\tp$ is $\tp\to \bp W^*$. Again 
$\bp\to q W^-$ where $q=u,c,t$.

\item\label{scD} If $m_{\bp}>m_{\tp}+m_W$ in which case $\bp\to\tp W^-$ 
and $\tp\to q W^+$ where $q=d,s,b$ (with branching ratio depending on 
the CKM elements).

\item\label{scE} If $m_{\tp}+m_W>m_\bp >m_\tp$ and $V_{i \bp}$ is 
relatively large, the dominant decay of $\bp$ is $\bp\to q W^-$ 
with a small branching ratio to the three body decay $\bp\to \tp W^*$. 
Again $\tp\to q W^+$ where $q=d,s,b$.

\item\label{scF} If $m_{\tp}+m_W>m_\bp>m_\tp$ but $V_{i \bp}$ is small 
enough, the dominant decay of $\bp$ is $\bp\to \tp W^*$. Again 
$\tp\to q W^-$ where $q=d,s,b$.

\end{enumerate}

In the scenarios where the dominant $\bp$ decay is $\bp\to q W^-$, there 
is an important distinction between the case where $q=t$ and $q=u,c$. In 
the $q=u,c$ case the quark will just manifest as a single jet while in 
the $q=t$ case, the top subsequently decays to $t\to b W$ and the $W$ 
may in turn decay leptonically or hadronically. The collider signature 
will thus depend on the nature of the $W$ decay.

If the CKM matrix is not unitary (i.e. the fourth generation is not a 
genuine sequential generation of the SM) or if there is a further fifth 
generation (thus rendering the $4\times 4$ CKM submatrix non-unitary) 
then 
there 
are some possible modifications to the above scenarios. If $V_{\tp\bp}$ 
is small, then in scenarios \ref{scA} and \ref{scD} it might not be the 
case that $\tp\to\bp W$ or $\bp\to\tp W$ are the dominant decay modes 
and these cases will resemble scenarios \ref{scB} and \ref{scE} 
respectively.

In the case $m_{\bp}+m_W>m_\tp>m_\bp$ where $V_{\tp i}$ is small enough 
to suppress the two body decay mode as in scenario \ref{scC}, it could 
happen that $V_{q \bp}$ for $q=u,c,t$ is sufficiently large that the 
mode $\tp\to q W^+W^-$ becomes important. If the $4\times 4$ CKM matrix 
is 
unitary then:

\begin{eqnarray}
\sum_{i=1,3}\left | V_{\tp i}\right |^2
=
\sum_{i=1,3}\left | V_{i \bp}\right |^2
\end{eqnarray}

\noindent therefore a large $V_{q\bp}$ will imply a large $V_{\tp i}$ 
for some $i=d,s$ or $b$ and so the two body mode must dominate. The 
analogous argument also applies for the decay $\bp\to q W^+W^-$ in 
scenario \ref{scF}.

\subsection{Decay Rates}

Let us now consider the two body decay of a heavy quark $q_1\to q_2 W$ 
where $q_1$ is a fourth generation quark and $q_2$ is either the other 
fourth generation quark or a 1-3 generation quark.

The total decay rate at tree level is:

\begin{eqnarray}
\Gamma(q_1\to q_2 W)
&=&
|V_{12}|^2\Gamma_2(m_1)
\Delta(1,x_{21},x_{W1})
\left (
\Delta(1,x_{21},x_{W1})^2
+(3+2x_{21}-3x_{W1})x_{W1}
\right )
\label{two_bdy_tree}
\end{eqnarray}

where

\begin{eqnarray}
x_{21}&=&\left(\frac{m_2}{m_1}\right)^2
\ \ \ \ \ 
x_{W1}=\left(\frac{m_W}{m_1}\right)^2
\nonumber\\
\Delta(a,b,c)&=&
\sqrt{\left|
a^2+b^2+c^2
-2ab-2bc-2ca
\right|}
\nonumber\\
\Gamma_2(m_1)&=&
\frac{G_F}{8\pi\sqrt{2}}m_1^3
\label{two_bdy_tree_defs}
\end{eqnarray}

Note that in the limit that $m_2<<m_1$ which would be the case if 
$q_2=u,d,c,s,b$, we can approximate Eqn.~(\ref{two_bdy_tree}) 
by:

\begin{eqnarray}
\Gamma(q_1\to q_2 W)
&\approx&
|V_{12}|^2\Gamma_2(m_1)
(1-x_{1W})^2
(1+2x_{1W})
+O(x_{21})
\label{two_bdy_tree_approx1}
\end{eqnarray}

Conversely if $x_{21}$ is not small (i.e. for decays between the two 
heavy quarks or to $tW$) one can expand in this expression in $x_{1W}$:

\begin{eqnarray}
\Gamma(q_1\to q_2 W)
&\approx&
|V_{12}|^2\Gamma_2(m_1)
\left (
(1-x_{21})^3
+x_{21}x_{W1}(1-x_{21})
\right )
+O(x_{W1}^2)
\label{two_bdy_tree_approx2}
\end{eqnarray}

Let us now consider the three body decay $q_1\to q_2 ff'$ where, in this 
paper, we will generally consider $q_1$ and $q_2$ to be the fourth 
generation quarks and $ff'$ are light fermion pairs which arise from 
the virtual $W$ so 
$ff'=ud$, $cs$, $us$, $cd$ , $e\nu$, $\mu\nu$ or $\tau \nu$.

At tree level, 

\begin{eqnarray}
\Gamma(q_1\to q_2 ff')
=
\left |V_{12}\right |^2
\left |V_{ff'}\right |^2
N_c(ff')
\Gamma_3(m_1)
I(x_{21},x_{W1})
\label{three_bdy_tree}
\end{eqnarray}

\noindent
$N_c(ff')=3$ for quark pairs and 1 for leptons; 
$V_{ff'}$ is the appropriate CKM element for quark pairs and 1 for 
lepton pairs and 

\begin{eqnarray}
\Gamma_3(m_1)&=&
\frac{G_F^2}{192 \pi^3}m_1^5.
\label{three_bdy_tree_defs1}
\end{eqnarray}

\noindent
The factor 
$I(x_{21},x_{W1})$
is given by

\begin{eqnarray}
I(x_{21},x_{W1})
&=&
12x_{1W}
\bigg(
\frac13 (1-x_{21})(2V^2-6x_{W1}U+x_{W1}W)
+x_{W1}^2 U \log\frac{1}{x_{21}}
\nonumber\\
&&+x_{W1}^2 \frac{2U^2-x_{21}}{V}
\left [
\arctan\frac{1-U}{V}   
+
\arctan\frac{U-x_{21}}{V}
\right ]
\bigg)
\label{three_bdy_tree_defs2}
\end{eqnarray}

\noindent
where

\begin{eqnarray}
U&=& \frac12 (1+x_{21}-x_{W1})
\nonumber\\
V&=&\frac12 \Delta(1,x_{21},x_{W1})
\nonumber\\
W&=& \frac12 (1+x_{21}+x_{W1})
\label{three_bdy_tree_defs3}
\end{eqnarray}

In the scenarios we are most interested in where the splitting between 
the two heavy quarks is not very large, the expression in 
Eqn.(\ref{three_bdy_tree_defs2})
is well approximated by:

\begin{eqnarray}
I(x_{21},x_{W1})
&=&
(1-x_{12})^5\left(
\frac25+\frac15 
(1-x_{12})
+\frac{4x_{W1}+3}{35x_{W1}}
(1-x_{12})^2
+O( (1-x_{12})^3)
\right)
\label{three_bdy_tree_approx1}
\end{eqnarray}

\begin{figure}
\centerline{
\includegraphics[angle=-90, 
width=0.6\textwidth]{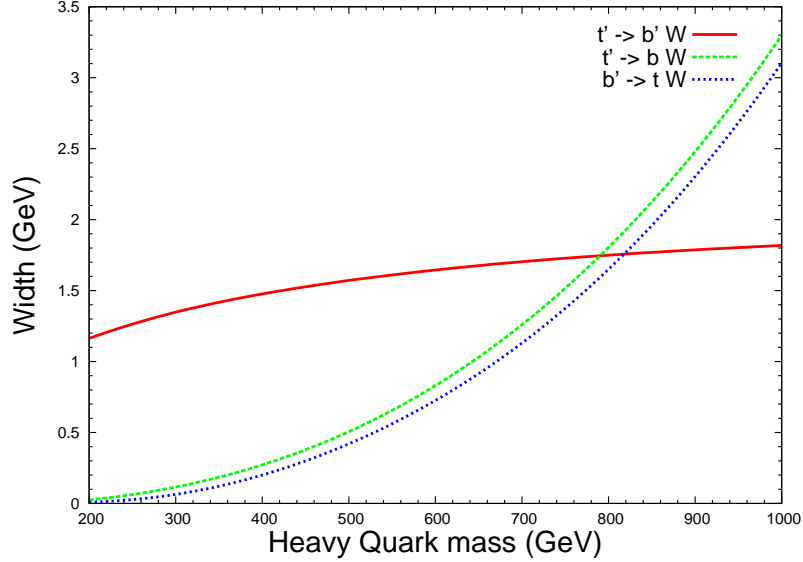}
}
\caption{
The rate of heavy quark decay to specific decay modes as a function of 
heavy quark mass. The solid curve is $\tp\to\bp W$ where 
$m_\tp-m_\bp=100$~GeV and $|V_{\tp \bp}|=1$. The dashed curved is 
$\bp\to 
tW$ where $|V_{t\bp}|=0.1$. The dotted curve is $\tp\to bW$ where 
$|V_{\tp b}|=0.1$.
}
\label{two_body_width_plot}
\end{figure}

\begin{figure}
\centerline{
\includegraphics[angle=-90, 
width=0.6\textwidth]{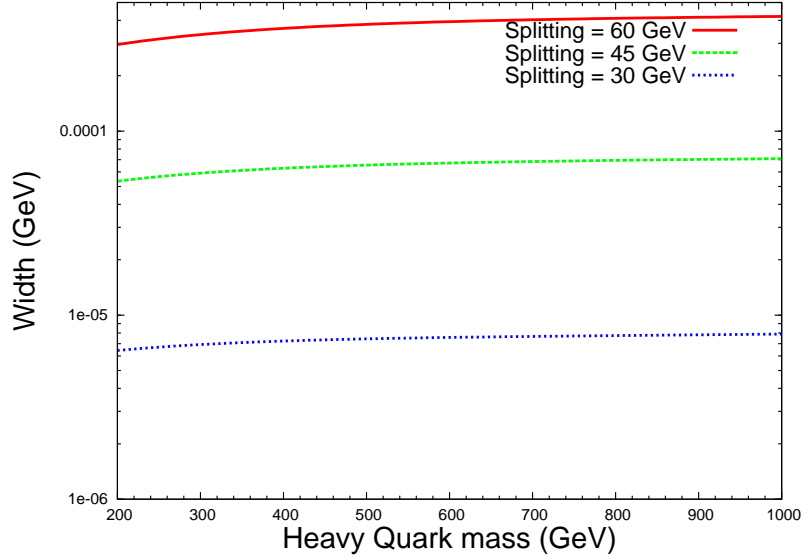}
}
\caption{
The decay rate for $\tp\to\bp e \nu$ as a function of heavy quark mass 
where $|V_{\tp b}|=1$. The solid curve is for $m_\tp-m_\bp=60$~GeV, the 
dashed curve is for $m_\tp-m_\bp=45$~GeV, and the dotted curve is for 
$m_\tp-m_\bp=30$~GeV,
}
\label{three_body_width_plot}
\end{figure}

\subsection{Decay Kinematics and Lepton Energy Spectrum for Two Body 
Decays}

The two body kinematics together with the V-A structure of the W 
couplings determine the energy spectrum of the 
lepton arising from the two body 
$\tp\to W^+ b$
and $\bp\to W^- t$.

In the case of $\bp\to t [W^+\to \ell^+\nu] $ the kinematic limits of 
the lepton energy in the rest frame of the $\bp$ are:

\begin{eqnarray}
E^{\ell}_{max} &=& \frac{1}{4m_{\bp}}
\left [
m_\bp^2+m_W^2-m_t^2+\Delta(m_\bp^2,m_t^2,m_W^2)
\right ]
\nonumber\\
E^{\ell}_{min} &=& \frac{1}{4m_{\bp}}
\left [
m_\bp^2+m_W^2-m_t^2+\Delta(m_\bp^2,m_t^2,m_W^2)
\right ]
\label{el_minmax}
\end{eqnarray}

\noindent
Within this kinematic region, the distribution is given by:

\begin{eqnarray}
\frac{d\Gamma}{dE_\ell}
&\propto&
E_\ell(E_0-E_\ell)
\label{el_dist}
\end{eqnarray}

\noindent
where $E_0=(m_\bp^2-m_t^2)/2/m_\bp$.

From this we can calculate the average lepton energy:

\begin{eqnarray}
\overline E_\ell
&=&
\frac{m_\bp}{4}
\frac
{1-3x_t+x_W+3x_t^2+x_W^2-3x_W^3-x_t^3-x_Wx_t^2+5x_tx_W^2}
{1+x_W-2x_t+x_t^2+x_tx_W-2x_W^2}
\nonumber\\
&=&
\frac{m_\bp}{4}\left (1-x_t   \right ) 
+O(x_t^2,x_W^2)
\end{eqnarray}

\noindent
where 
$x_W=m_W^2/m_\bp^2$ 
and 
$x_t=m_t^2/m_\bp^2$.

When a $\bp$ pair is produced, the matrix element and the structure 
functions tend to drive it to lower $\hat s=(p_\bp+ 
p_{\overline {\bp}})^2$. 
The transverse momentum distribution of the lepton should thus be well 
approximated by the transverse energy distribution of the $\bp$ at rest. 
The average transverse momentum is thus given by:

\begin{eqnarray}
\overline P_T^\ell
&=&\frac{\pi}{4}\overline E_\ell
\approx \frac{\pi}{16}m_\bp
\end{eqnarray}

\noindent
The $P_T$ distribution in the limit where the $\bp$ is at rest is:

\begin{eqnarray}
\frac{d\Gamma}{d P_T}
&=&
\left\{
\begin{array}{lr}
P_T\left(
E_0 \cosh^{-1}\frac{E_{max}}{P_T}
-
\sqrt{E_{max}^2-P_T^2}
\right )
&
{\rm if\ }E_{min}<P_T<E_{max}
\\
P_T\left(
E_0 \cosh^{-1}\frac{E_{max}}{P_T}
-E_0 \cosh^{-1}\frac{E_{min}}{P_T}
-
\sqrt{E_{max}^2-P_T^2}
+\sqrt{E_{min}^2-P_T^2}
\right )
&
{\rm if\ }0<P_T<E_{min}
\end{array}
\right .
\end{eqnarray}

In $\bp$-quark decay a lepton of the opposite sign may also be produced 
through the cascade through the top quark, $\bp\to W^- [t\to b [W^+\to 
\ell^+\nu]]$. Again if the $\bp$ is at rest, we can develop an analytic 
expression for the energy spectrum of this lepton. 

It is useful to divide the spectrum into three segments by the energies:

\begin{eqnarray}
E_0 &=& \frac12 m_W e^{-\left (\theta_t+\theta_W\right )} \nonumber\\
E_1 &=& \frac12 m_W e^{-\left |\theta_t-\theta_W\right |} \nonumber\\
E_2 &=& \frac12 m_W e^{+\left |\theta_t-\theta_W\right |} \nonumber\\
E_3 &=& \frac12 m_W e^{+\left (\theta_t+\theta_W\right )} \nonumber\\
\end{eqnarray}

\noindent where

\begin{eqnarray}
\theta_W&=&\arccosh\frac{x_t+x_W}{2\sqrt{x_W}} \nonumber\\
\theta_t&=&\arccosh\frac{1+x_t-x_W}{2\sqrt{x_t}}.
\end{eqnarray}

The energy spectrum is thus

\begin{eqnarray}
\frac{d\Gamma}{dE_\ell}
&\propto&
\left \{
\begin{array}{ll}
F\left( y,z_{max}\right)
-
F\left( y,\frac{x_W+4y^2}{4y}\right)
&
{\rm if}\ E_0< E_\ell < E_1 
\\ 
F\left( y,z_{max}\right)
-
F\left( y,z_{min}\right)
&
{\rm if}\ E_1< E_\ell < E_2 
\\ 
F\left( y,z_{max}\right)
-
F\left( y,\frac{x_W+4y^2}{4y}\right)
&
{\rm if}\ E_2< E_\ell < E_3  
\end{array}
\right .
\end{eqnarray}

\noindent 
where 

\begin{eqnarray}
z_{max}&=&\sqrt{x_W}\cosh{\theta_t+\theta_W}\nonumber\\
z_{min}&=&\sqrt{x_W}\cosh{\theta_t-\theta_W}
\end{eqnarray}

\noindent
and

\begin{eqnarray}
F(y,z)&=&
\frac{y}{4\eta}
\Big(
(2yx_t-(2y+6yx_t+x_t)z+8x_t(1+y)z^2+4x_tz^3)
\nonumber\\
&&+x_W((1+6y+11x_t+1-yx_t)-(3-6yx_t+5x_t)z-4(1-x_t)z^2) 
\nonumber\\
&&+x_W^2((4-6yx_t-10y-7x_t) +(5+2yx_t+4y+3x_t)z+4z^2)
\nonumber\\
&&+x_W^3((2y-6-x_t)-(1+2y)z)
+x_W^4
\Big)
\nonumber\\
&&-\frac{y}{2}\arccosh( y)
\Big (x_t(1+4y)+2x_W(1+x_t+y-yx_t)-x_W^2(3+x_t+2y) \Big)
\end{eqnarray}

\noindent
and $\eta=\sqrt{z^2-x_w}$

\section{Event Samples}\label{samples}

In order to find evidence for heavy quark pair production at hadronic 
colliders, we use the decay scenarios discussed above to suggest which 
signals should be searched for. In this discussion we would like to 
highlight three issues which lead to acceptance cuts that apply all the 
signal classes we consider. First of all, we will introduce a set of 
basic cuts which will preserve the signal but reduce the SM background 
to a manageable level (about $10\times$ signal). Next, we will consider 
extracting a reconstructed $m_Q$ from the kinematics of the events that 
pass the basic cuts. We will see that a histogram in the reconstructed 
$m_Q$ generally separates the signal from background and allows the 
determination of the heavy quark mass. In order to most effectively use 
this method, however, it is useful to reduce the combinatorial 
background, particularly if the signal consists of a large multiplicity 
of jets. As we will show, this combinatorial background can be greatly 
reduced due to the fact that most of the jets are in pairs resulting 
from W-boson decay.

Let us first consider the possible heavy quark decay modes and then turn 
our attention to the analysis of the cases which are likely to be of 
greatest experimental interest. 

In general,
if we assume that the heavy quark decays dominantly through a two body 
decay mode, the net decay  of the heavy quark will be to a light quark 
plus 1-3 of W-bosons

If $Q$ is the heavy quark (either a $\tp$- or $\bp$- quark) and  $q$ 
one 
of 
the five lightest
quarks $u,d,s,c,b$, the decay chain will assume one of the 
following forms:

\begin{eqnarray}
\begin{array}{rl}
(1)& Q \to q W\\
(2)& Q \to q W W\\
(3)& Q \to q W W W\\  
\end{array}
\label{decay_forms}
\end{eqnarray}

In particular, 
decay channel (1) occurs when a $\tp$ decays to a $b$, $s$ or $d$ quark 
and a W-boson or when a $\bp$ decays to a $c$ or $u$ and a W-boson. Case 
(1) would be the dominant $\bp$ decay mode in the mixing scenario where 
$V_{t\bp}$ was much smaller than $V_{c\bp}$ and/or $V_{u\bp}$.

Conversely, 
the $\bp$ quark will decay dominantly through 
decay channel (2) 
if the first decay in its cascade is $\bp\to t W$ where the top quark 
then decays to $bW$.
The $\tp$ could also decay through a channel like this in the scenario 
where
$m_\tp>m_\bp+m_W$ and the $\bp$ decayed 
through channel (1), thus $\tp\to W\bp\to WWc$ or $WWu$. 

Decay channel (3) would apply if the $\tp$ cascaded down to the $\bp$ 
which in turn decayed via channel (2). Thus $\tp\to \bp W\to t WW\to b 
WWW$. 

If the mixing between the fourth and third generations is sufficiently 
small and the quark mass splitting is less than $m_W$ then two other 
decay modes involving three body decay channels may be important:

\begin{eqnarray}
\begin{array}{rl}
(4)& Q \to Q'W^*\to qWW^*\\
(5)& \tp \to \bp W^*\to t WW^*\to b WWW^*\\
\end{array}
\label{decay_forms2}
\end{eqnarray}

\noindent
($W^*$=virtual W-boson)
where in channel (4) $Q$ is the heavier fourth generation quark and $Q'$ 
is the lighter fourth generation quark.

Since the fourth generation quark $Q$ is pair produced, depending on 
which of the channels above controls the decay, there are potentially 
up to 6 W-bosons plus 2 light quark jets in the final state. If all 
those W-bosons decayed hadronically, that would result in a final state 
with up to 14 jets. In any case the QCD background to a purely hadronic 
final state is likely overwhelming so we must consider cases where at 
least one of the W-bosons decays leptonically.

In particular, we will consider the prospect of signals where one or two 
of the W-bosons decay leptonically. In such a case, the signal will be a 
final state with one or two hard lepton(s) and significant missing 
momentum.

Depending on which of the channels 1-5 is dominant for each quark 
species, it is not unreasonable to suppose that signals of this type 
will received contributions from both species of quarks. This is a 
natural situation if the two species are roughly degenerate and so both 
$\tp$ and $\bp$ quarks will be produced at roughly comparable rates, 
especially if the masses of the $\bp$- and $\tp$-quarks are around 
400-600~GeV and the collisions are at LHC energies, $\sqrt{s}\simeq 
O(7)$~TeV. Such a state of affairs can be helpful in building a signal 
indicating a fourth generation even before the individual contributions 
from $\tp$ and $\bp$ are separately identified.

The case where one or two W-bosons decay leptonically therefore leads to 
three signal channels for the fourth generation of quarks. We will 
consider the signals in each of these channels in order to determine the 
signal to background ratio and how the signal may be used to reconstruct 
the mass of the quark. These issues are related in that the difference 
in kinematics between SM backgrounds and the heavy quark signals means 
that if the mass of the quark can be reconstructed, this will provide a 
good mechanism for separating signals from background.

The three event samples which we consider are as follows:

\begin{enumerate}

\item
Single lepton sample: The signature of this sample is $\ell + nj + 
\slp_T$ where $\ell$ is a lepton, either $e$ or $\mu$, $nj$ 
means $n$ jets and $\slp_T$ means missing transverse momentum.

\item Like sign di-lepton sample: The signature of this sample is 
$\ell_1^\pm\ell_2^\pm + nj + \slp_T$

\item
Opposite sign di-lepton sample: The signature of this sample is 
$\ell_1^\pm\ell_2^\mp + nj + \slp_T$. Note that for some of the 
potential SM backgrounds, it might be helpful to consider $\ell_1\neq 
\ell_2$. Only heavy quarks that cascade to at least two W-bosons will 
contribute to this sample. 

\end{enumerate}

In our analysis we will consider mainly the scenario where the splitting 
between the two heavy quark masses is less than $m_W$ and the CKM 
element between the fourth generation quarks and lighter quarks is large 
enough that the dominant decay mode is the two body decay to lighter 
quarks. Thus the decay modes for the $\tp$ and $\bp$ quarks we are 
mainly considering are:

\begin{eqnarray}
\tp\to b W\nonumber\\
\bp\to t W\to b WW
\label{mainDecayModesConsidered3}
\end{eqnarray}

\noindent which are the modes that apply if the dominant mixing of the 
fourth generation is with the third. The analysis we carry out however 
easily generalizes to the case where this assumption is weakened. Thus 
if the decay in fact proceeds through

\begin{eqnarray}
\tp\to d\ {\rm or}\ s W\nonumber\\
\bp\to u\ {\rm or}\ c W\nonumber\\
\label{mainDecayModesConsidered12}
\end{eqnarray}

\noindent the kinematics in these cases will be identical to the $\tp$ 
decay to $bW$ and so the analysis we discuss below will apply to all of 
these cases. In addition, if b-tagging can be carried out then we can 
distinguish the $bW$ final state from $u,c,s,{\rm\ or\ }d\ W$.

For our event analysis we first generated the signal and background
events with the aid of 
{\tt MadGraph}~\cite{Stelzer:1994ta}
and later interfaced these to {\tt
PYTHIA6}~\cite{pythia} for further analysis including decays of the top
and W's.

For the $Q$-pair event generation, we wrote the {\tt MadGraph} 
model files to incorporate the $t^\prime$ and $b^\prime$ and their 
interactions. We use {\tt CTEQ6L}~\cite{Pumplin:2002vw} to evaluate 
parton densities.  The renormalization scale, $\mu_R$, and the 
factorization scale, $\mu_F$ are fixed at 

\begin{eqnarray} 
\mu_R = 
\sqrt{\hat{s}} = \mu_F 
\end{eqnarray}

Jet formation has been done using default {\tt PYTHIA} scheme 
implemented through {\tt PYCELL}. We also incorporate effects of initial 
state radiation (ISR) and final state radiation (FSR) using the same 
simulation package.

The basic cuts that apply in these tables
on leptons, $l = e, 
\mu$ and, jets, j (including b's) which consists of

\begin{itemize}
 
\item Lepton should have $p_{T_l} > 25$~GeV and
$\left|\eta_{_l}\right| \leq 2.7$, to ensure that they lie within the
coverage of the detector.
 
\item jets should have $p_{T_j} > 25$~GeV and $\left|\eta_{_j}\right|
\leq 2.7$
 
\item Spatial resolution between {\em lepton - lepton}, {\em lepton -
jet}, and, {\em jet - jet} should be $\Delta{R}_{ll} \geq 0.4$,
$\Delta{R}_{lj} \geq 0.4$, $\Delta{R}_{jj} \geq 0.4$ respectively,
(where $\Delta{R}_{ik} = \sqrt{{\Delta{\eta}_{_{ik}}}^2
+{\Delta{\phi}_{_{ik}}}^2 }$, $\Delta{\eta}_{_{ik}} = \eta_{_i} -
\eta_{_k}$, $\Delta{\phi}_{_{ik}} = \phi_{_i} - \phi_{_k}$), such that
the leptons and jets are well separated in space.
    
\item A missing transverse energy cut, $\not\!\!E_T > 30~{\rm GeV}$ to
enhance the likelihood that 
leptons are due to $W$ decay.
\end{itemize}  

In addition to the cuts mentioned above, we apply the following cuts to 
reduce the background further:

\begin{itemize} 

\item A minimum cut on 
the scalar sum of transverse 
momenta ($H_T$) of the final state lepton, jets, and the missing 
transverse energy 
of 350~GeV. $H_T$ is defined to be:

\begin{eqnarray}
H_T = p_{T_{visible}} + \not\!\!E_T = 
\sum\limits_{i = l, j} p_{T_i} + \not\!\!E_T.
\label{HTdef}
\end{eqnarray}

\end{itemize}

\begin{table}[h]
\centering
\begin{tabular}{|c|c|c|c|c|c|c|}\hline
Quark& $\sqrt{s}$ (TeV) &
cuts& $m_{Q}=300$~GeV & $m_{Q}=450$~GeV & $m_{Q}=600$~GeV &SM 
background\\\hline
$\tp$ & 14  &
$Basic$                      &6469, 552, 0&824, 73, 0
&170, 15, 0&221833, 16479, 8.8\\
$\tp$ & 14  &
$Basic + H_T > 350$~GeV      &5571, 464, 0&809, 71, 0
&169, 14, 0&~46846, ~3472, 6.4\\
\hline
\hline
$\tp$ & 10  &
$Basic$                      &2404, 188, 0&272, 22, 0& 
49, 5, 0&63609, 4467, 4.2\\
$\tp$ & 10  &
$Basic + H_T > 350$~GeV      &2074, 158, 0&265, 21, 0& 49, 
5, 0&12013, ~847, 3\\
\hline
\hline
$\tp$ & 7  &
$Basic$                      
&785, 61, 0&69, 6, 0& 10, 1, 0&22847, 1621, 1.7\\
$\tp$ & 7  &
$Basic + H_T > 350$~GeV      &668, 50, 0&67, 6, 0& 10, 1, 0&~4054, ~275, 
1.2\\
\hline
\hline
$\bp$ & 14  &
$Basic$                      &8948, 1210, 625&1092, 166, 
86&224, 35, 18&221833, 16479, 8.8\\
$\bp$ & 14  &
$Basic + H_T > 350$~GeV      &7293,  960, 582&1057, 159, 
84&221, 35, 17&~46846, ~3472, 6.4\\
\hline
\hline
$\bp$ & 10  &
$Basic$                      &3312, 457, 220&370, 54, 
28&65, 10, 6
&63609, 4467, 4.2\\
$\bp$ & 10  &
$Basic + H_T > 350$~GeV      &2654, 358, 212&356, 52, 
27&64, 10, 6
&12013, ~847, 3\\
\hline
\hline
$\bp$ & 7  &
$Basic$                      &1060, 145, 74&94, 13, 7&14, 
2, 1
&22847, 1621, 1.7\\
$\bp$ & 7  &
$Basic + H_T > 350$~GeV      &841, 113, 70&90, 13, 7&13, 
2, 1
&~4054, ~275, 1.2\\
\hline
\end{tabular}
\caption{
Number of signal and background events for a number of scenarios. In 
each case, the three numbers indicate the single lepton; opposite sign 
dileptons (OSD) and same sign dileptons (SSD) events from the $\tp$- and 
$\bp$-pair production at the LHC for $\sqrt{s}=$ 14, 10 and 7~TeV and 
$\int {\cal L} dt = 1$ fb$^{-1}$ without the requirement of isolation on 
jets. The basic cuts are: $p_{T_{l, j}} > 25$~GeV, $\left|\eta_{_{l, 
j}}\right| \leq 2.7$; $\Delta{R}_{l,l},\Delta{R}_{l,j} \geq 0.4$ and 
$\not\!\!E_T > 
30~{\rm GeV}$. 
} 
\label{big-table-1a}
\end{table}

Effects of the aforementioned cuts are shown in Table \ref{big-table-1a} 
where results for both quark species are considered and results are 
given in the cases of $\sqrt{s}=14$, $10$ and $7$~TeV. In all cases we 
consider the signals for heavy quark masses $m_{Q} = 300, 450$ and 
$600$~GeV. In Table \ref{big-table-1a} we did not put any isolation cuts 
on the jets. In Table \ref{big-table-2a} we further demand that all the 
jets are separated with $\Delta R_{jj} > 0.4$. We see that this latter 
requirement does not alter the numbers very much.

\begin{table}[h]
\centering
\begin{tabular}{|c|c|c|c|c|c|c|}\hline
Quark& $\sqrt{s}$ (TeV) &
cuts& $m_{Q}=300$~GeV & $m_{Q}=450$~GeV & $m_{Q}=600$~GeV &SM 
background\\\hline
$\tp$ & 14  &
$Basic$                      &5906, 509, 0&747, 66, 0&152, 13, 0
&204180, 14716, 7.3\\
$\tp$ & 14  &
$Basic + H_T > 350$~GeV      &5066, 416, 0&731, 64, 0&151, 13, 0
&~42625, ~3071, 5.8\\
\hline
\hline
$\tp$ & 10  &
$Basic$                      &2182, 172, 0&246, 20, 0& 44, 4, 0
&57858, 4071, 3.4\\
$\tp$ & 10  &
$Basic + H_T > 350$~GeV      &1876, 145, 0&239, 19, 0& 44, 4, 0
&10890, ~771, 2.7\\
\hline
\hline
$\tp$ & 7  &
$Basic$                      &704, 54, 0&62, 5, 0& 9, 1, 0
&20378, 1448, 1.4\\
$\tp$ & 7  &
$Basic + H_T > 350$~GeV      &597, 44, 0&60, 5, 0& 9, 1, 0
&~3557, ~233, 1.1\\
\hline
\hline
$\bp$ & 14  &
$Basic$                      &7952, 1073, 548&982, 147, 78&201, 32, 16
&204180, 14716, 7.3\\
$\bp$ & 14  &
$Basic + H_T > 350$~GeV      &6468,  843, 520&951, 142, 76&200, 32, 15
&~42625, ~3071, 5.8\\
\hline
\hline
$\bp$ & 10  &
$Basic$                      &2952, 403, 195&330, 49, 25&59, 9, 5
&57858, 4071, 3.4\\
$\bp$ & 10  &
$Basic + H_T > 350$~GeV      &2353, 315, 189&319, 47, 24&58, 9, 5
&10890, ~771, 2.7\\
\hline
\hline
$\bp$ & 7  &
$Basic$                      &935, 128, 65&83, 12, 6&12, 2, 1
&20378, 1448, 1.4\\
$\bp$ & 7  &
$Basic + H_T > 350$~GeV      &738, ~99, 61&80, 12, 6&12, 2, 1
&~3557, ~233, 1.1\\
\hline
\end{tabular}
\caption{
This table shows the number of events with the same cuts as in 
Table~\ref{big-table-1a} with the addition of the jet isolation cut that 
all the jets are separated with $\Delta R_{jj} > 0.4$.
}
\label{big-table-2a} 
\end{table}

In our approach, to further enhance the signal to background ratio and 
characterize the fourth generation quarks, we will first consider the 
reconstruction of the heavy quark mass from the kinematics of the event. 
In this endeavor we must deal with the combinatorial background that 
results from the high jet multiplicity; in particular, the kinematic 
role of each of the jets in the event is not a priori known. This 
problem is most acute in in the high jet multiplicities which result 
from $\bp$-pair events. We then discuss the various methods which result 
in the reduction of this combinatorial background. We now detail our 
analysis in each of the signal types.

\subsection{Analysis of Single Lepton Sample}

Under our assumptions, both $\tp$ and $\bp$ decays can contribute to 
signals of this type. 

For each event of this type, the key to determining the kinematics and 
therefore the heavy quark mass is the partitioning of the jets between 
the two $Q$'s in the initial state. In particular, in this case one of 
the heavy quarks decays only to hadrons while the others decay includes 
the one lepton observed. We will denote these two heavy quarks by $Q_h$ 
and $Q_\ell$ respectively. Thus if there are $n$ jets, of which $k$ 
should originate from the $Q_\ell$ then an initial $\left 
(^{n}_{k}\right)$ combinations must be tried where only one of the 
possible partitions will be "correct".

In the correct partition, it is possible to determine the momentum of 
the unobserved neutrino. If we denote by $j_h$ the total 4-momentum of 
all the jets assigned to the $Q_h$ side of the event and $j_\ell$ all 
the jets assigned to the $Q_\ell$ side of the event then the following 
five constraints apply to the four undetermined neutrino 4-momentum:

\begin{eqnarray}
\begin{array}{rl}
(1)&(\nu+\ell)^2=m_W^2 \\
(2)&(\nu+\ell+j_\ell)^2=j_h^2\ \ \ (=m_Q^2)  \\
(3)&(\nu)^2=0                 \\
(4)& \nu_x=\slp_x             \\
(5)& \nu_y=\slp_y          \\
\end{array}
\label{one_lepton_constraints}
\end{eqnarray} 

\noindent
Note that if we combine equation (1) and (2) with (3) we can rephrase 
these two conditions as:

\begin{eqnarray}
\begin{array}{rl}
(1')&\nu\cdot\ell=\frac12 m_W^2 \\
(2')& 2\nu\cdot(\ell+j_\ell)+(\ell+j_\ell)^2=j_h^2\ \ \ (=m_Q^2)  \\
\end{array}
\end{eqnarray} 

\noindent
which are linear in $\nu$

Generally, we can use two of conditions 1-3 to solve for the $t$- and 
$z$-components of the neutrino momentum while 4 and 5 give the $x$- and 
$y$-components. The remaining condition acts as a check to ensure that 
we have a consistent partitioning of the jets. For example if for a 
given partition of the jets in each event we solve $\{1',2',3,4\}$ then 
(because the equations are linear) we will have a unique determination 
of $\nu$ and therefore $|\nu|^2$ and $m_Q^2=|j_h|^2$. Only for the 
correct partitions of events will the reconstruction give 
$|\nu|^2\approx 0$. Thus if we plot all possible reconstructions on a 
scatter plot of $|\nu|^2$ versus $|j_h|^2$ and accept only those events 
in a strip near $|\nu|^2\approx 0$ we should find an accumulation of 
events near the real value of $m_Q$. Those events outside of the strip 
are presumably background or wrongly partitioned events. If there were 
data from two (or more) quarks mixed together in the sample, then there 
would be multiple peaks corresponding to the masses of each of the 
quarks present. Since all of the decay channels 1-5 can contribute to 
this sample, this method will ultimately determine all of the fourth 
generation masses regardless of which mixing scenario applies.

\begin{figure}[htb]
\centerline{
\includegraphics[angle=0, width=.38\textwidth]{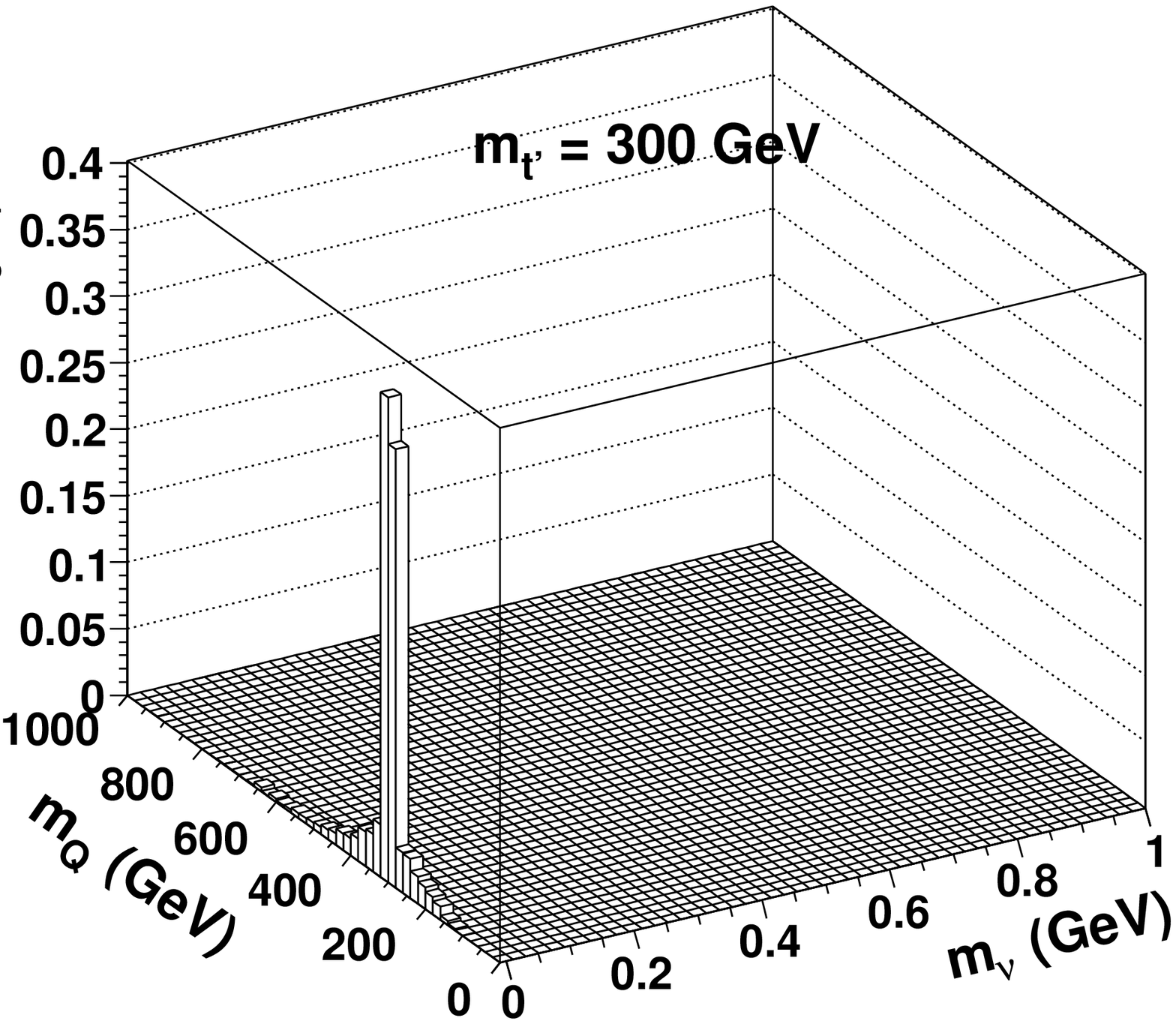}
\includegraphics[angle=0, width=.38\textwidth]{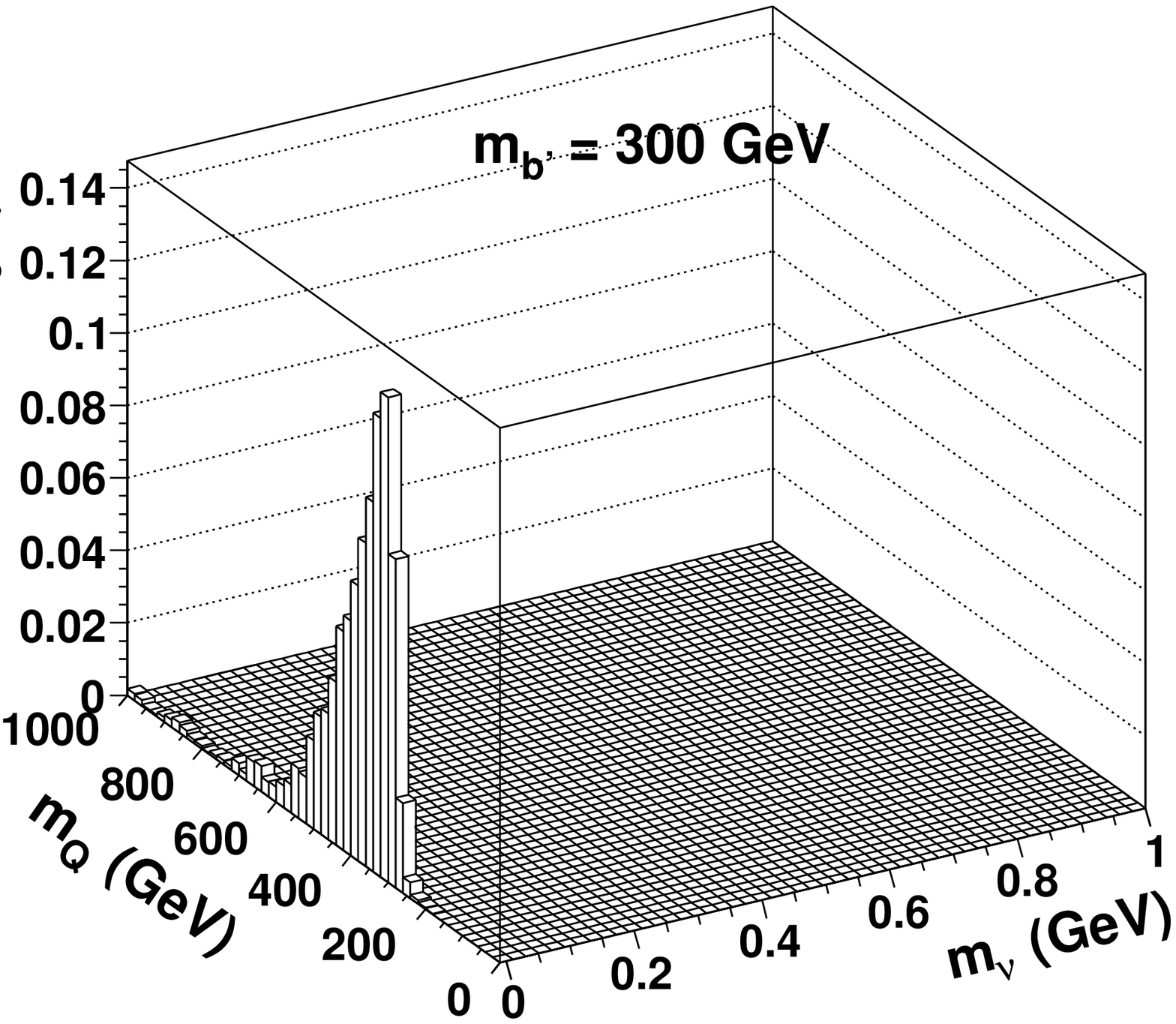}
}
\centerline{
\includegraphics[angle=0, width=.38\textwidth]{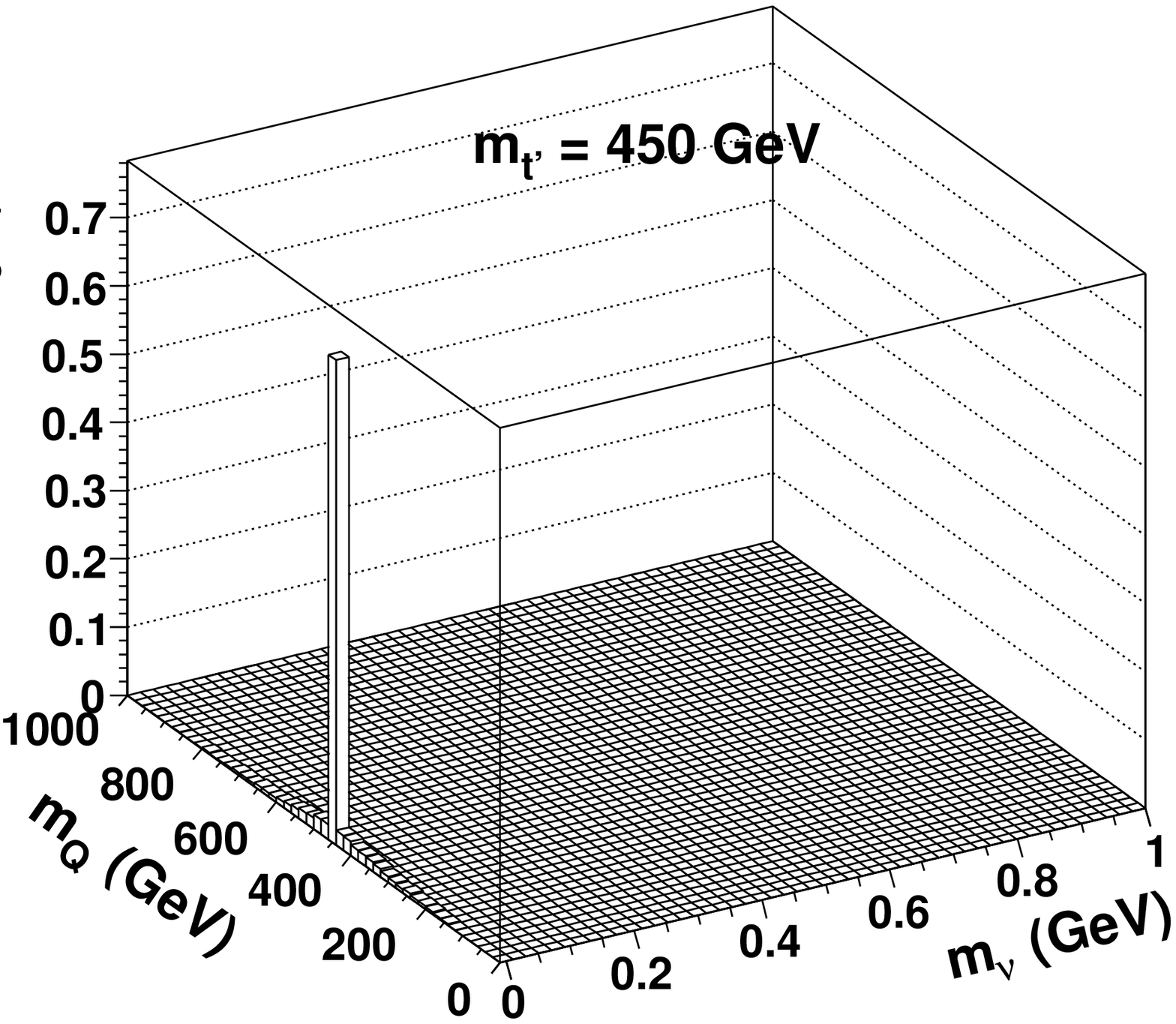}
\includegraphics[angle=0, width=.38\textwidth]{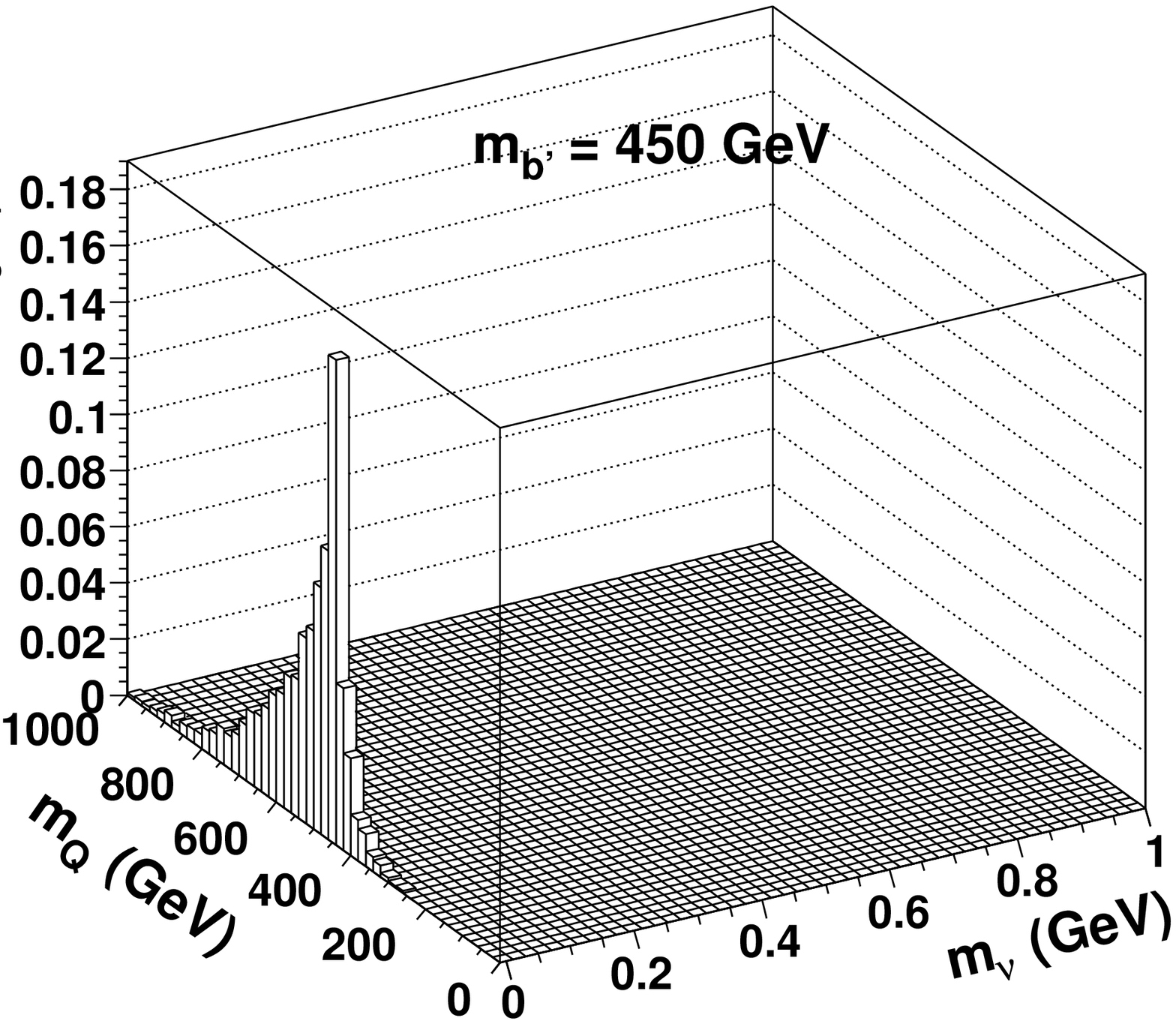}
}
\centerline{
\includegraphics[angle=0, width=.38\textwidth]{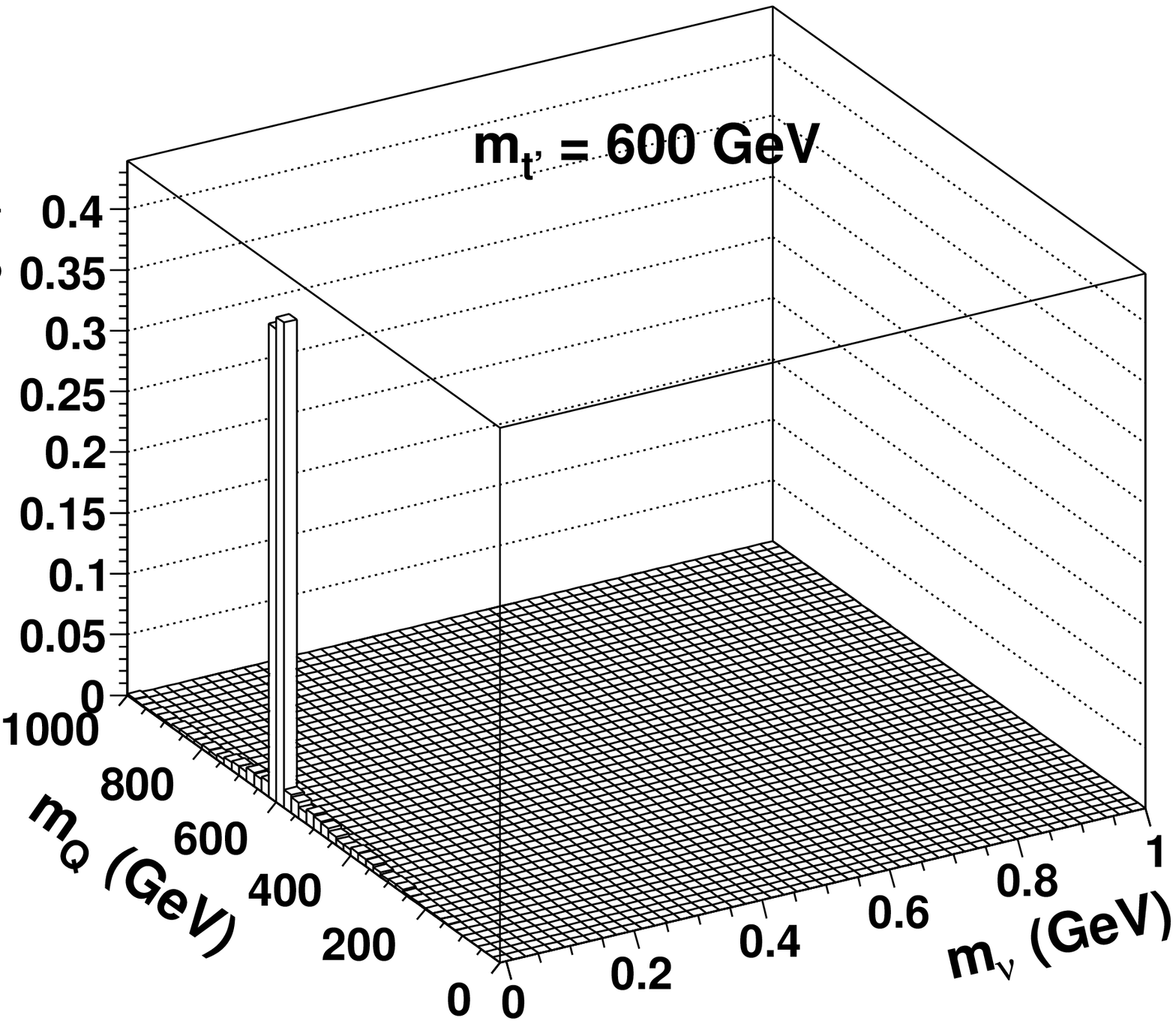}
\includegraphics[angle=0, width=.38\textwidth]{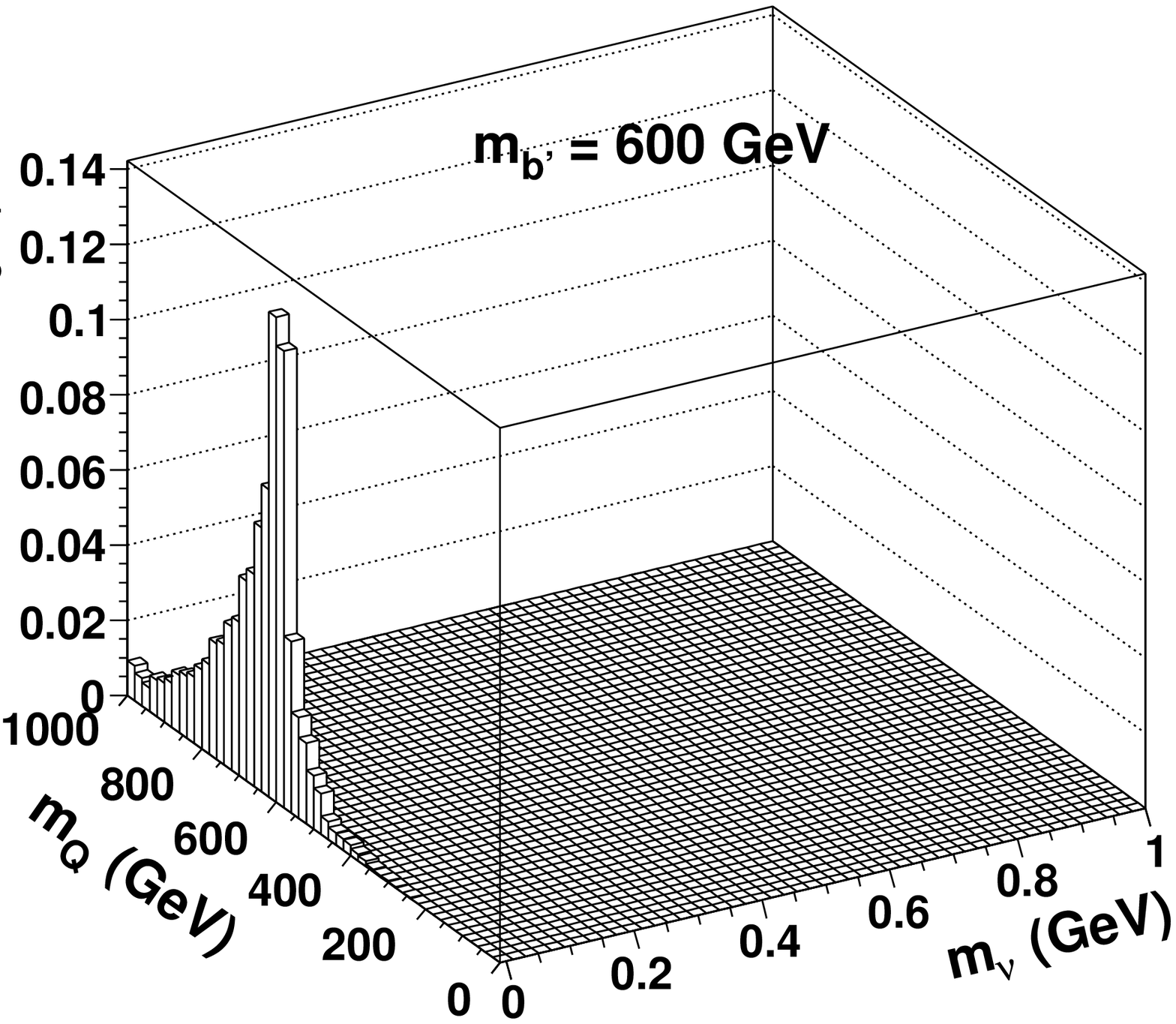}
}
\caption{2D histogram for reconstructed $m_\nu$ 
and  $m_Q$ from single lepton signal after selecting events. 
Left plot is for~$\tp$-quarks and right is~for~$\bp$-quarks.}
\label{fig:2dhistqn}
\end{figure}

In Figure \ref{fig:2dhistqn} we show Leggo plots of the reconstructed 
$m_Q$ versus reconstructed $m_\nu$ for a number of different scenarios. 
For the correct partitioning of the jets the events would indeed be at the 
physical values of $m_Q$ and $m_\nu$ but not so for the incorrect 
partitions.

Returning to Eqn.~\ref{one_lepton_constraints} we can also use two 
slightly different approaches to reconstructing the quark mass which may 
offer some advantages. If we start with constrains $\{1,3,4,5\}$ then on 
each event we extract the apparent masses of each side of the event: 
$m_{Q1}^2=(\ell+\nu+j_\ell)^2$ and $m_{Q2}^2=j_h^2$. We then check 
condition 2 by constructing an $m_{Q_1}^2$ versus $m_{Q_2}^2$ scatter 
plot. The correct partitions will be near the diagonal $m_{Q2}^2\approx 
m_{Q1}^2$, implementing condition 2, and the quark mass or masses will 
be revealed as accumulations in $\overline m_Q=\frac12 (m_{Q1}^2+ 
m_{Q_2}^2)$. In this method, for each event the neutrino momentum is 
determined independently of the partitioning of the jets. However, since 
the equations are quadratic, there is a two fold ambiguity in the 
solution so on the scatter plot two points must be plotted for each 
partitioning of each event increasing the combinatorial background.

\begin{figure}[htb]
\centerline{
\includegraphics[angle=0, width=.38\textwidth]{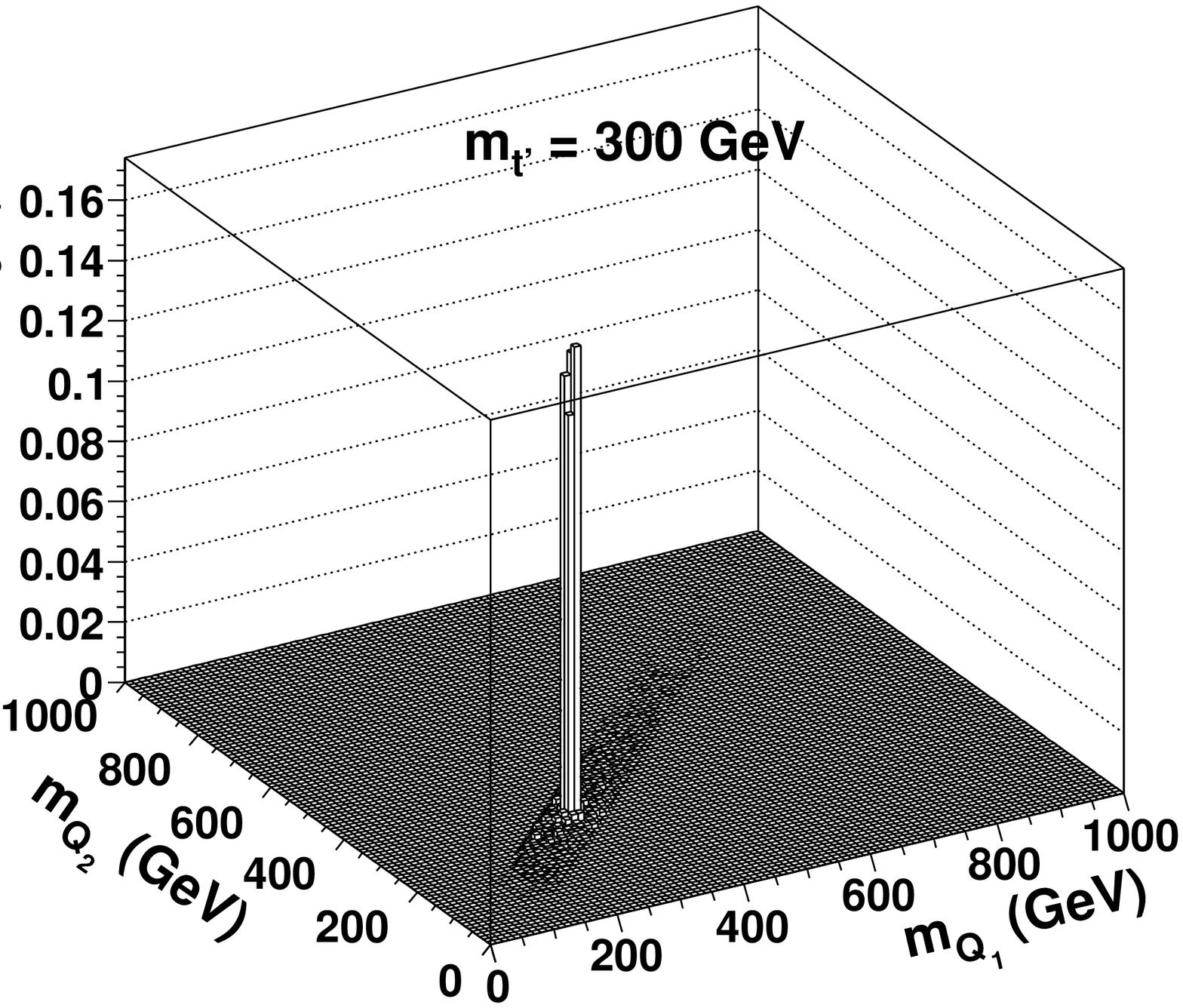}
\includegraphics[angle=0, width=.38\textwidth]{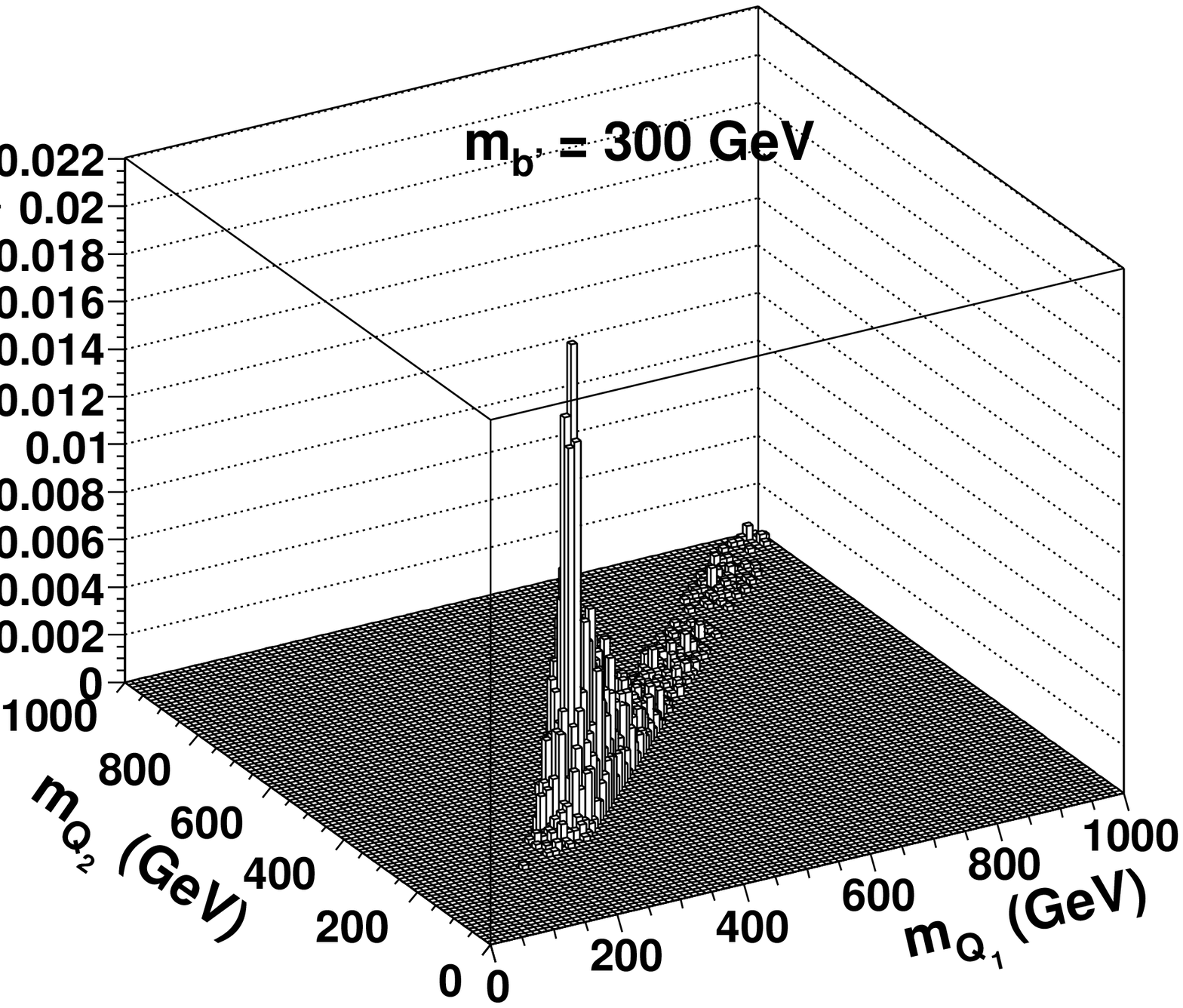}
}
\centerline{
\includegraphics[angle=0, width=.38\textwidth]{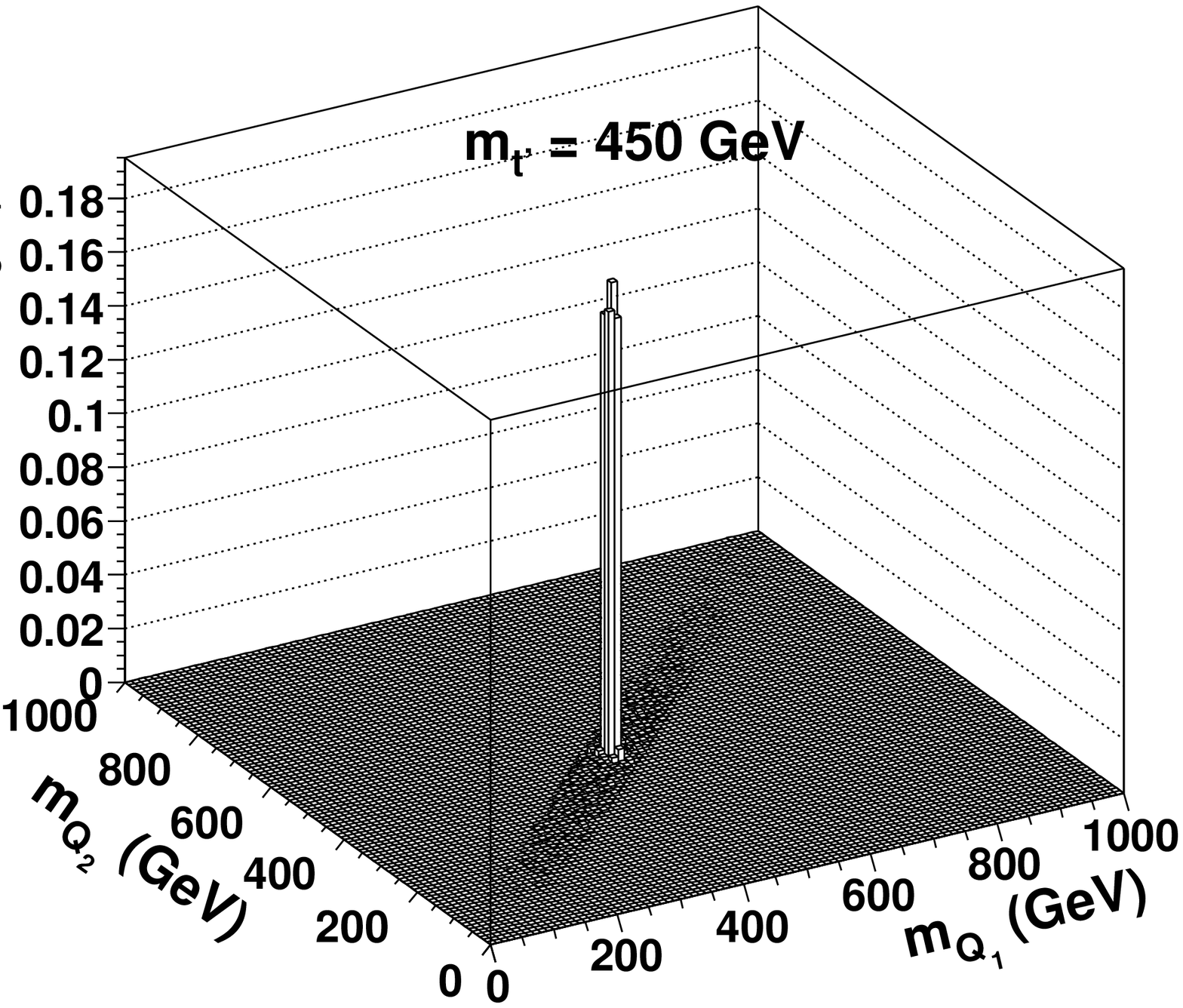}
\includegraphics[angle=0, width=.38\textwidth]{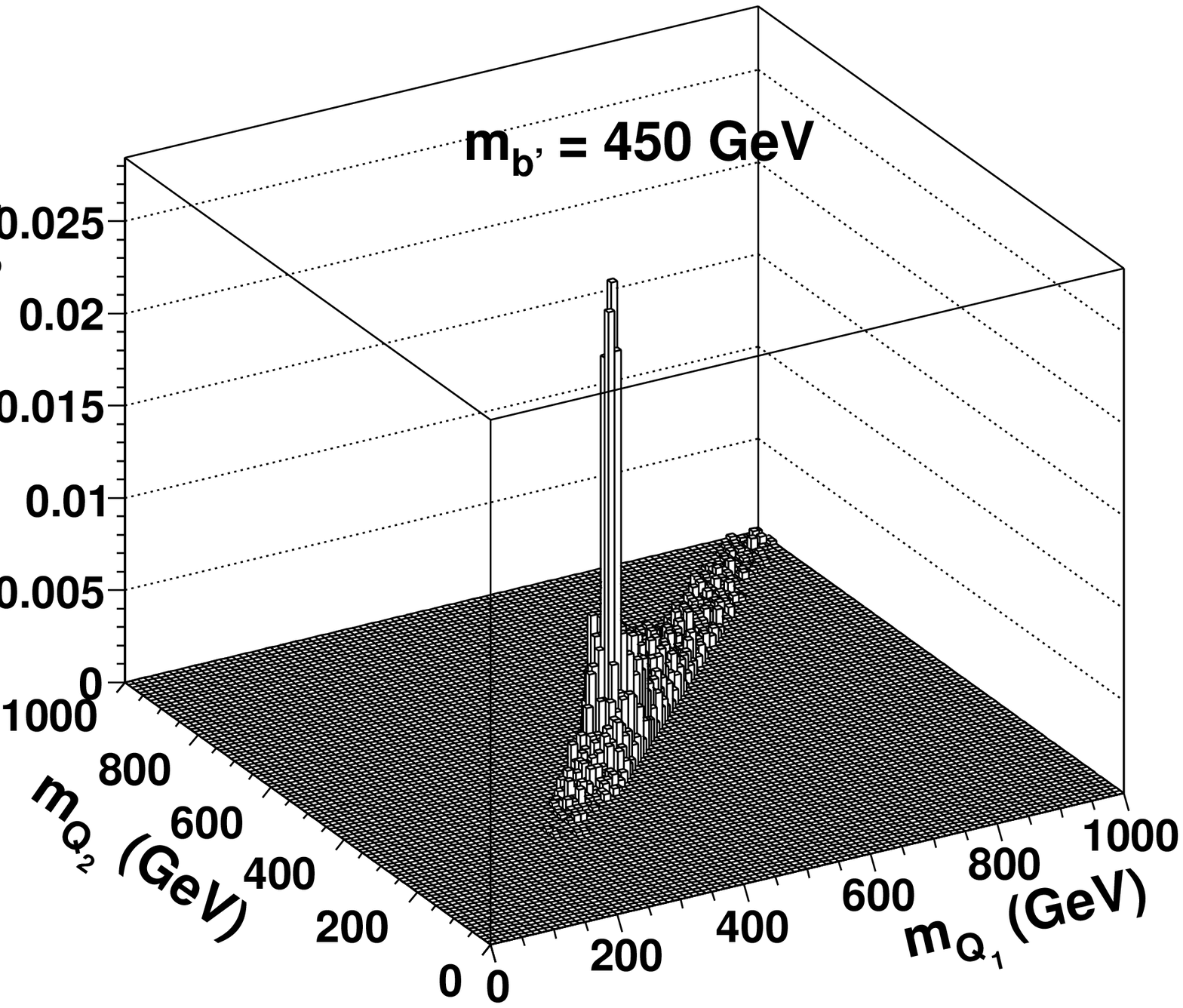}
}
\centerline{
\includegraphics[angle=0, width=.38\textwidth]{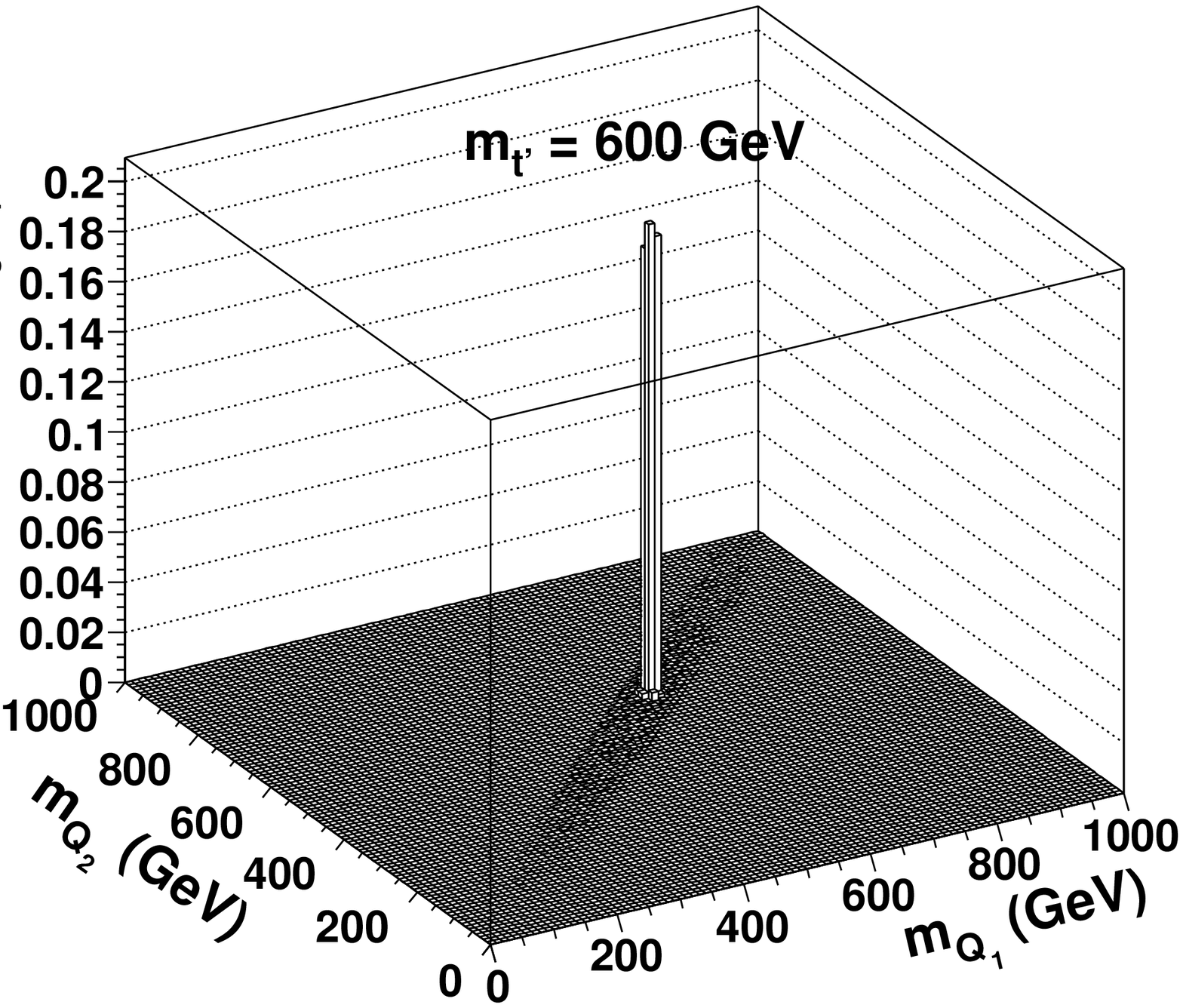}
\includegraphics[angle=0, width=.38\textwidth]{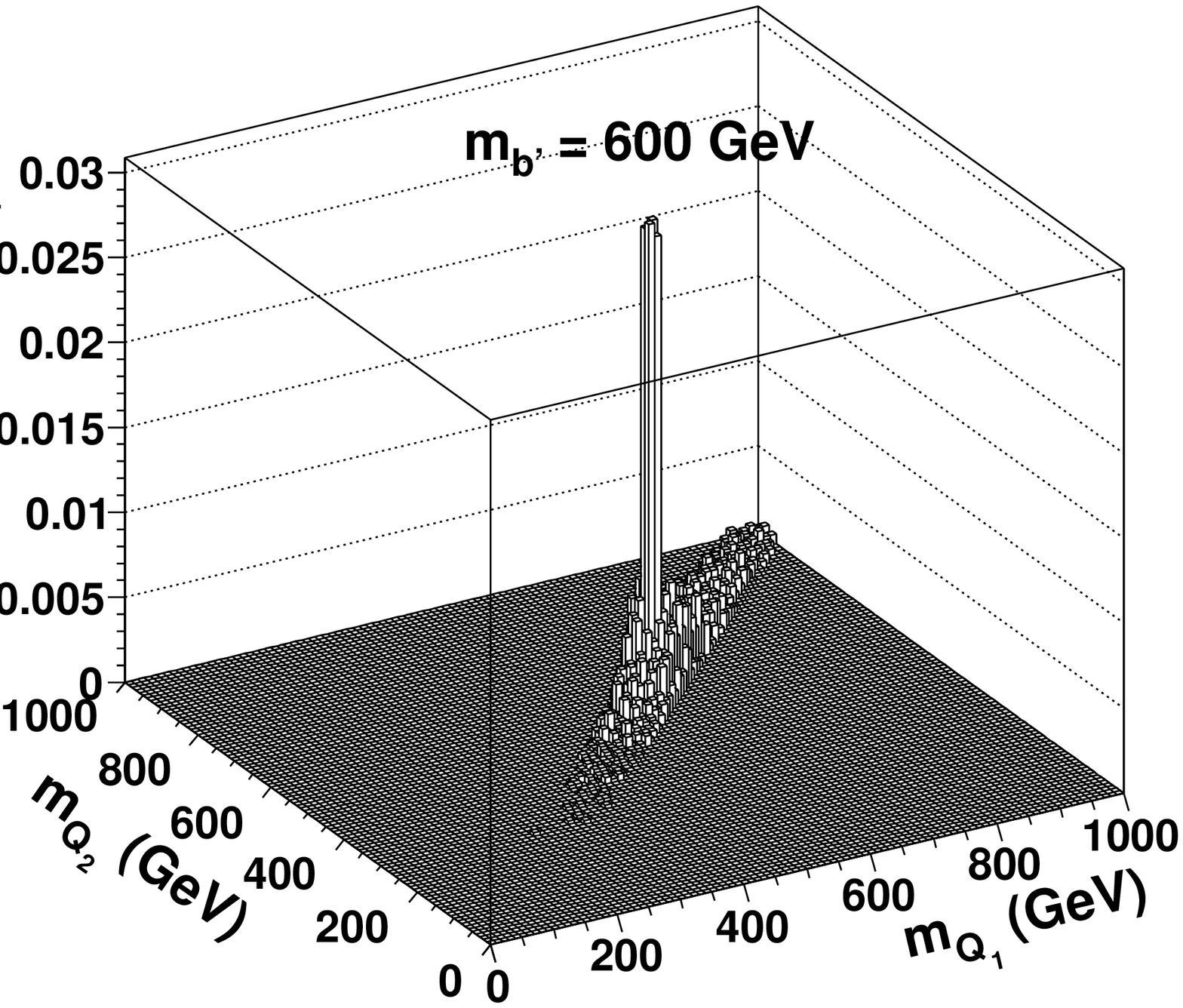}
}
\caption{2D histogram for reconstructed $m_{Q_h} (m_{Q_2})$ and  
$m_{Q_\ell} (m_{Q_1})$ from single lepton signal after selecting events. 
Left plot is for~$\tp$-quarks and right is~for~$\bp$-quarks. 
}
\label{fig:2dhistqq}
\end{figure}

In Figure \ref{fig:2dhistqq} we show a histogram of $m_{Q1}$ versus 
$m_{Q_2}$ for the $\tp$ and $\bp$ cases. Again the peaks at the correct 
value of $m_Q$ correspond to correct partitioning of the jets.

\begin{figure}[htb] 
\centerline{
\includegraphics[angle=0, width=.38\textwidth]{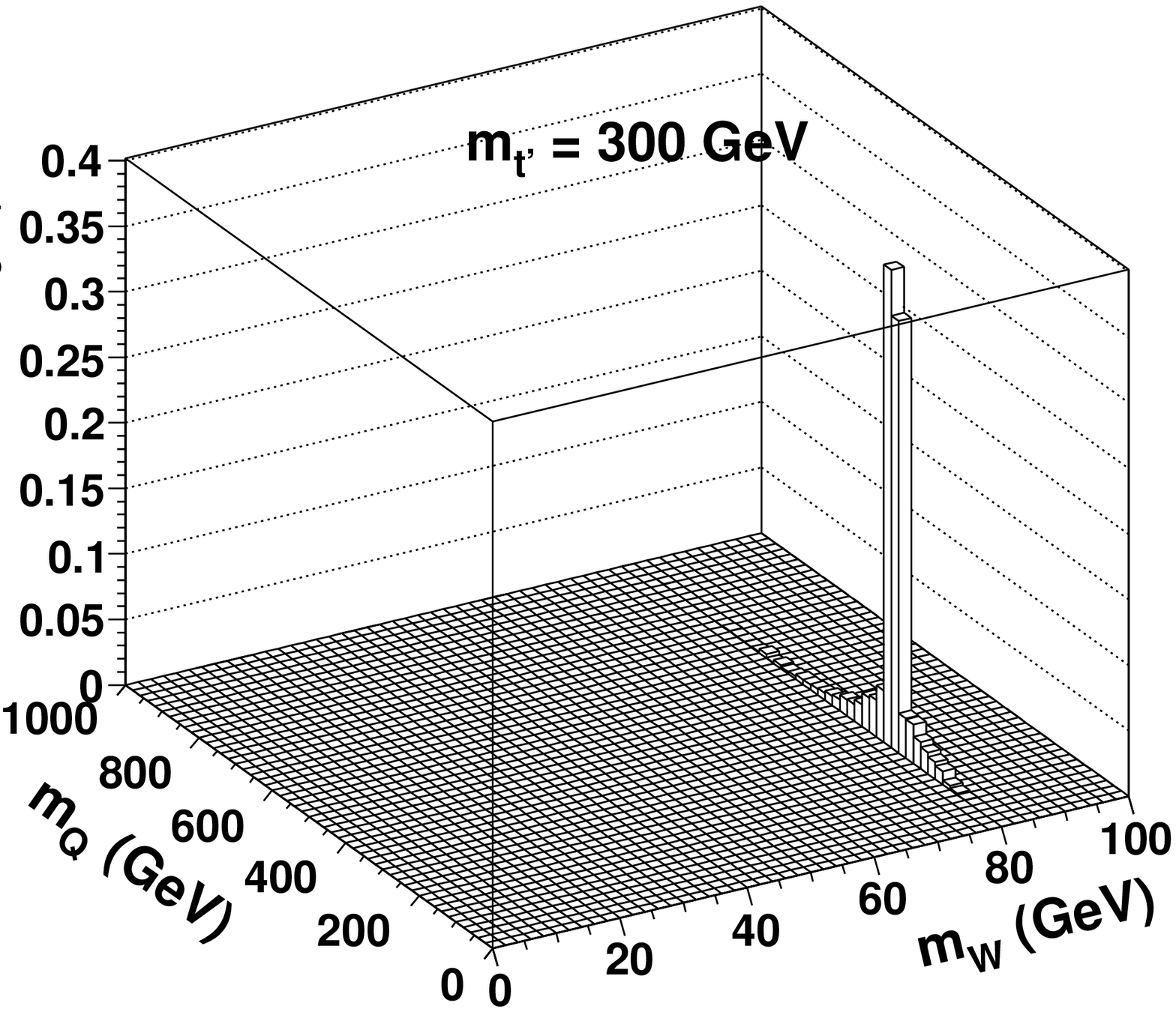}
\includegraphics[angle=0, width=.38\textwidth]{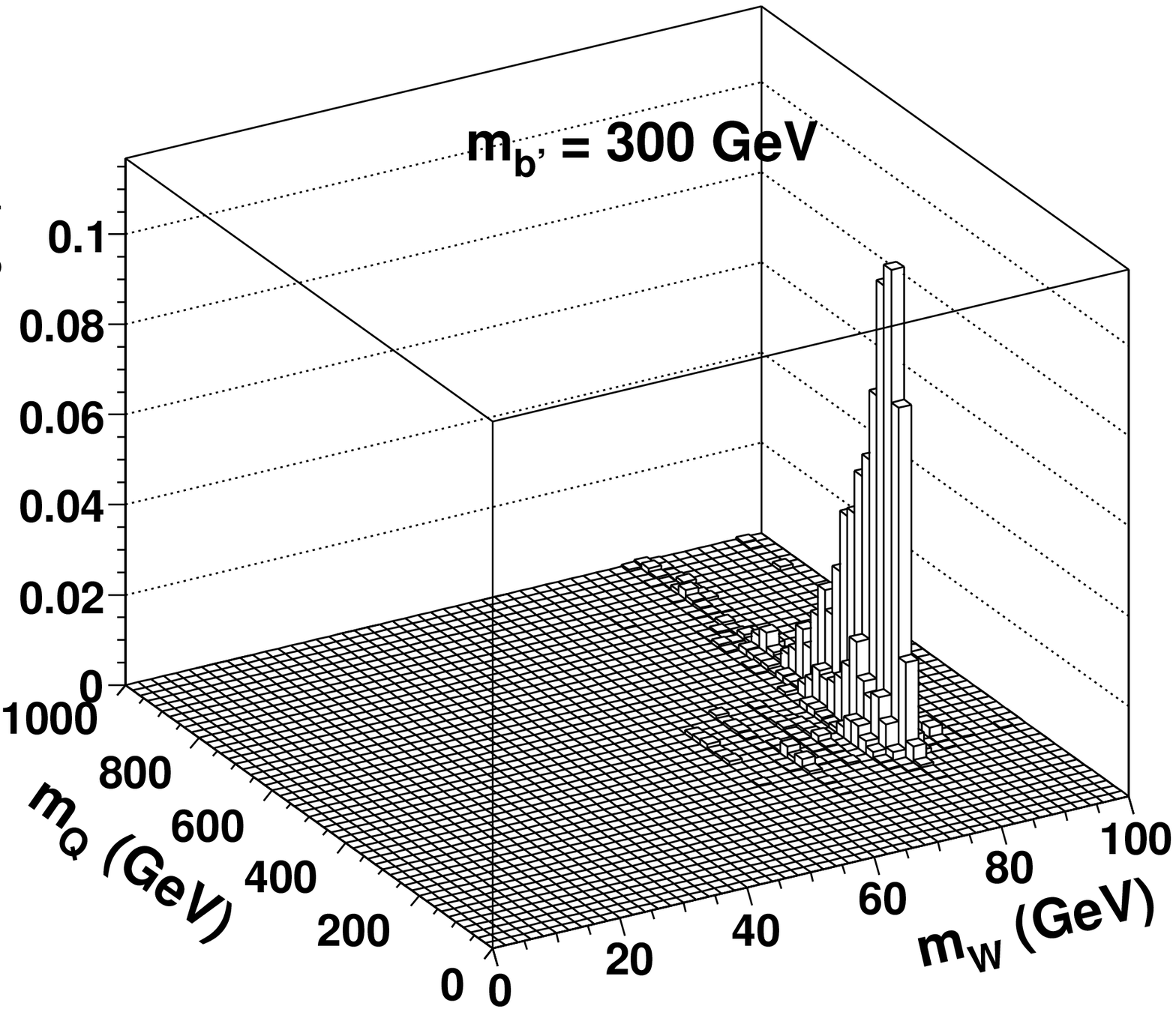}  
}
\centerline{
\includegraphics[angle=0, width=.38\textwidth]{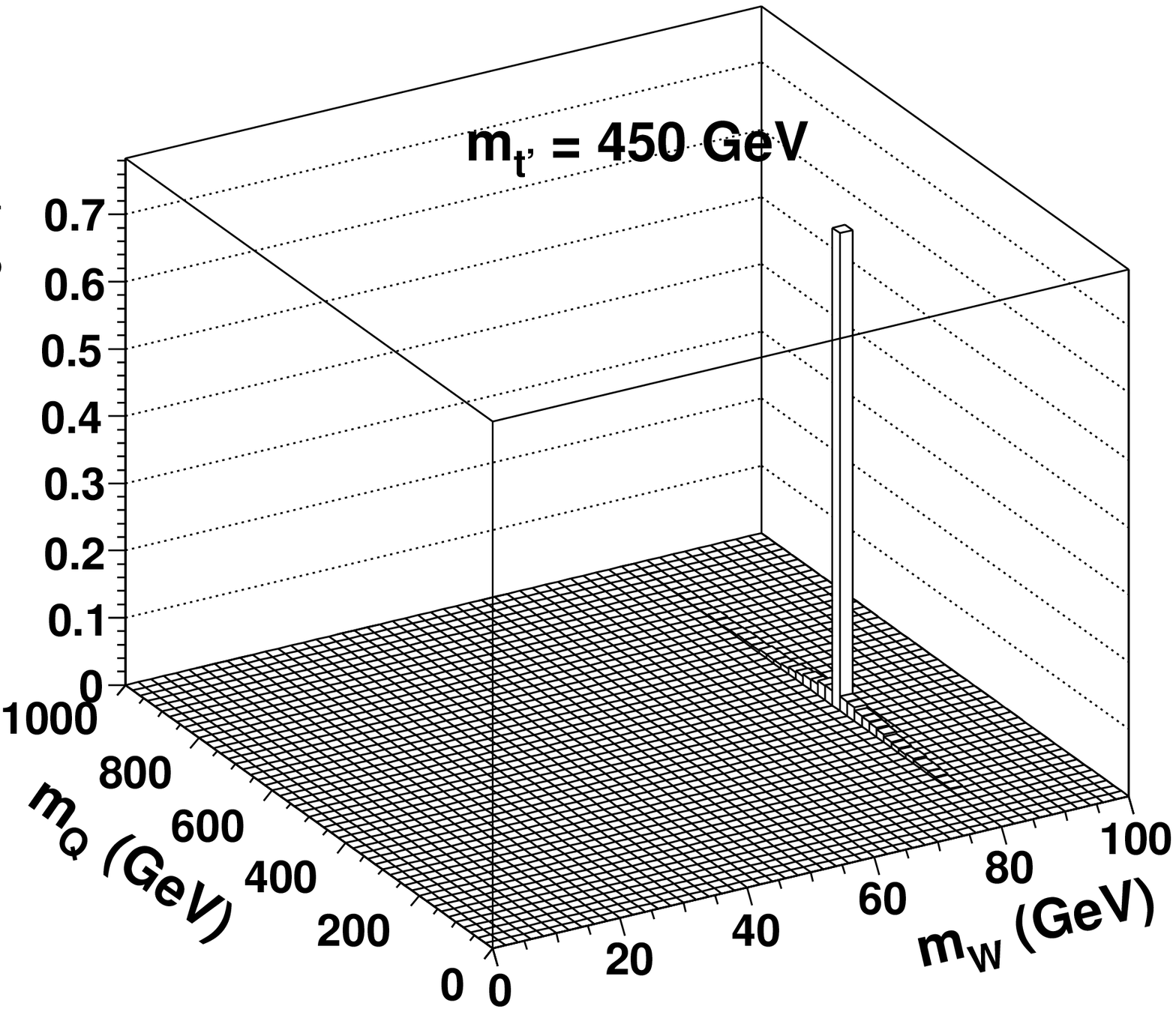}
\includegraphics[angle=0, width=.38\textwidth]{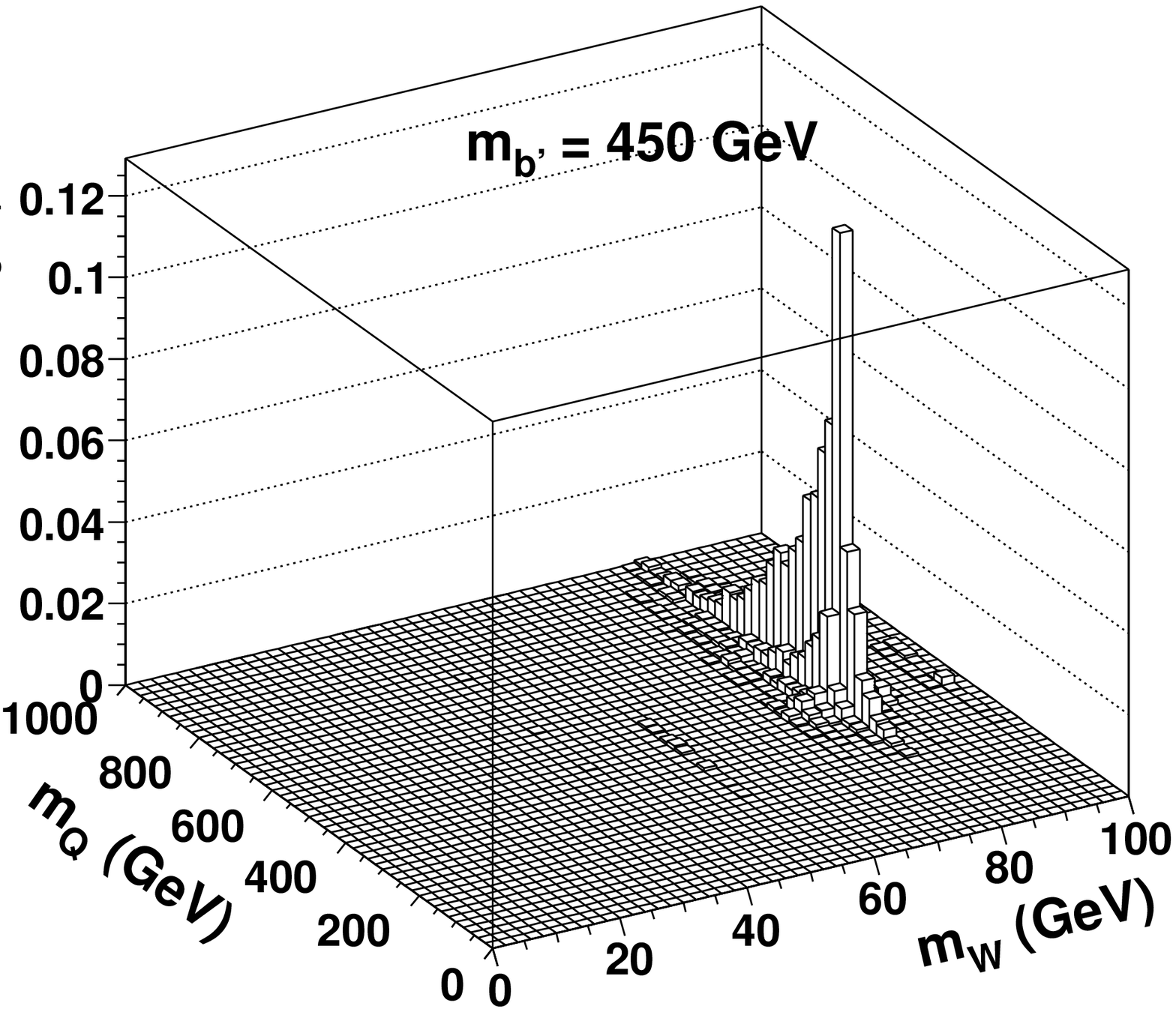}
}
\centerline{
\includegraphics[angle=0, width=.38\textwidth]{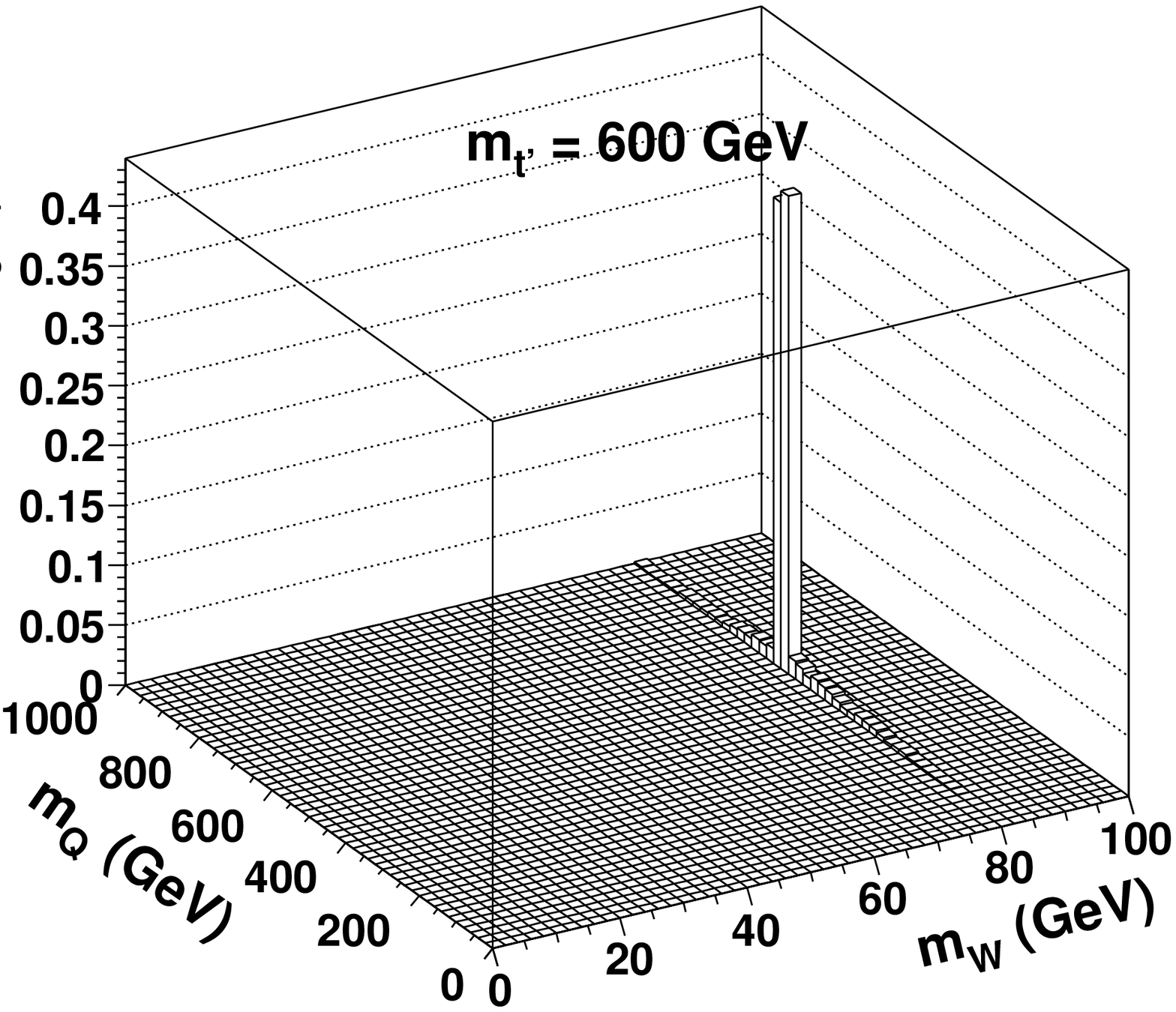}
\includegraphics[angle=0, width=.38\textwidth]{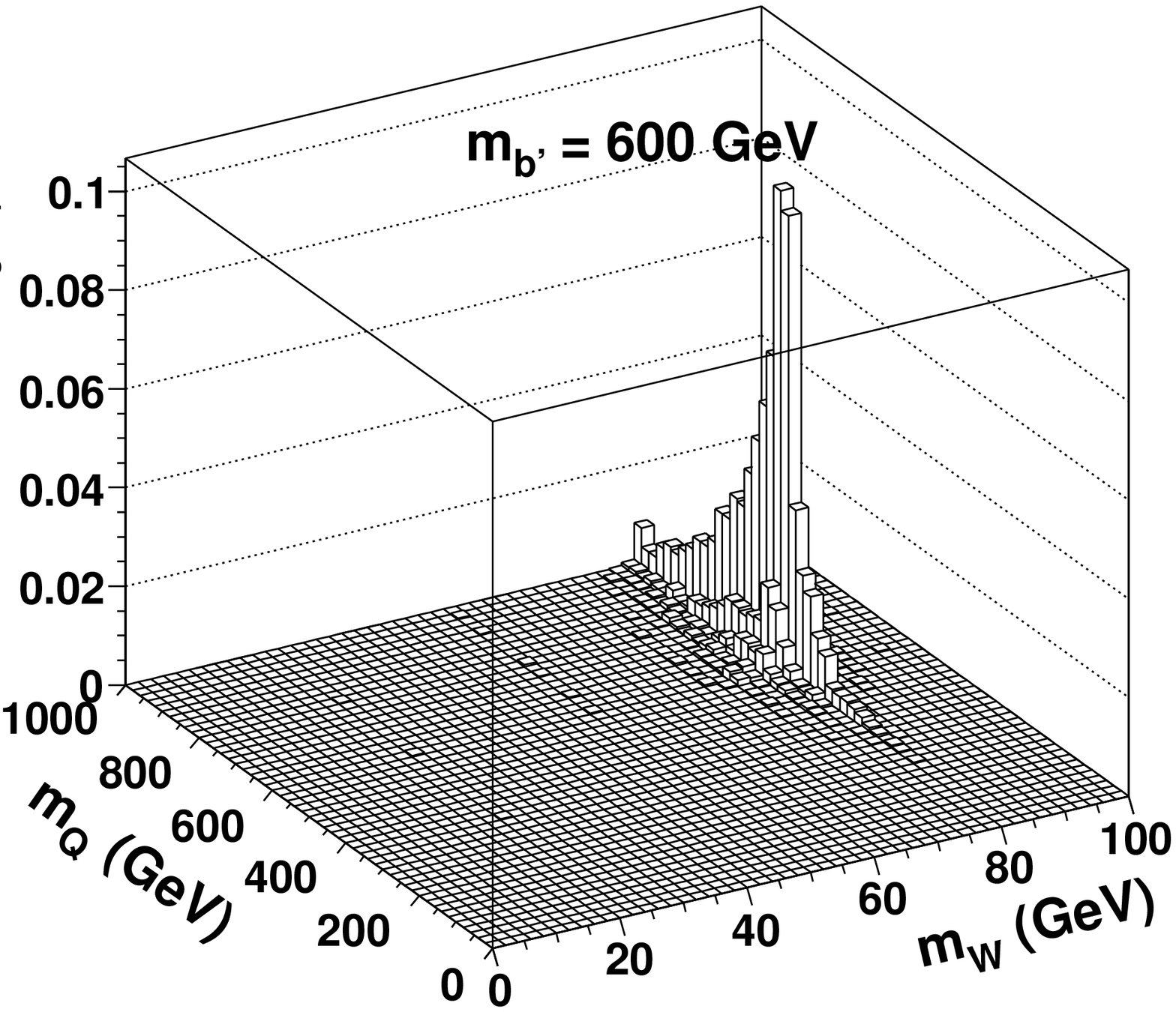}
}
\caption{2D histogram for 
reconstructed $m_W$ and  $m_Q$ from single lepton 
signal after selecting events. 
Left plot is for $\tp$ and right is for $\bp$. }
\label{fig:2dhistqw}
\end{figure}

In Figure \ref{fig:2dhistqw} we illustrate another approach which is to 
solve the equations $\{2,3,4,5\}$ and then reconstruct the $W$ mass as 
$(m_W^{recon.})^2=(\nu+\ell)^2$. As with the above method the equations 
here are quadratic giving an additional 2 fold ambiguity. In this 
approach the resulting scatter plot will be in $(m_W^{recon})^2$ and 
$j_h^2$ where the correct partition will be in a strip near 
$m_W^{recon}\approx m_W$. The heavy quark mass or masses can be 
extracted from accumulations in $j_h^2$. This approach has the advantage 
that if there were another (beyond the SM) particle playing the role of 
the $W$-boson in the $Q$ decays such as a charged Higgs, an additional W 
boson or a Kaluza-Klein excitation of the W-boson, this would be evident 
in additional accumulations of events in the $m_W^{recon}$ variable.

Conversely, using this method, if decay channels (4) or (5) are 
significant, the case where virtual W-boson decays to $\ell \nu$ will 
lead to additional points on the scatter plot with the correct value of 
$j_h^2$ but with $m_W^{recon}<m_Q-m_{Q'}<m_W$.

In enumerating the various jet partitions for a given event, it is 
useful to combine equations (1') and (2') which leads to the inequality:

\begin{eqnarray}
j_h^2\geq m_W^2+2j_\ell\cdot \ell+j_\ell^2
\end{eqnarray}

\noindent
where this inequality will eliminate at least half of the possible 
partitions.

Let us now see how to use such kinematics to enhance the signal to 
background. For this, we will start with the method where we solve 
1,3,4,5 to determine $m_{Q1}$ and $m_{Q2}$. For a given partition of the 
jets, we define $\Delta m_Q=|m_{Q1}-m_{Q2}|$. For a given event, we will 
select the reconstructed value of $m_Q$, $m_Q^{recon}$ to be the value 
of $\overline m_Q$ corresponding to the partition with the minimum value 
of $\Delta m_Q$.

Another cut which may be helpful in limiting the combinatorial 
background is to first pair up the jets in pairs with roughly the W 
mass. For example in the $\tp$ case there must exist one pair of jets 
which results from the decay of a W-boson and therefore should have an 
invariant mass of $m_W$. Let us denote the deviation of such a jet pair 
from $m_W$ by, $\Delta m_W=m_W-m_{2jets}$, so using this cut, we would 
only consider partitions of jets where $|\Delta m_W|$ was smaller than 
some threshold. In the case of $\bp$ this cut is more constraining since 
it would apply to three different jet pairs in a given jet assignment.

\begin{figure}[htb]
\centerline{
\includegraphics[angle=0, width=.5\textwidth]{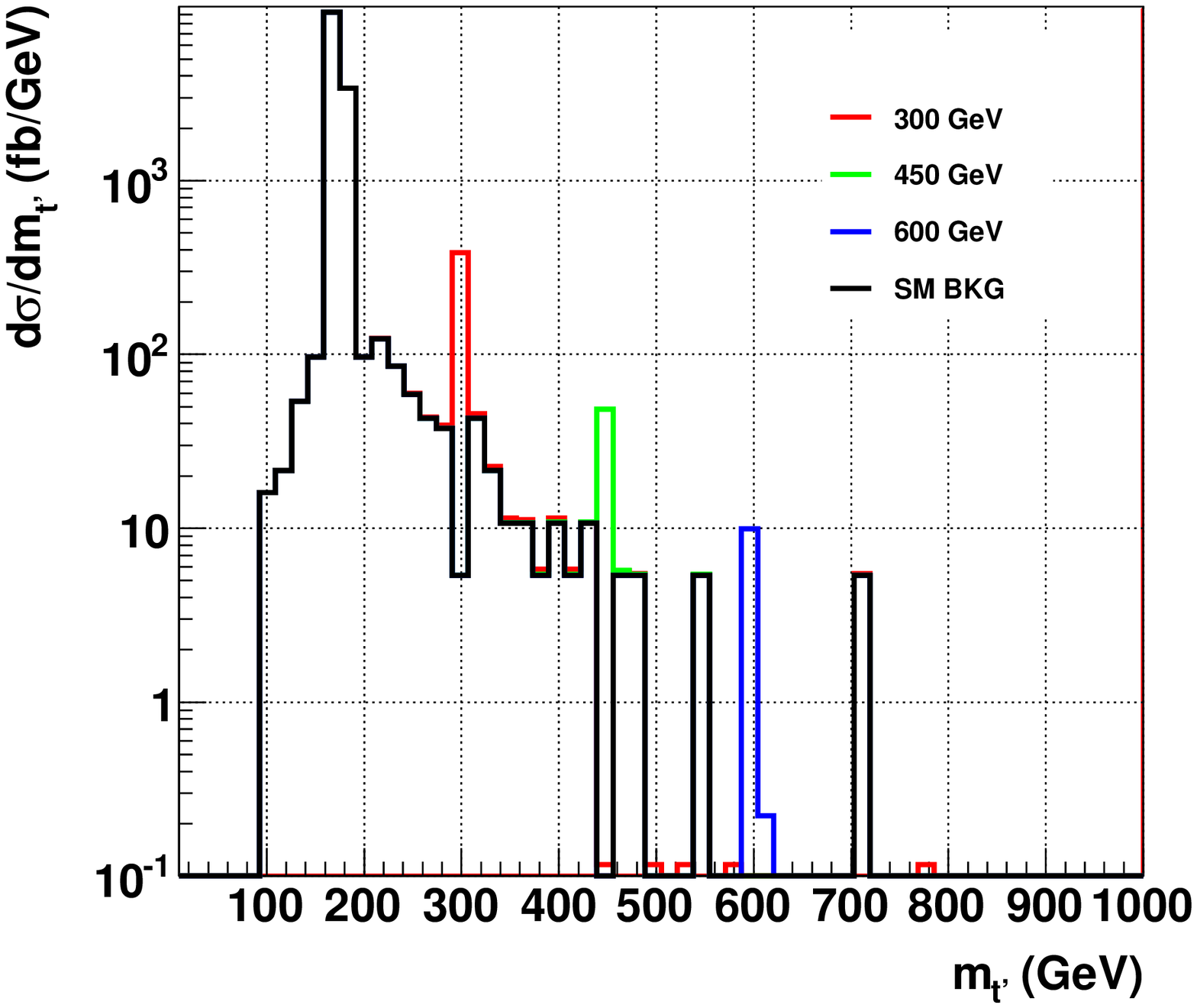}
\includegraphics[angle=0, width=.5\textwidth]{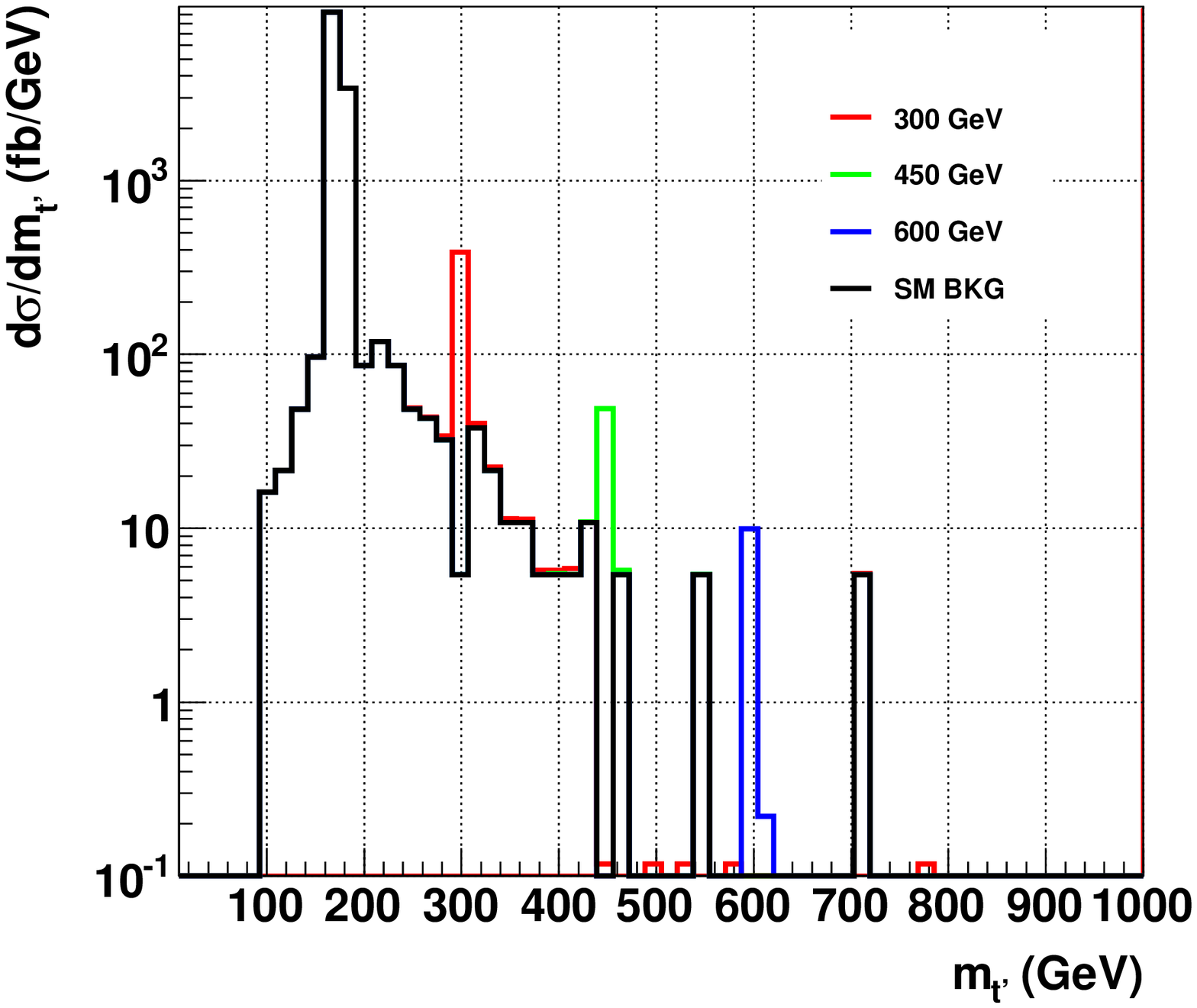}
}
\centerline{
\includegraphics[angle=0, width=.5\textwidth]{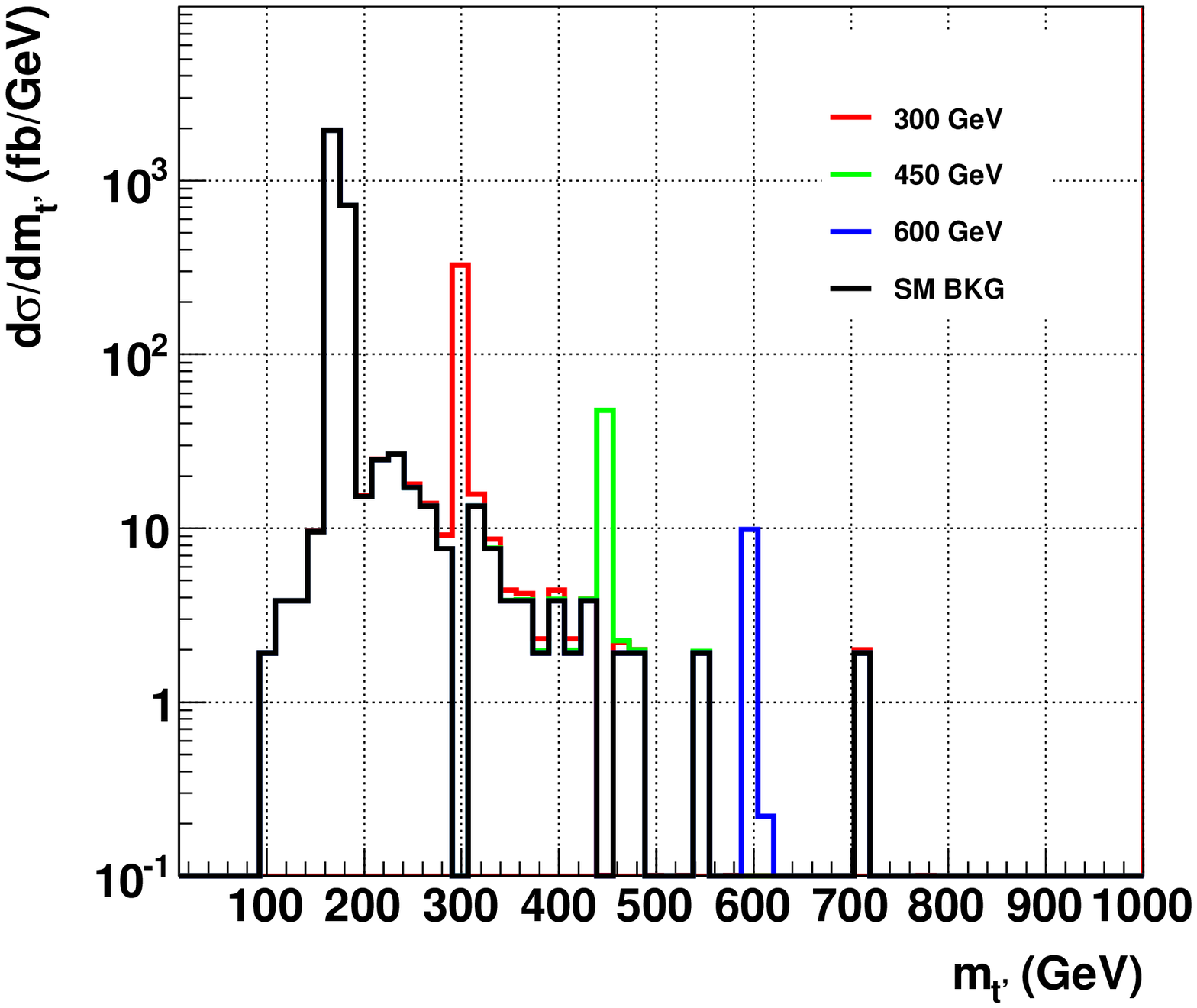}
\includegraphics[angle=0, width=.5\textwidth]{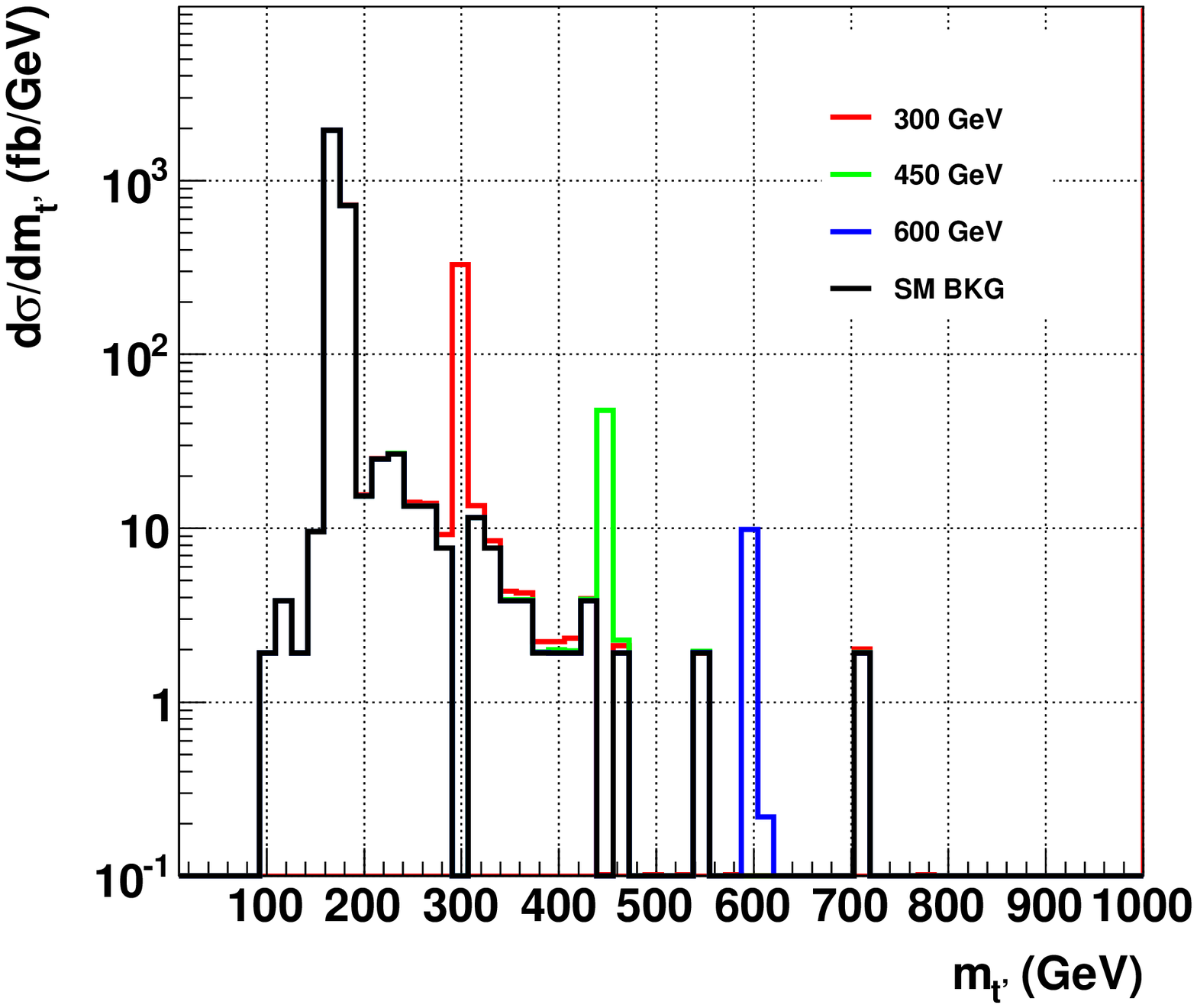}
}
\caption{
Reconstructed $t^\prime$ masses from single lepton case. Signal plots 
also include the SM background. Upper two plots are with only basic cuts
while the lower two plots are with basic $+ H_T$ 
(i.e. $H_T$=
the scalar sum of transverse 
momenta of the final state lepton, jets, and the missing 
transverse energy)
cut. In each case right
plots assume W reconstruction with $|\Delta m_W| < 15 $~GeV while the
left plots are without W reconstruction. In all the plots we choose only
that permutation in an individual event where the $|m_{Q_1} - m_{Q_2}|$
is minimum.}
\label{fig:recosd1a}
\end{figure}

In Figure~\ref{fig:recosd1a} we show a histogram of the reconstructed 
$\tp$ mass using this method. In the cut on the upper left we just use 
the basic cuts. In the two plots on the right we use the cut $\Delta 
m_W<15$~GeV while in the lower two plots we impose the $H_T$ cut. Thus 
the graph on the lower right has both cuts imposed. The plots are shown 
for the SM background and for $m_{\tp}=300$, $450$ and $600$~GeV. 
Clearly the signal peaks well above background and the $H_T$ cut appears 
helpful in enhancing this further.

\begin{figure}[htb]
\centerline{
\includegraphics[angle=0, width=.5\textwidth]{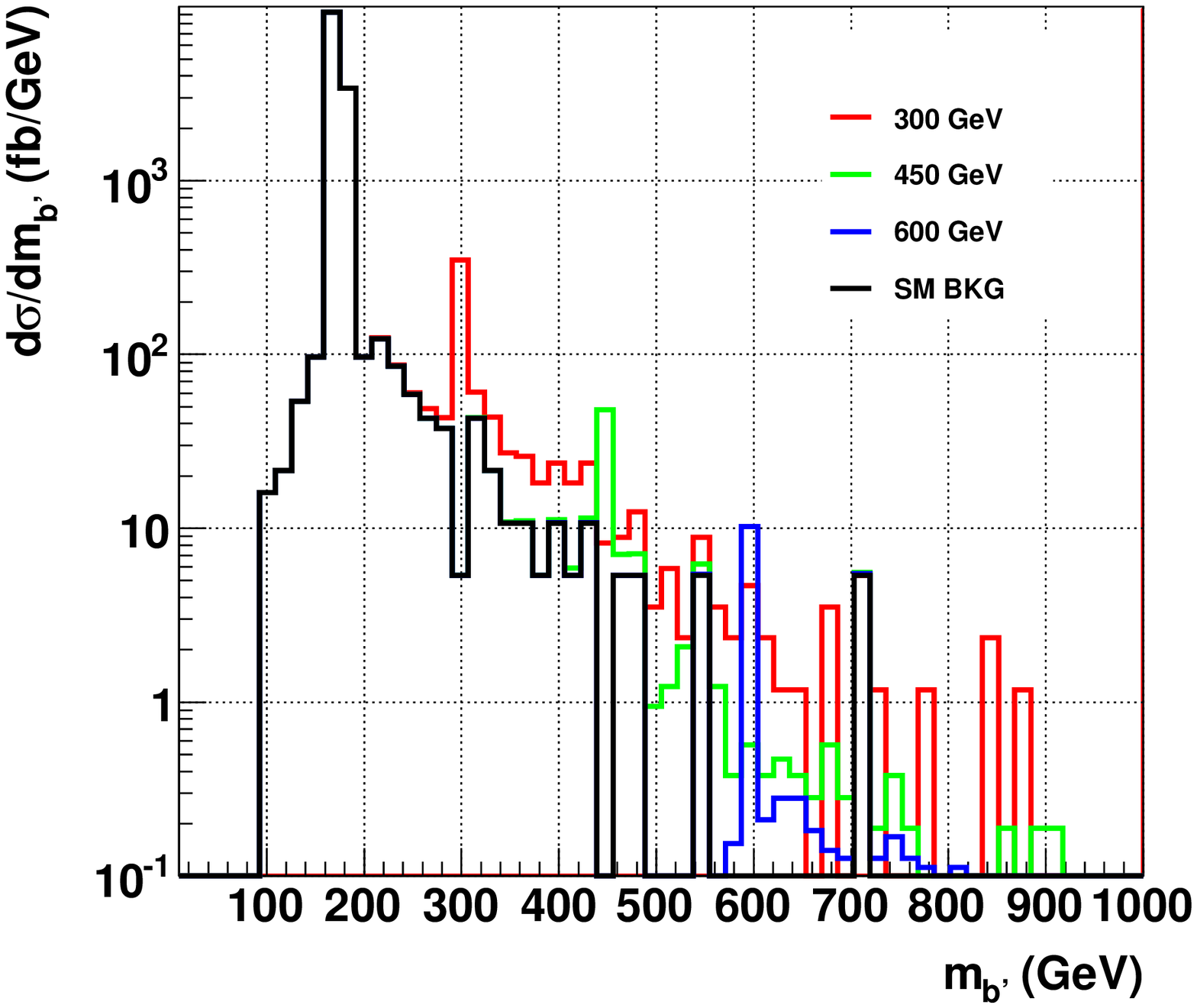}
\includegraphics[angle=0, width=.5\textwidth]{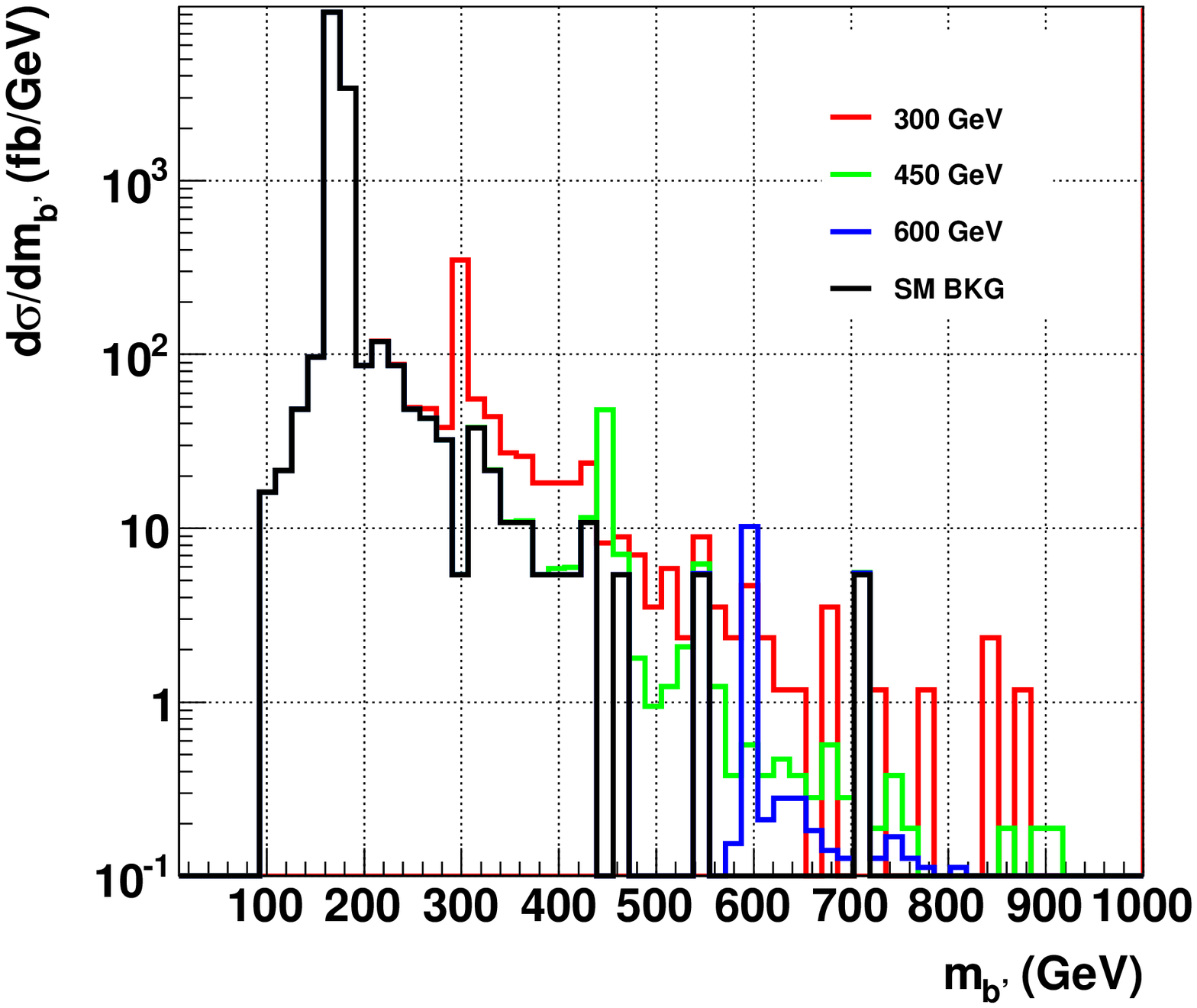}
}
\centerline{
\includegraphics[angle=0, width=.5\textwidth]{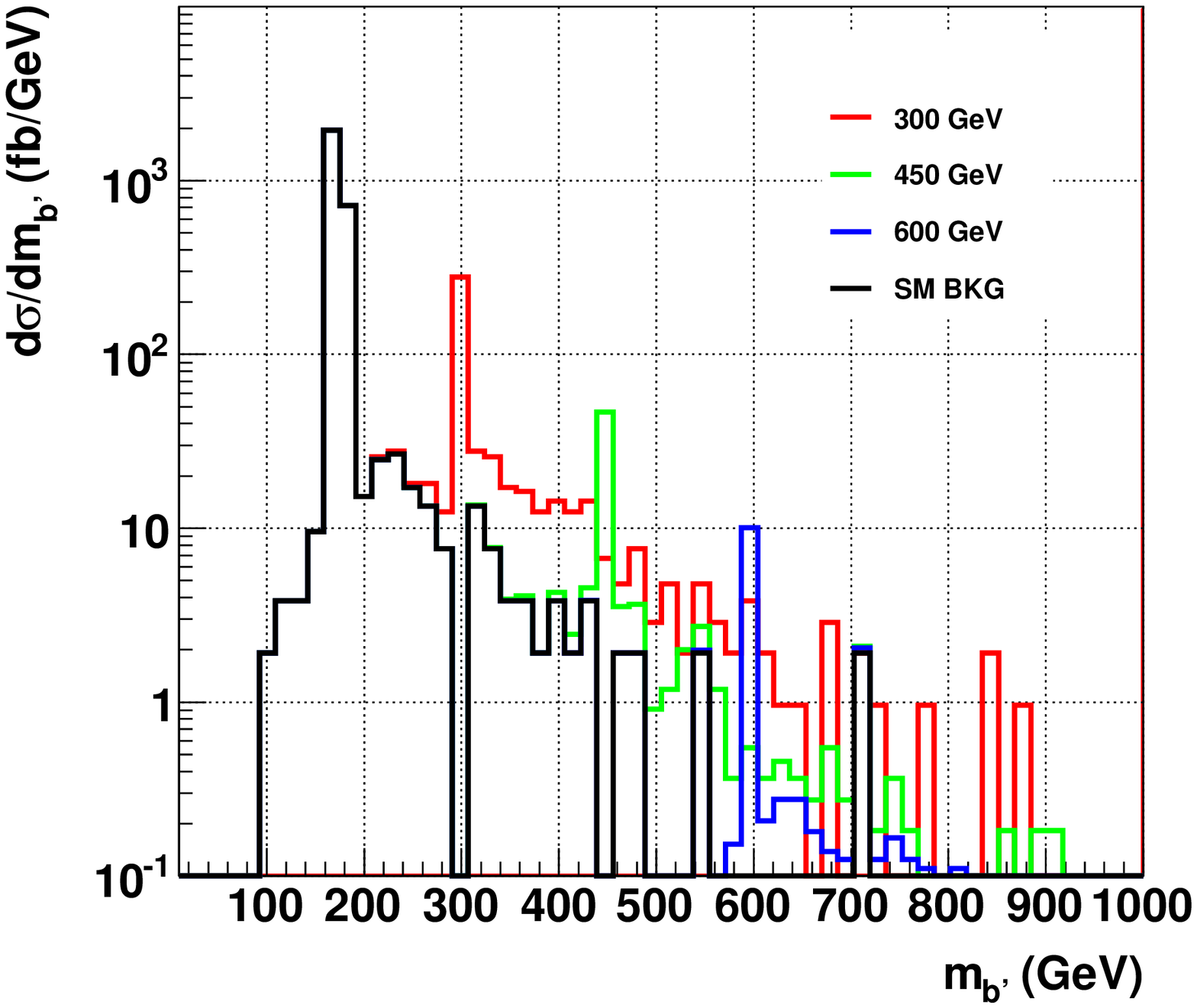}
\includegraphics[angle=0, width=.5\textwidth]{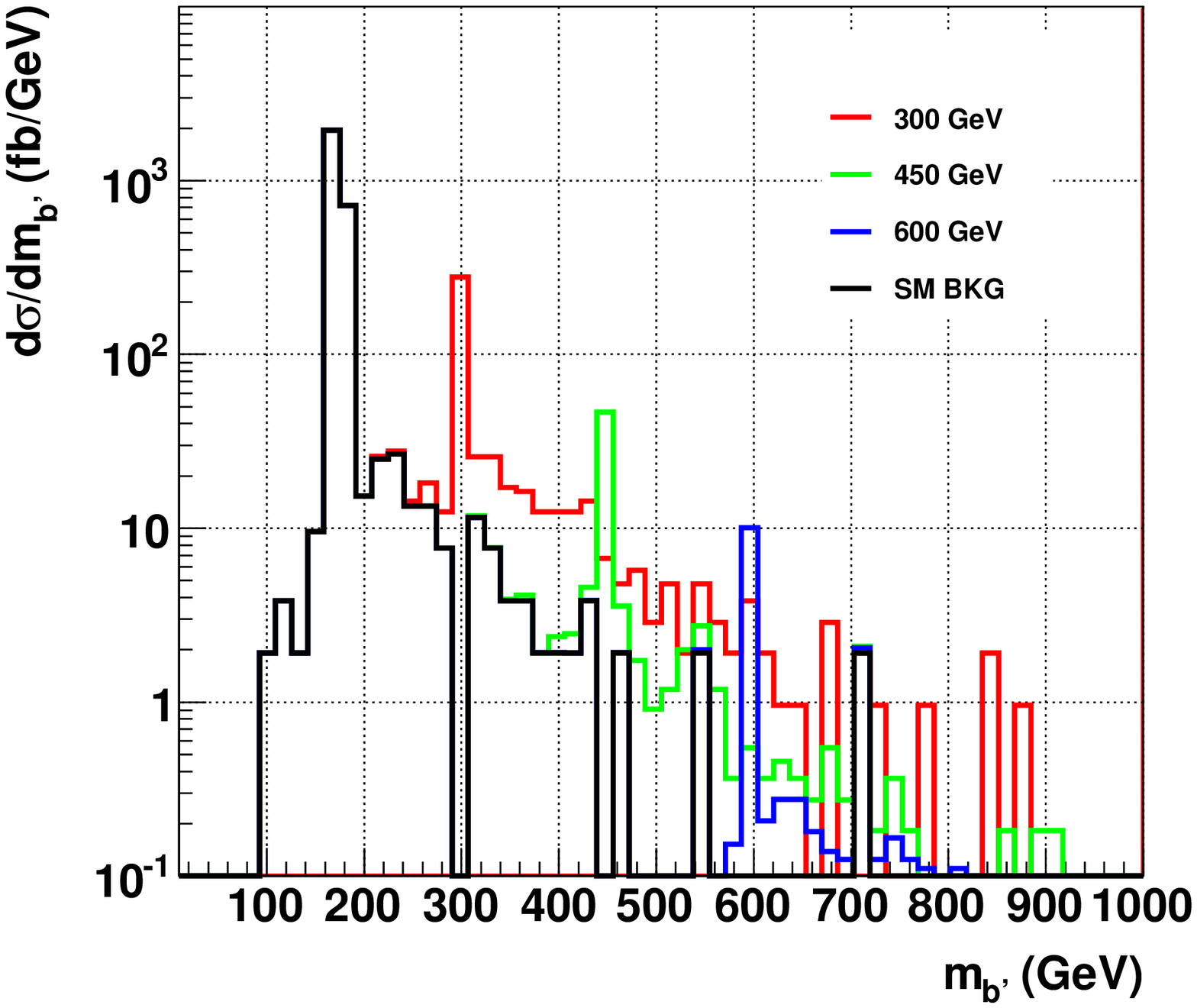}
}
\caption{
Reconstructed $b^\prime$ masses from single lepton case. Signal plots 
also include the SM background. Upper two plots are with only basic cuts
while the lower two plots are with basic $+ H_T$ 
(i.e. $H_T$=
the scalar sum of transverse 
momenta of the final state lepton, jets, and the missing 
transverse energy)
cut. In each case right
plots assume W reconstruction with $|\Delta m_W| < 15 $~GeV while the
left plots are without W reconstruction. In all the plots we choose only
that permutation in an individual event where the $|m_{Q_1} - m_{Q_2}|$
is minimum.}
\label{fig:recosd1b}
\end{figure}

\begin{figure}[htb]
\centerline{
\includegraphics[angle=0, width=.5\textwidth]{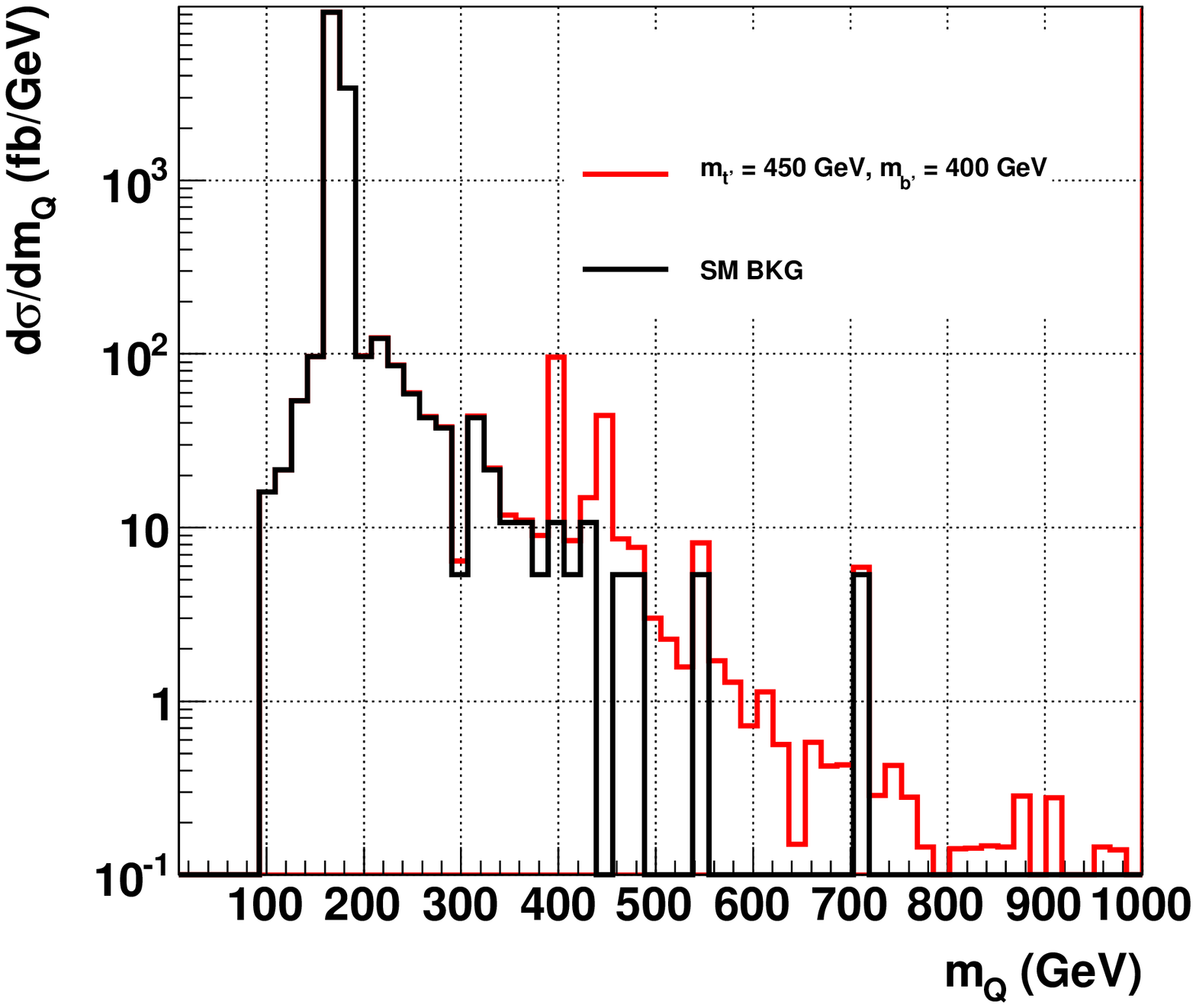}
\includegraphics[angle=0, width=.5\textwidth]{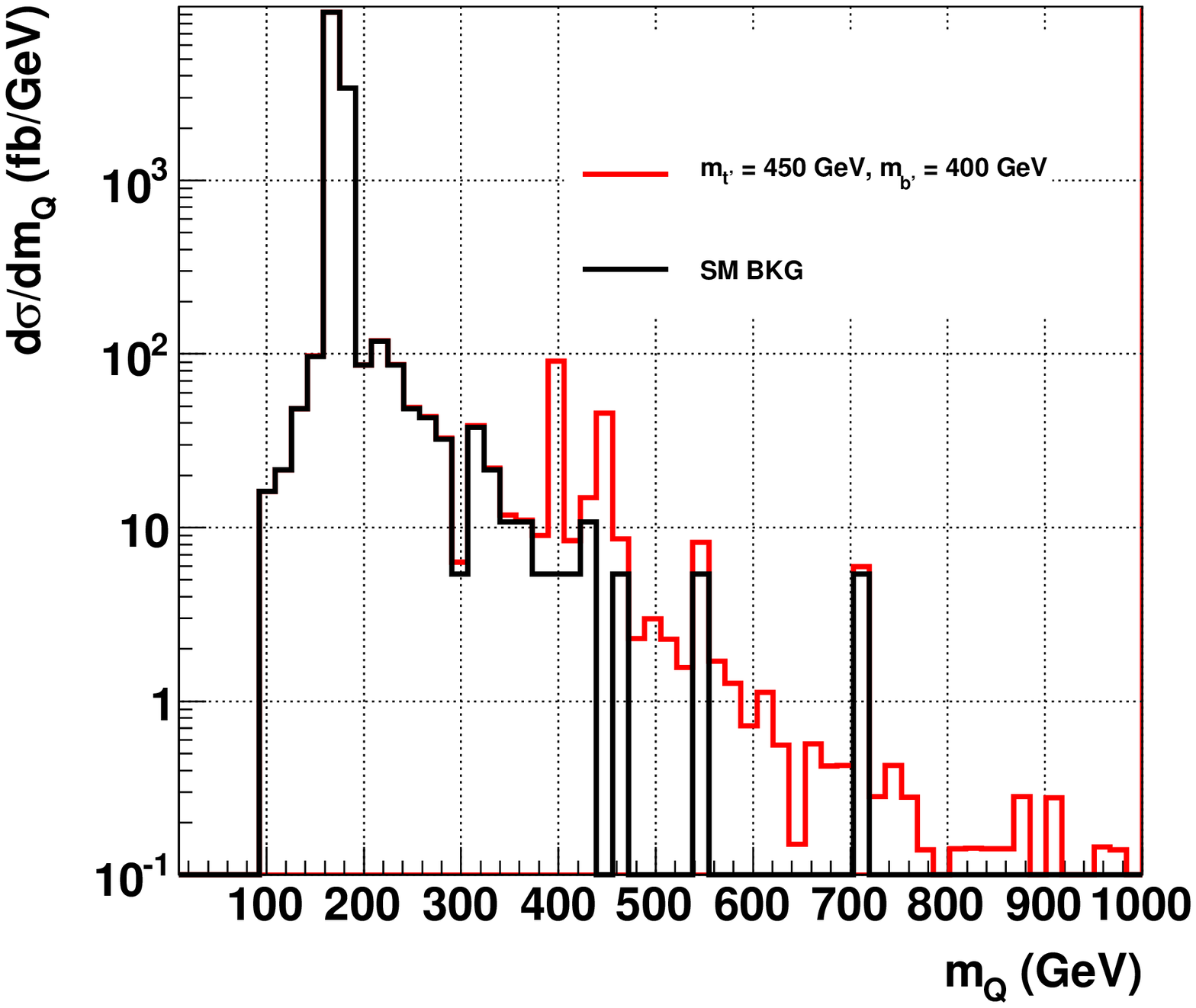}
}
\centerline{
\includegraphics[angle=0, width=.5\textwidth]{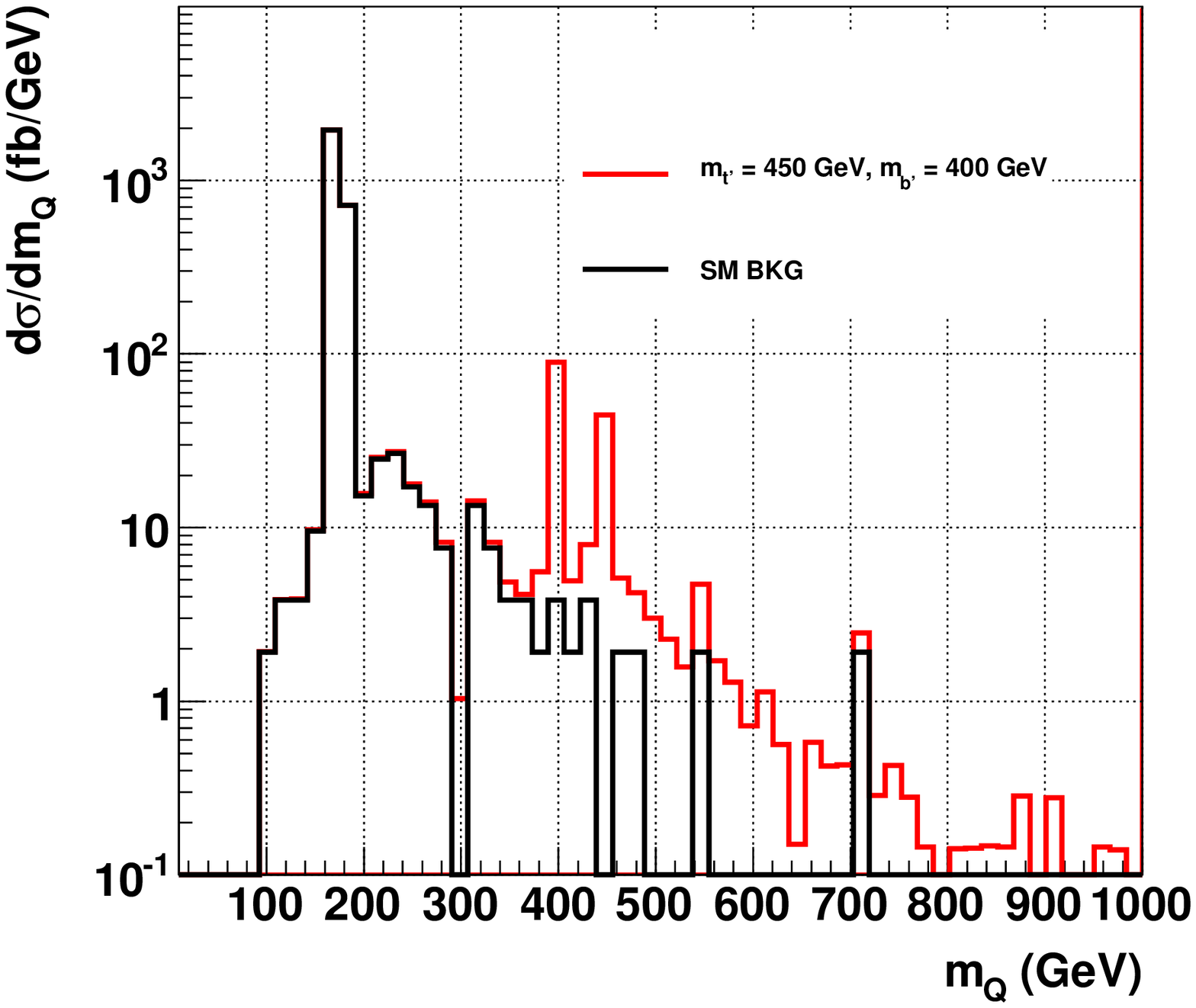}
\includegraphics[angle=0, width=.5\textwidth]{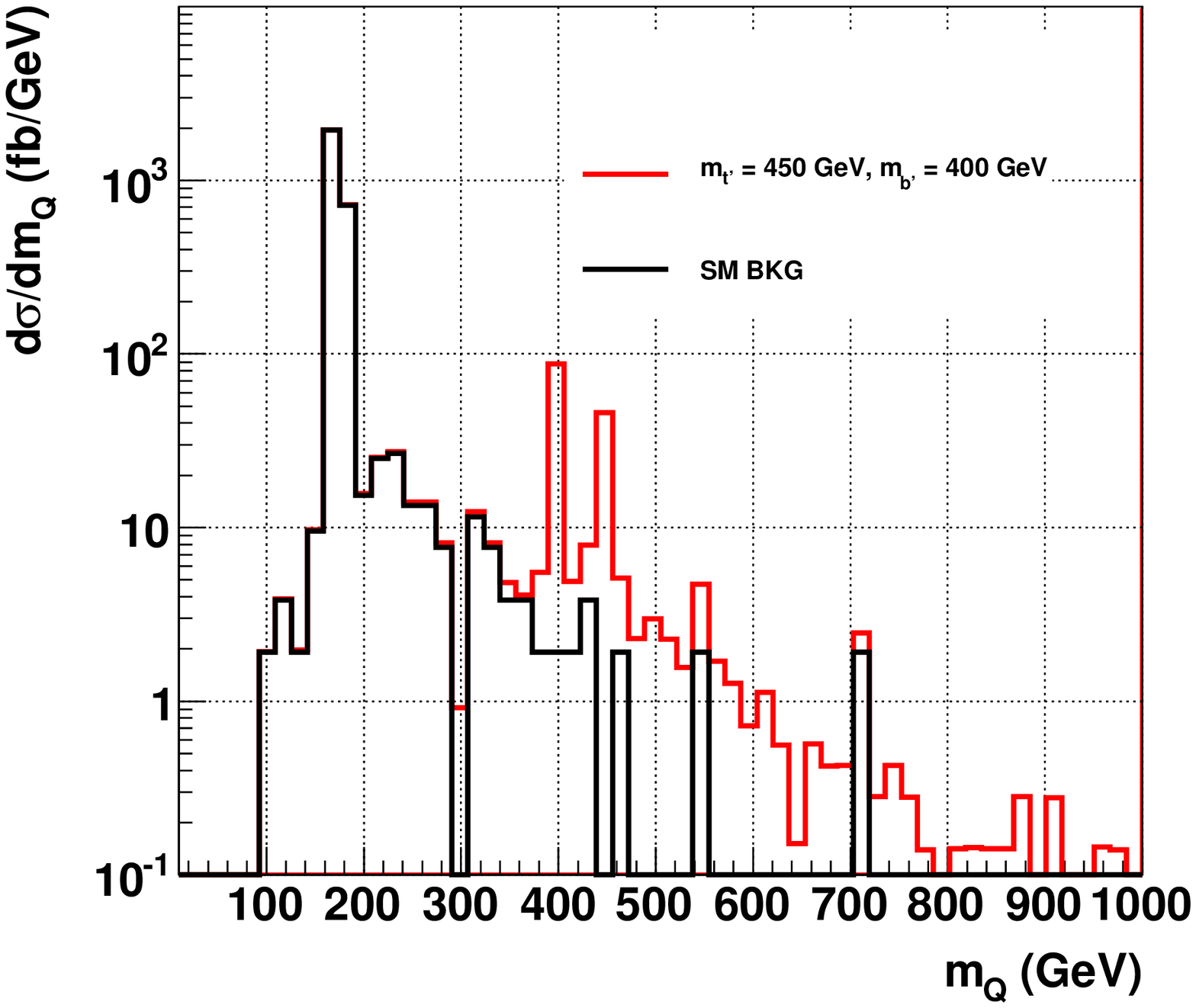}
}
\caption{
Reconstructed quark mass from the combined signals of a $\tp$- and 
$\bp$-quark in the single lepton case. 
Signal plots also include the SM background. Upper two plots are with
only basic cuts while the lower two plots are with basic $+ H_T$ 
(i.e. $H_T$=
the scalar sum of transverse 
momenta of the final state lepton, jets, and the missing 
transverse energy)
cut. In
each case right plots assume W reconstruction with $|\Delta m_W| < 15 
$~GeV while the left plots are without W reconstruction. $m_{t^\prime} =
450$~GeV and $m_{b^\prime} = 400$~GeV are assumed here. In all the plots
we choose only that permutation in an individual event where the
$|m_{Q_1} - m_{Q_2}|$ is minimum.}
\label{fig:recosd1c}
\end{figure}

In Figure~\ref{fig:recosd1b} we apply the same method to the case of 
$\bp$ and again the mass peak is well above the background and the 
signal is further enhanced by the $H_T$ cut. In 
Figure~\ref{fig:recosd1c} we consider the same method in the case where 
we consider the total signal where both species contribute. 
As an illustration here we are assuming that 
$m_\bp=450$~GeV and $m_\tp=500$~GeV. The close mass of the two quarks 
gives a signal which is larger than we would get if just one quark 
contributed
(see also~\cite{Holdom:2011uv}).

\subsection{Dilepton signals}

If two of the W-bosons in the decay chains we are considering decay 
leptonically there will consequentially be two leptons in the final 
state. If those two W-bosons are of opposite sign then the lepton pair 
will therefore be of opposite sign while if the two W-bosons are of the 
same sign then the pair will likewise be of the same sign. 
Due to the simplifying assumptions we are making,
this latter case only occurs in the decay chain of $\bp$-quark pairs. 
Below we will consider the problem of extracting the heavy quark mass 
from the kinematics in a dilepton signal while here we will consider the 
characteristics of the signal itself.

Hard leptons are often part of the signal of new physics. In this case, 
the dileptons would be produced in association with jets and missing 
momentum. To study this signal (for opposite sign and like sign), we 
selected Montecarlo events passing the basic cuts described above and 
also imposed the cut that $H_T>350$~GeV. 

\begin{figure}[htb]
\centerline{
\includegraphics[angle=0, width=.5\textwidth]{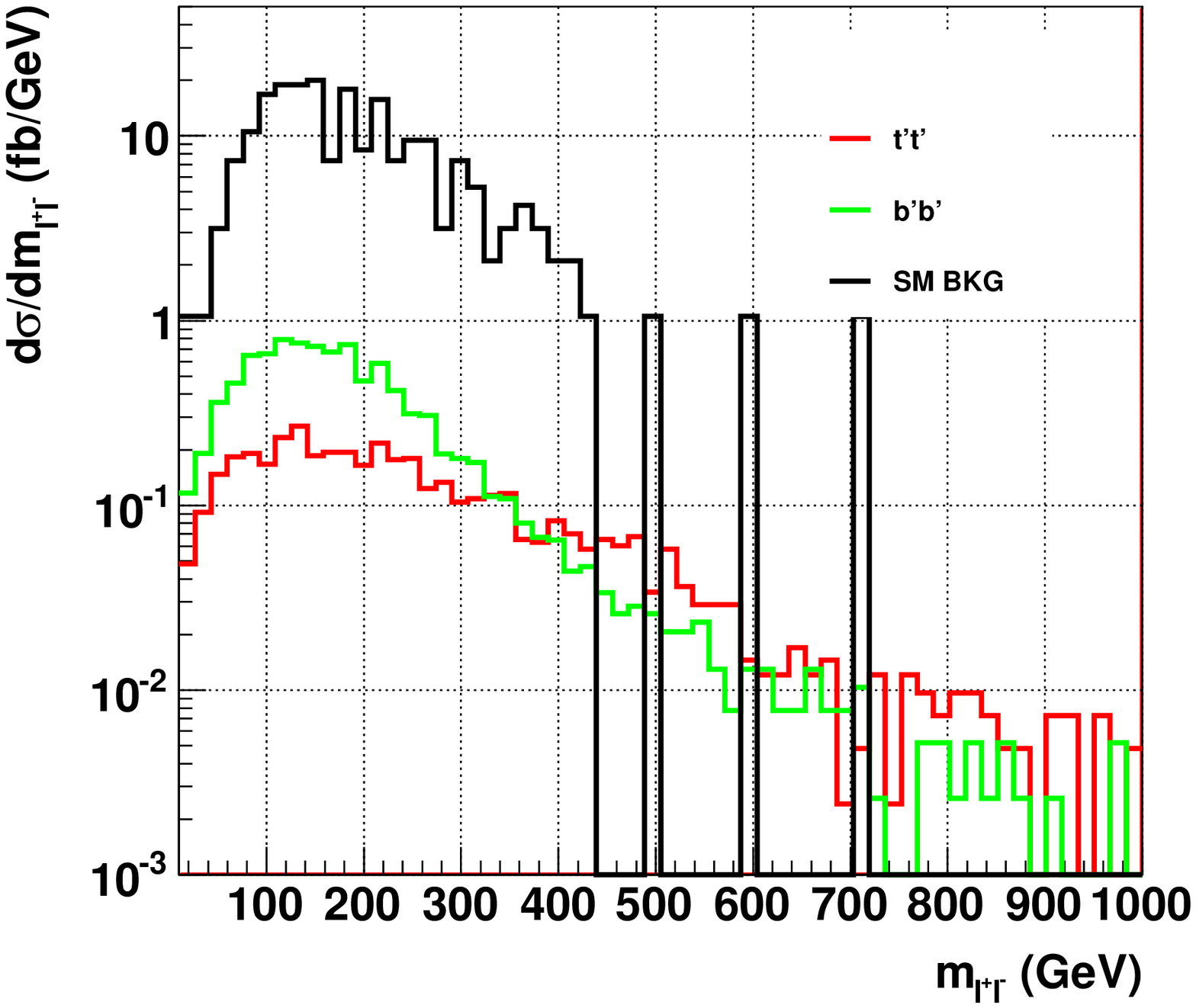}
\includegraphics[angle=0, width=.5\textwidth]{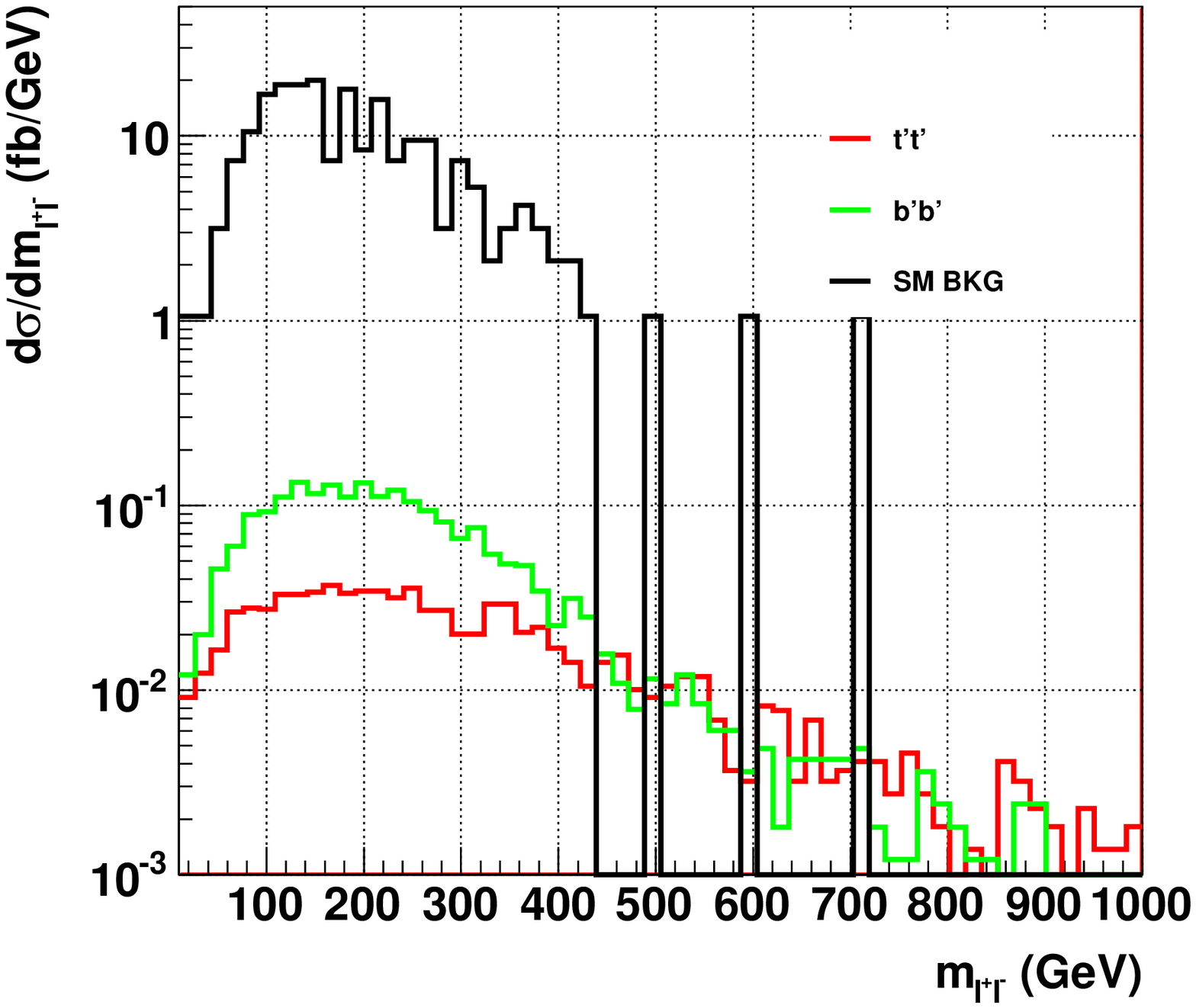}
}
\centerline{
\includegraphics[angle=0, width=.5\textwidth]{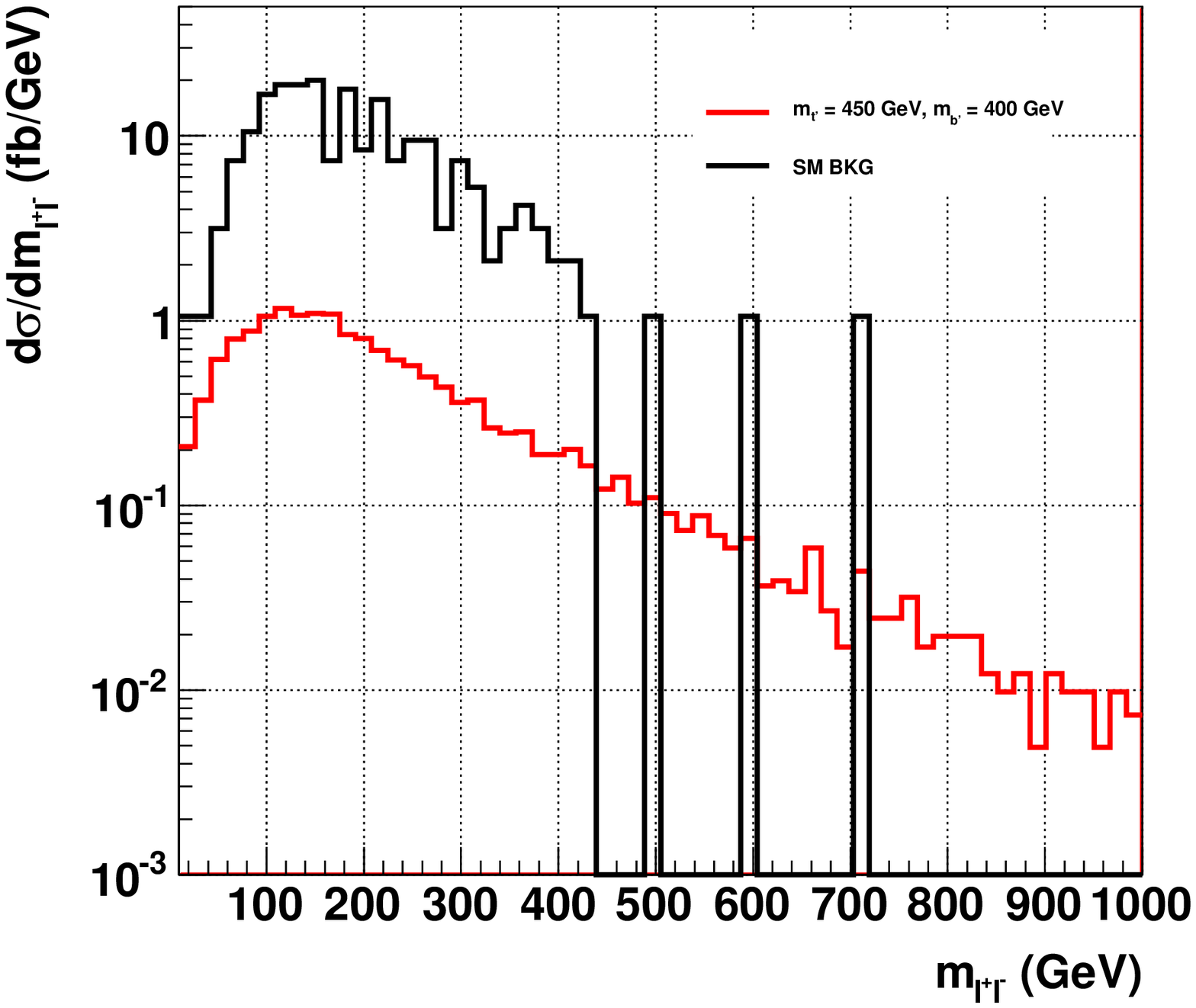}
}
\caption{ 
$m_{l^{+}l^{-}}$ 
distributions for OSD (opposite sign dileptons) cases with $m_Q = 450$ 
(top left), $m_Q = 600$ (top right) and combined case 
with $m_{t'} = 450$ and $m_{b'} = 400$~GeV (bottom).}
\label{OSD-mll} 
\end{figure}

In Figure 
\ref{OSD-mll} 
we plot the invariant mass spectrum for opposite 
sign dilepton pairs in the case of $\bp$- and $\tp$-quarks of mass 450 
and 600~GeV as well as the case with both species present where 
$m_\bp=400$~GeV and $m_\tp=450$~GeV.

Note that the $\tp$ decay gives rise to harder leptons because both of 
the leptonic W-bosons arise from the primary decay while in the $\bp$ 
case the spectrum is softer because some of the leptonic W-bosons arise 
from secondary top decay. Likewise the overall rate for the $\bp$-quark 
signal is large since there are 3 times as many pairs of W-bosons which 
could decay leptonically to give the signal. Unfortunately, in all cases 
the SM3 background, largely from top pairs, is so large that it obscures 
this signal by about an order of magnitude with the cuts we used in 
these plots.

\begin{figure}[htb]
\centerline{
\includegraphics[angle=0, width=.5\textwidth]{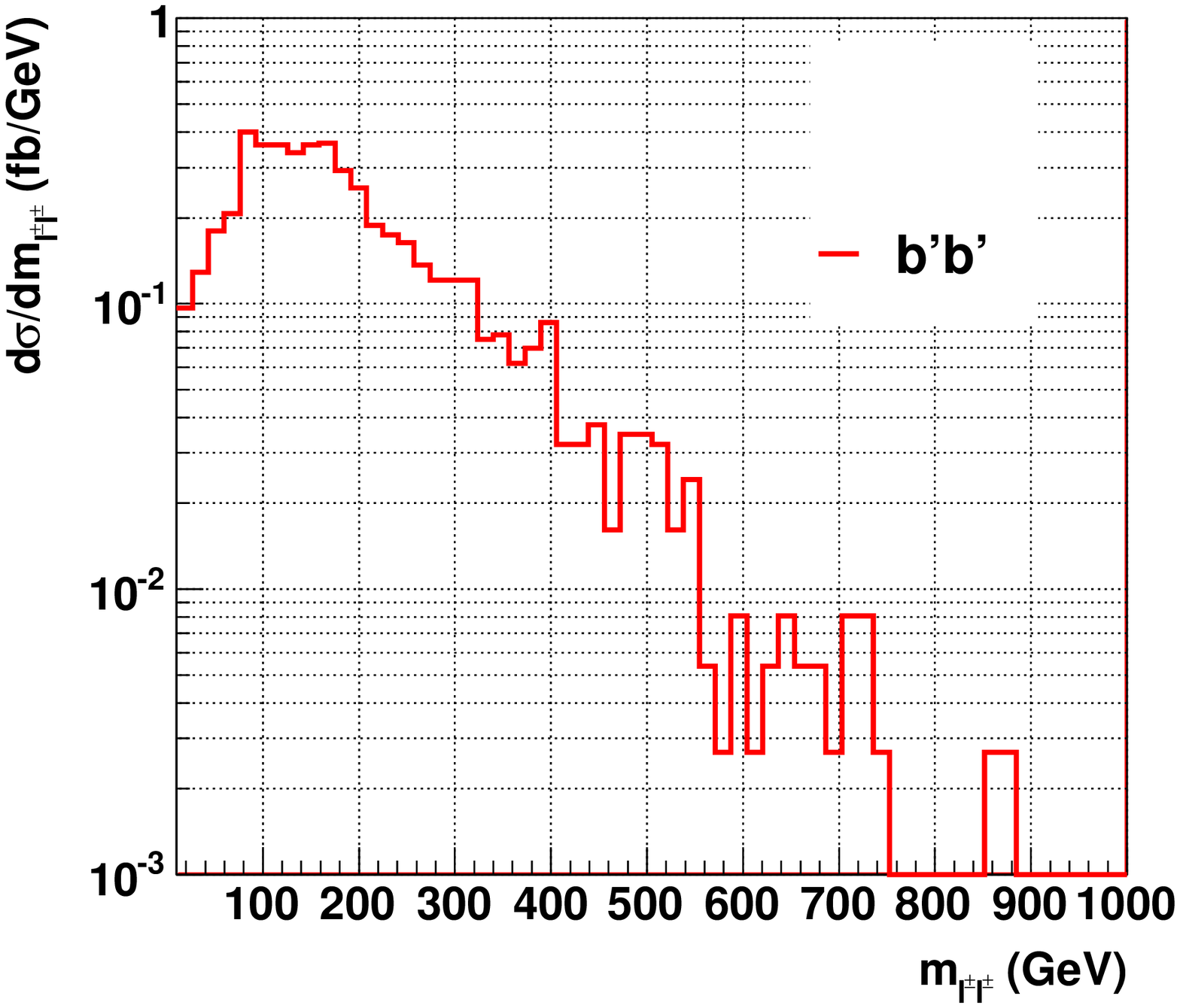}
\includegraphics[angle=0, width=.5\textwidth]{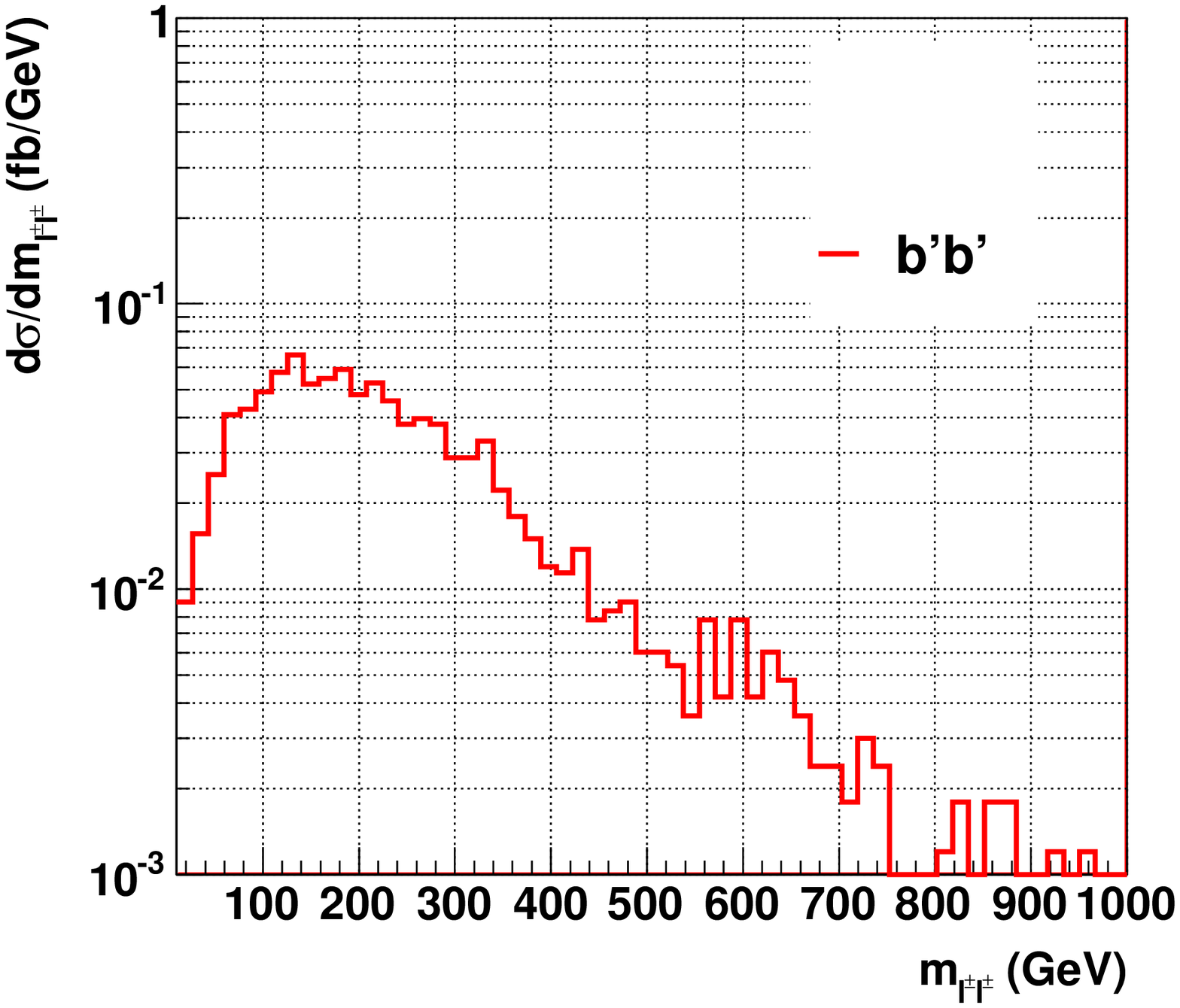}
}
\caption{
$m_{{l^\pm}{l^\pm}}$ 
distributions for SSD (same sign dilepton) cases with 
$m_Q = 450$ (left), $m_Q = 600$ (right). 
}
\label{SSD-mll}
\end{figure}

In Figure \ref{SSD-mll} we plot the invariant mass spectrum for same 
sign dilepton pairs in the case of $\bp$-quarks of mass 450~GeV and 
600~GeV. As discussed in the previous section, the SM background here is 
not an issue.

The same sign dilepton signal should thus be a prominent signal for 
$\bp$-quarks and from the cross-section and shape of the invariant mass 
spectrum, one should be able to infer the $\bp$ mass without considering 
the kinematics discussed below. In Table \ref{mll-table} we show the 
average invariant mass of the dilepton pairs, $\overline{ m_{ll}}$ in 
the cases considered above. We find that the ratio $\overline{ 
m_{ll}}/m_Q$ is stable as a function of $m_Q$ for each type of decay, in 
particular $\overline{ m_{ll}}/m_Q\approx 0.54$ for OSD in $\tp$-quark 
pairs; 
$\overline{ m_{ll}}/m_Q\approx 0.40$ for OSD in $\bp$-quark pairs and 
$\overline{ m_{ll}}/m_Q\approx 0.42$ for SSD in $\bp$-quark pairs.

\begin{table}[h]
\centering
\begin{tabular}{|c|c|c|c|c|c|c|}\hline
Case & 
$\tp$ 450~GeV&
$\tp$ 600~GeV&
$\bp$ 450~GeV&
$\bp$ 600~GeV&
$\bp$ 400~GeV and 
$\tp$ 450~GeV &
SM Background\\ \hline \hline
OSD & 255 & 313 & 182 & 236 & 220 & 197 \\\hline
SSD & N/A    & N/A    & 195 & 245 & N/A    &  N/A   \\\hline
\end{tabular}
\caption{
The average dilepton mass 
for OSD (opposite sign dilepton) and SSD (same sign dilepton) cases. 
For the OSD case
we consider 
$\tp$-pair and $\bp$-pair production
as well as 
the Standard Model background. In the SSD case the background is 
negligible.
}
\label{mll-table} 
\end{table}

\subsection{Analysis of Like Sign Dilepton Sample}

Heavy quark production can contribute to this sample if it decays 
through channels 2-5 (See Eqns.~\ref{decay_forms} 
and~\ref{decay_forms2}). Under the assumptions we are using, this signal 
will arise from channel (2) which is the case when a $\bp$ undergoes two 
body $tW$ decay and the cascade of the top quark gives the second 
W-boson.

In this case, we can in principle reconstruct both of the unobserved 
neutrino momenta provided we have the correct assignment of jets. To see 
how this works, consider the underlying topology of such an event:

\begin{eqnarray}
\bp_1&\to& W_1t_1 \nonumber\\
&& \hookrightarrow W_1\to \ell_1\nu_1\nonumber\\
&& \ \ \ \hookrightarrow t_1\to jets
\nonumber\\
\bp_2&\to& W_2t_2 \nonumber\\
&& \hookrightarrow W_2\to jets\nonumber\\
&& \ \ \ \ \ \hookrightarrow t_2\to b_2 W_3\nonumber\\
&& \ \ \ \ \ \ \ \ \ \hookrightarrow W_3\to \ell_2\nu_2
\end{eqnarray}

Assuming that we have correctly assigned jets or sets of jets to the 
momenta $t_1$, 
$W_2$ and $b_2$, then the kinematic constraints are:

\begin{eqnarray}
m_W^2&=& (\ell_1+\nu_1)^2=(\ell_2+\nu_2)^2\nonumber\\
m_t^2&=&(\ell_2+\nu_2+b_2)^2\nonumber\\
0&=&\nu_1^2=\nu_2^2\nonumber\\
(\ell_1+\nu_1+t_1)^2&=&(\ell_2+\nu_2+b_2+W_2)^2\nonumber\\
\slp_{x,y}&=&(\nu_1+\nu_2)_{x,y}
\end{eqnarray}

These kinematic constraints provide 8 equations for the 8 unknown 
components of $\nu_1$ and $\nu_2$. In solving these equations there is 
in general a two fold or four fold ambiguity in additional to the 
combinatorial ambiguity. 

\begin{figure}[htb]
\centerline{
\includegraphics[angle=0, width=.5\textwidth]{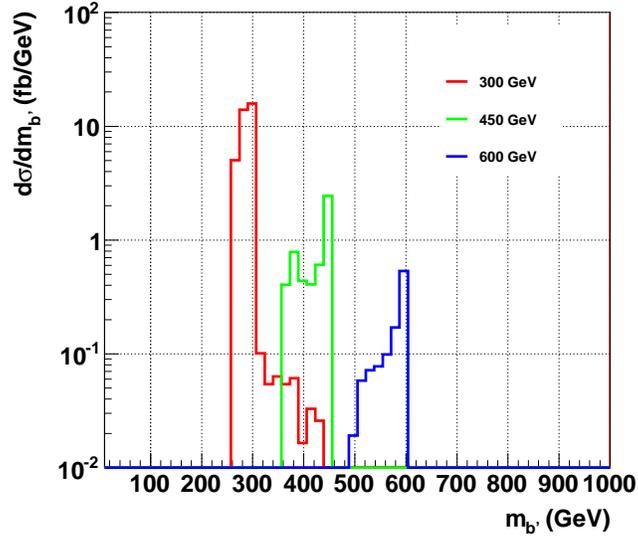}
}
\caption{
The reconstructed $\bp$ masses 
from SSD (same sign dilepton)   
signal case at $\sqrt{s}=14$ TeV~\cite{sqrtsNote}.
}
\label{fig:recssd}
\end{figure}

If we take the events from this sample and analyze them in this way, 
then there will be a peak at the correct value of $m_\bp$. Figure 
\ref{fig:recssd} shows a histogram of the $\bp$ mass reconstructed in 
this way for a number of different masses of the $\bp$ 
at $\sqrt{s}=14$ TeV.

\subsection{Analysis of Opposite Sign Dilepton Sample}

For the signal where opposite sign dileptons are produced, it is useful 
to distinguish three different scenarios, 
for how the leptons arise,
which we will denote OSD1, 
OSD2 and OSD3.
These scenarios vary in terms of the number of kinematic constraints 
which are available to reconstruct the mass of the heavy quark.

{\bf OSD1:} 
The first scenario applies only in the case of $\bp$ production. In 
this case, it could happen that 
both of the leptons arise from top decay, for instance in the 
topology:

\begin{eqnarray}
\bp_1 &\to& W_1 t_1\to \ell_1 \nu_1 b_1   \nonumber\\
&&\hookrightarrow W_1\to jets   \nonumber\\
\bp_2 &\to& W_2 t_2\to \ell_2 \nu_2 b_2   \nonumber\\
&&\hookrightarrow W_2\to jets   \nonumber\\
\end{eqnarray}

\noindent
Then the reconstruction of the event is overdetermined as in the case of 
the single lepton signal.

To see this, let  us list 
the kinematic constraints:

\begin{eqnarray}
(b_1+\ell_1+\nu_1+W_1 )^2&=&
(b_2+\ell_2+\nu_2+W_2 )^2\ \ \ \  (=m_Q^2)
\nonumber\\
m_W^2&=&
(\ell_1+\nu_1)^2=
(\ell_2+\nu_2)^2
\nonumber\\
m_t^2&=&
(\ell_1+\nu_1+b_1)^2=
(\ell_2+\nu_2+b_2)^2
\nonumber\\
(\nu_1+\nu_2)_x&=&\slp_x, 
\ \ \ \ \ 
(\nu_1+\nu_2)_y=\slp_y
\nonumber\\
0&=&\nu_1^2=\nu_2^2
\end{eqnarray}

\begin{figure}[htb]
\centerline{
\includegraphics[angle=0, width=.5\textwidth]{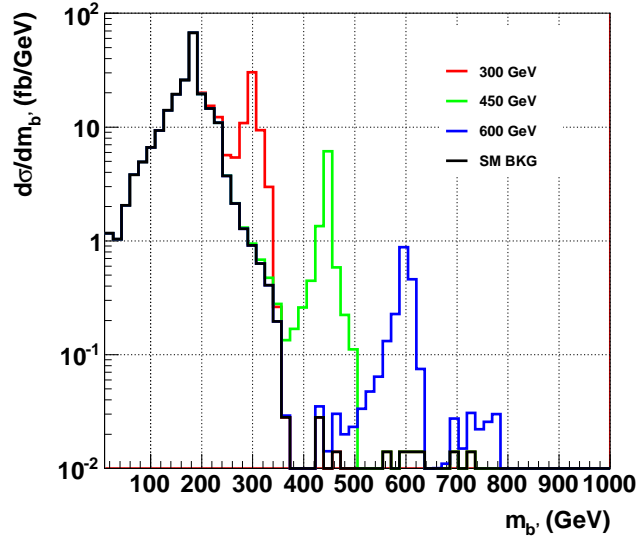}
}
\caption{ 
Reconstructed $\bp$ 
mass from OSD (opposite sign dilepton) 
signal case 1 (OSD1) at $\sqrt{s}=14$ TeV\cite{sqrtsNote}. 
SM background is also presented.
}
\label{fig:recosd1}
\end{figure}

\noindent
Here there are 9 equations to determine the 8 unknown components of 
$\nu_1$ and $\nu_2$ which is helpful in deciding if a given event is 
consistent with this topology. In Figure 
\ref{fig:recosd1}
we show a histogram of events 
reconstructed in this scenario at $\sqrt{s}=14$ TeV.

{\bf OSD2:} The second opposite sign dilepton scenario (OSD2) also 
applies to $\bp$ production where we assume that one of the $\bp$ quarks 
decays to jets and both of the leptons arise from the decay chain of the 
other.  In the first case, if there is no merging between jets from the two 
heavy quarks, then for one of the partitions of the jets into $j_h$, the 
jets from the hadronic $Q$ and $j_\ell$, the jets from the leptonic $Q$, 
$m_Q$ is given by $j_h^2=m_Q^2$. As in the single lepton sample we can 
somewhat reduce the combinatorial background by noting that  the correct 
partition of jets must satisfy the inequality:

\begin{eqnarray}
j_h^2>j_\ell^2+2m_W^2+2j_\ell\cdot(\ell_1+\ell_2)
\end{eqnarray}

\begin{figure}[htb]
\centerline{
\includegraphics[angle=0, width=.5\textwidth]{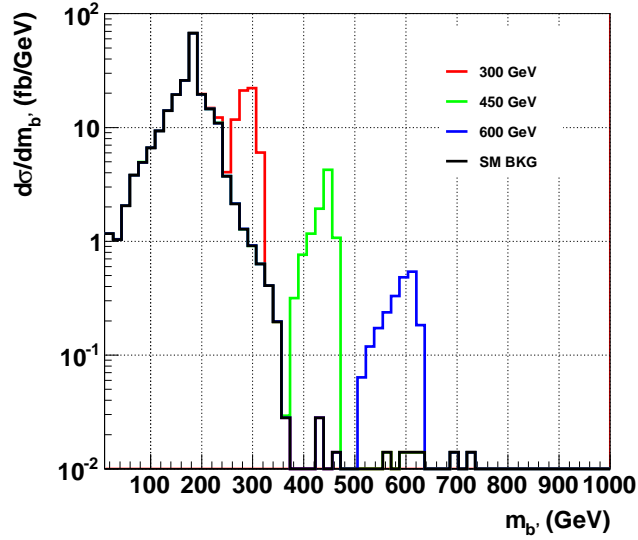}
}
\caption{
Reconstructed $\bp$ masses from opposite 
sign dilepton signal case 2 (OSD2)
at $\sqrt{s}=14$ TeV\cite{sqrtsNote}.  
SM background is also presented.
}
\label{fig:recosd2}
\end{figure}

\noindent
Thus if we construct a histogram of all jet partitions which satisfy 
this relation, there should be an accumulation at the correct value(s) 
of $m_Q$. In this analysis it is not necessary to decompose the missing 
momentum into the two individual neutrinos accompanying the two leptons. 
In Figure 
\ref{fig:recosd2}
we show a histogram of the reconstructed heavy quark mass in this 
scenario
at $\sqrt{s}=14$ TeV.

{\bf OSD3:} The third opposite sign dilepton scenario (OSD3) is the most 
general. In this case we assume that the leptons are produced promptly 
from the initial heavy quark decays, for instance

\begin{eqnarray}
\tp\to b [W\to \ell\nu] 
\nonumber\\
\tp\to b [W\to \ell\nu] 
\end{eqnarray}

This scenario would therefore apply to both $\tp$ and $\bp$ pair 
production since overall we just assume that $Q\to [W\to \ell \nu]+jets$ 
on both sides of the event. Since there are no particles with known 
masses in the decay chain besides the $W$-bosons which produce the 
leptons, this is the least kinematically constrained case.

In general to reconstruct the heavy quark mass we need to reconstruct 
the neutrino momenta but 
in this case
we do not have enough 
information to do so. This is evident because the kinematic 
constraints in this case are:

\begin{eqnarray}
(j_1+\ell_1+\nu_1)^2&=&
(j_2+\ell_2+\nu_2)^2\ \ \ \  (=m_Q^2)
\nonumber\\
m_W^2&=&(\ell_1+\nu_1)^2=(\ell_2+\nu_2)^2
\nonumber\\
0&=&\nu_1^2=\nu_2^2
\nonumber\\
(\nu_1+\nu_2)_x&=&\slp_x
\ \ \ \ \ 
(\nu_1+\nu_2)_y=\slp_y
\end{eqnarray}

\noindent
which provides only 7 constraints on the 8 unknown components of $\nu_1$ 
and $\nu_2$ and so we are one constraint short of what is required to 
fully reconstruct the event. 

\begin{figure}[htb]
\centerline{
\includegraphics[angle=0, width=.5\textwidth]{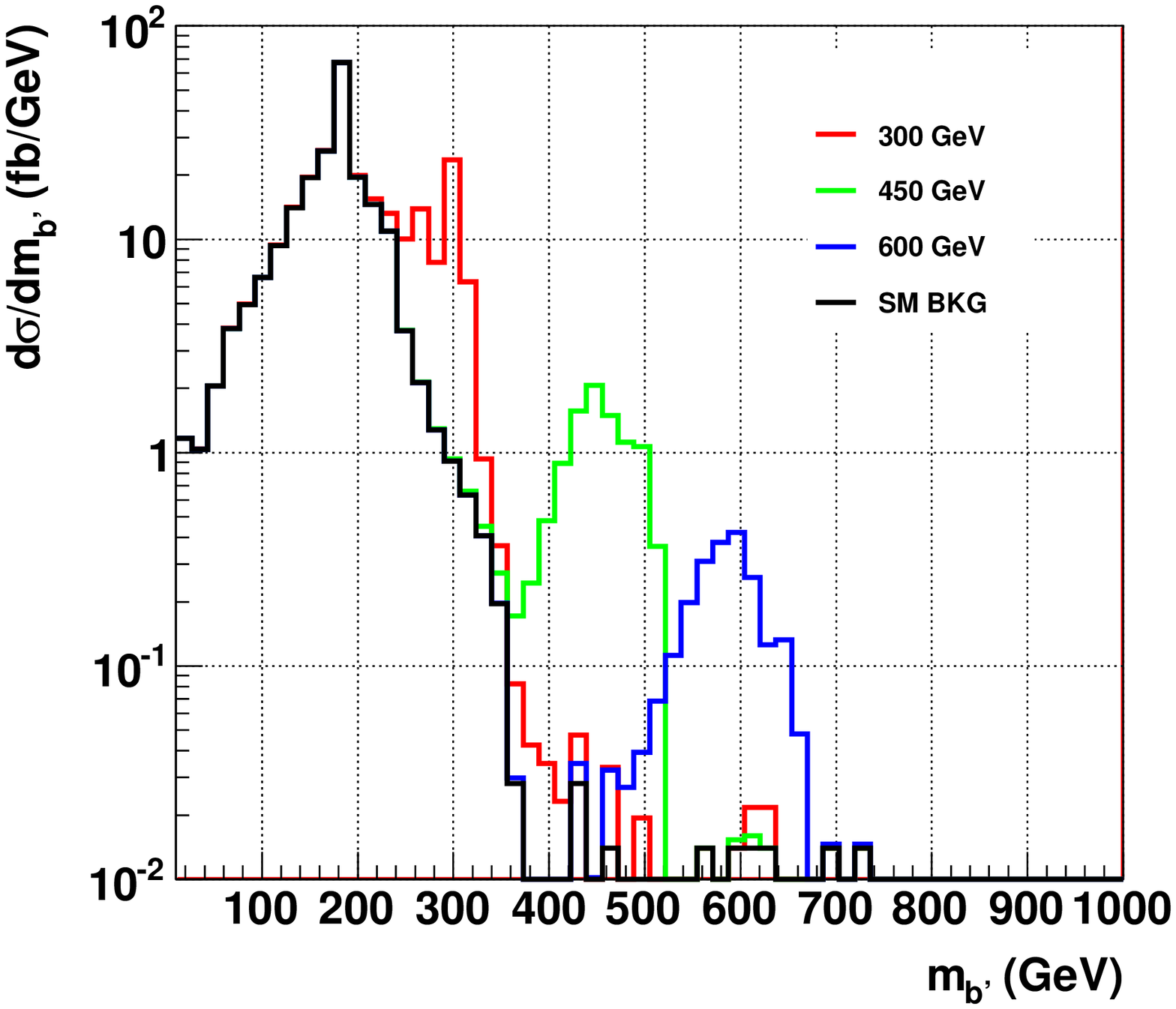}
\includegraphics[angle=0, width=.5\textwidth]{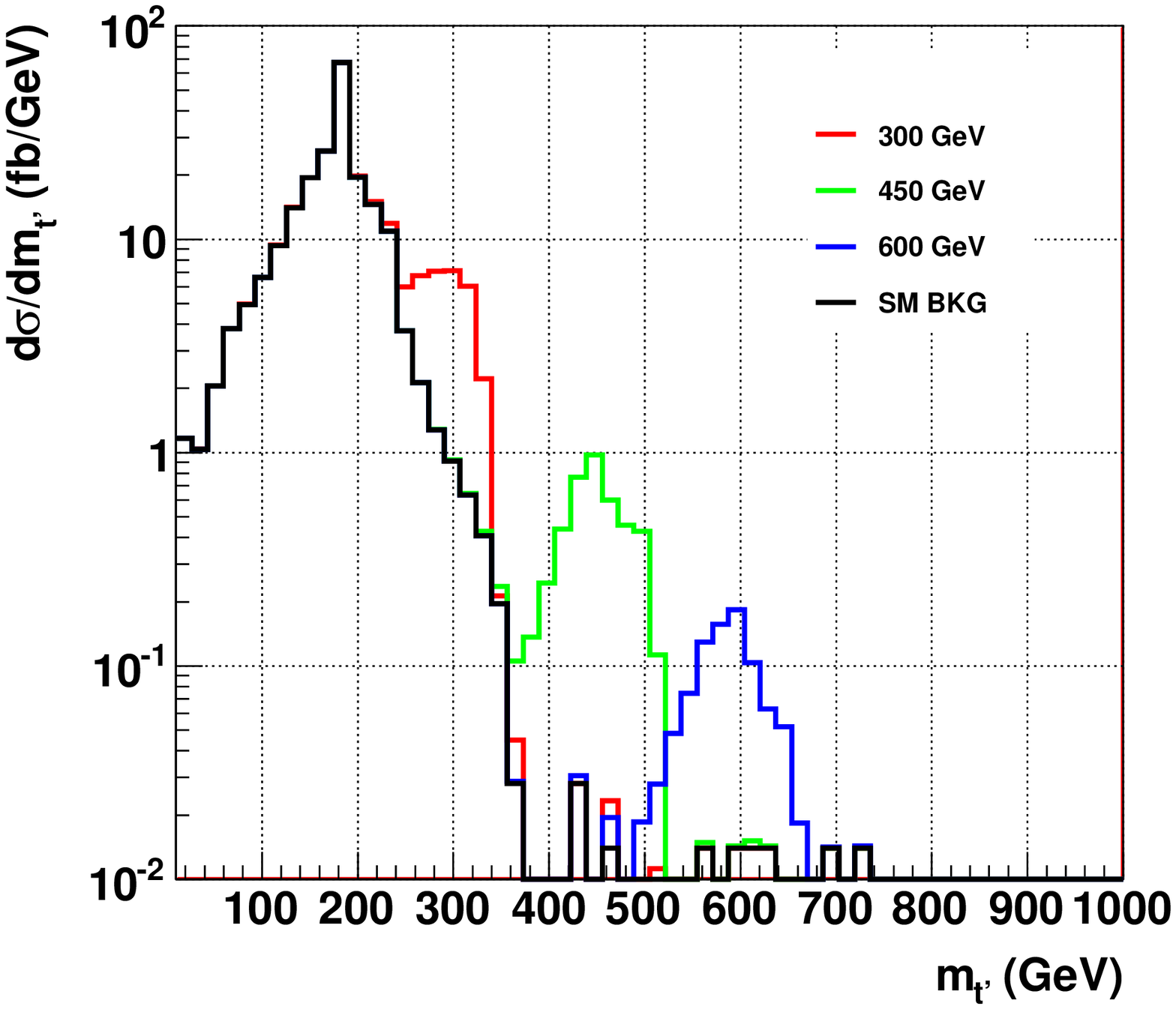}
}
\caption{
Reconstructed $\tp$ and $\bp$ masses 
for opposite sign dilepton case 3 (OSD3) at $\sqrt{s}=14$ TeV\cite{sqrtsNote}. 
SM background is also presented.
}
\label{fig:recosd3}
\end{figure}

We can, however, take advantage of the kinematics of the heavy quark 
decay and obtain an approximate value of $m_Q$ for an individual event. 
To do this, note that both the matrix element and structure functions of 
the 
proton tend to favor heavy quark production near threshold. If we make 
the approximation that the $Q$ pair is at threshold, we can replace the 
first condition above with the following two conditions:

\begin{eqnarray}
(j_1+\ell_1+\nu_1)_z&=&
(j_2+\ell_2+\nu_2)_z
\nonumber\\
(j_1+\ell_1+\nu_1)_t&=&
(j_2+\ell_2+\nu_2)_t
\end{eqnarray}

\noindent
which then gives a total of eight conditions for the eight unknown 
components of $\nu_1$ and $\nu_2$. If $\tilde m_Q$ is the reconstructed 
mass of the $Q$ based on this assumption, the accumulation of $\tilde 
m_Q$ will give an indication of the true $Q$ mass(es). Note that there 
is a two fold ambiguity in solving this system of equations. In Figure 
\ref{fig:recosd3} we show a histogram of the reconstructed $\tp$ and 
$\bp$ masses in this scenario at $\sqrt{s}=14$ TeV.

\begin{figure}[htb]
\centerline{
\includegraphics[angle=0, width=.5\textwidth]{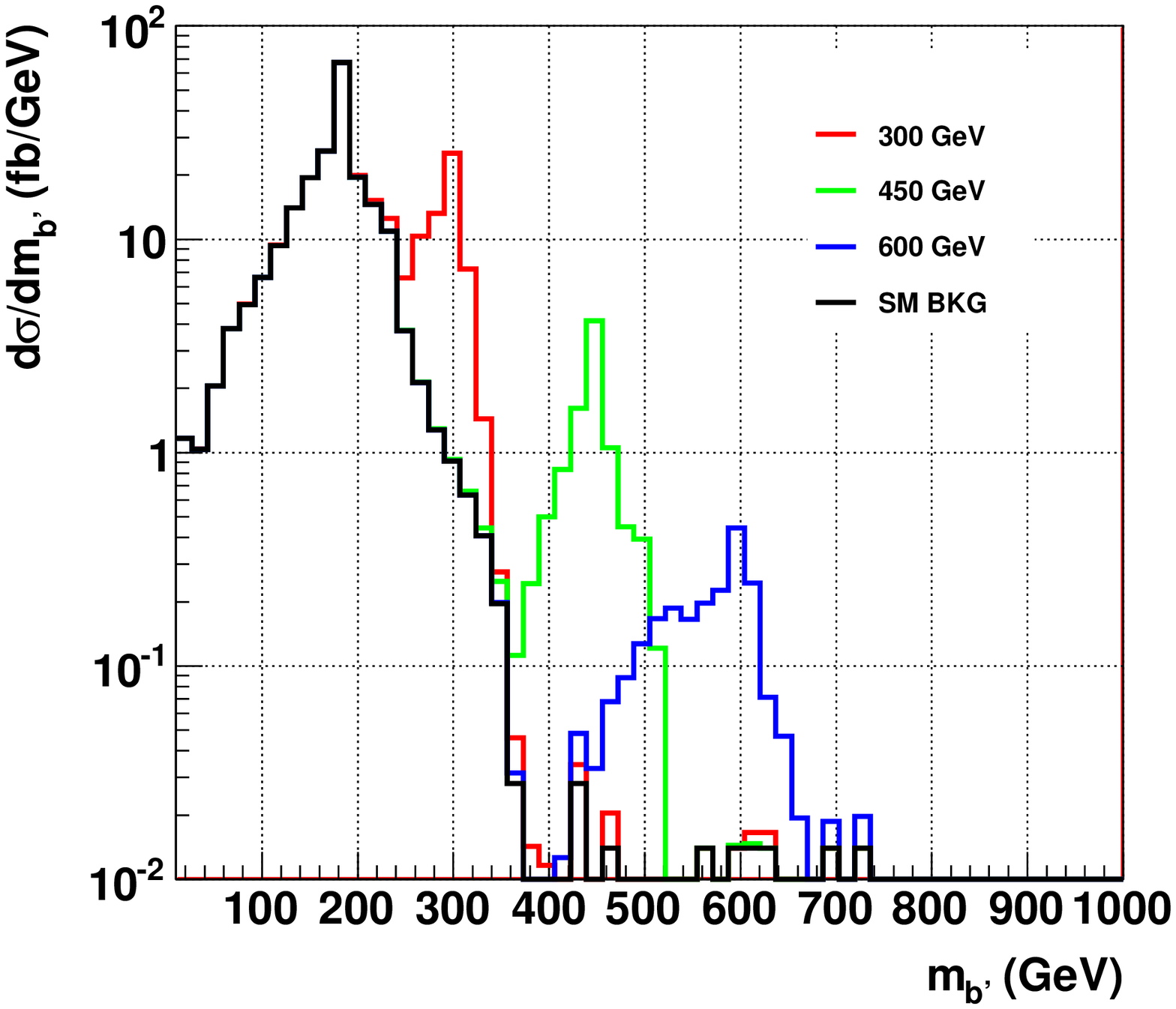}
}
\caption{
Reconstructed $\bp$ masses
where a mixture of OSD1, OSD2 and OSD3 events are analyzed
using the three different methods, i.e. assuming that the event has 
OSD1,
OSD2 and OSD3 topology at $\sqrt{s}=14$ TeV\cite{sqrtsNote}. 
SM background is also presented.
}
\label{fig:recon_osdbcombined}
\end{figure}

In the above three plots (Figures~\ref{fig:recosd1}-~\ref{fig:recosd3}), 
we have assumed that the topology which gives
rise to the input was known. In reality, of course, we would not know
which scenario might be producing a given event but rather we would have
an event with jets plus an opposite sign dilepton passing the initial
cuts. Our event sample would therefore be a mixture of OSD1, OSD2 and
OSD3 events.

To handle this situation let us consider the following approach: we take 
each event and analyze it as if it is, in turn, and OSD1 event, and OSD2 
event and an OSD3 event. Thus for each event we potentially have three 
sets of reconstructed $\bp$ masses. In practice much of the time only 
the correct mode of analysis produces physical results. In any case, in 
Figure \ref{fig:recon_osdbcombined} we have used each of the three forms 
of analysis on each event in a sample of $\bp$-quark pairs containing 
OSD1, OSD2 and OSD3 decay topologies and plotted a histogram of all the 
reconstructed masses which arise. A single event might contribute 
multiple points to the histogram. As can be seen, there is still 
reasonably strong peaking at the correct $\bp$ mass.

It is clear from the above Figures that the masses of $\tp$ and 
$\bp$ can be reconstructed within $10$ percent accuracy in all the 
cases.

\subsection{ $2b + 6W$ signal}

A very interesting signal arise when the $V_{t^\prime 
b^\prime}, V_{tb}
>> V_{t^\prime b}$. In this case the 
branching ratio for $t^\prime \to b^\prime W^\star$ will be $\simeq 1$. 
Now since $b^\prime$ decays into a top and a $W$ boson, in case of 
$t^\prime$ - pair production, our signal will consist of $2b + 6 W$'s. 
Below in Tables~\ref{tpbp},~\ref{tpbpr} we present event rates for the 
cases when one or two W's decay leptonically for two cases of the mass 
difference $\Delta m = m_{t^\prime} - m_{b^\prime} = 25$~GeV and 
$50$~GeV respectively.

In Tables~\ref{tpbp},~\ref{tpbpr} we note that event rate is lesser for 
smaller $\Delta m$. This is because the lepton/jets arising due to the 
decay of off-shell $W$'s will be relatively softer for smaller $\Delta 
m$ and therefore these leptons/jets will suffer more from the $p_T$ cuts 
on them.

\begin{table}[h]
\centering
\begin{tabular}{|c|c|c|c|c|}\hline
$\sqrt{s} (TeV)$&cuts& $m_{t'}=350$~GeV & $m_{t'}=450$~GeV & 
$m_{t'}=600$~GeV\\\hline\hline
&&$m_{t^\prime} - m_{b^\prime} = 25$~GeV&&\\\hline
7 &$Basic$                      &509, 84, 51&116, 19, 11&19, 3, 2\\
&$Basic + H_T > 350$~GeV      
&442, 72, 42&111, 18, 11&17, 3, 2\\\hline
10 &$Basic$                      
&1655, 272, 159&413, 72, 43&77, 13, 9\\
&$Basic + H_T > 350$~GeV      
&1439, 235, 136&286, 69, 30&76, 13, 
9\\\hline
14 &$Basic$                      
&4516, 736, 477&1222, 222, 127&259, 45, 28\\
&$Basic + H_T > 350$~GeV      
&3942, 648, 423&1177, 213, 119&256, 44, 
28\\\hline
\hline
&&$m_{t^\prime} - m_{b^\prime} = 50$ GeV&&\\\hline
7 &$Basic$                      &534, 137, 87&119, 30, 21&18, 5, 3\\
&$Basic + H_T > 350$~GeV      &445, 119, 75&114, 29, 20&18, 5, 3\\\hline
10 &$Basic$                      &1773, 432, 296&442, 118, 74&78, 21, 13\\
&$Basic + H_T > 350$~GeV      &1501, 379, 254&421, 114, 71&77, 21, 
13\\\hline
14 &$Basic$                      &4843, 1233, 754&1276, 344, 223&261, 72, 44\\
&$Basic + H_T > 350$~GeV      &4093, 1079, 648&1223, 331, 214&258, 71, 
44\\\hline
\hline
\end{tabular}
\caption{
Number of signal and background single lepton, opposite sign dileptons 
(OSD), and same sign dileptons (SSD) events from the $t'$-pair 
production at the LHC from $t^\prime t^\prime \to 2b + 6 W$ state for 
$m_{t^\prime} - m_{b^\prime} = 25$~GeV and $50$~GeV for $\sqrt{s} = 7, 
10$ and $14$~TeV and $\int {\cal L} dt = 1$ fb$^{-1}$ without the 
requirement of isolation on jets. The basic cuts are the same as in 
Table~\ref{big-table-1a}.
}
\label{tpbp}
\end{table}

\begin{table}[h]
\centering
\begin{tabular}{|c|c|c|c|c|}\hline
$\sqrt{s} (TeV)$&cuts& $m_{t'}=350$~GeV & $m_{t'}=450$~GeV & 
$m_{t'}=600$~GeV\\\hline\hline
&&$m_{t^\prime} - m_{b^\prime} = 25$~GeV&&\\\hline
7 &$Basic$                       &417, 68, 41&97, 16, 9&15, 3, 2\\
&$Basic + H_T > 350$~GeV         &361, 58, 35&93, 15, 9&15, 3, 2\\\hline
10 &$Basic$                      &1334, 214, 125&354, 61, 36&65, 11, 7\\
&$Basic + H_T > 350$~GeV         &1156, 184, 108&340, 59, 35&64, 11, 
7\\\hline
14 &$Basic$                      &3801, 629, 399&1042, 188, 109&225, 39, 24\\
&$Basic + H_T > 350$~GeV         &3317, 552, 354&1004, 181, 103&222, 38, 
23\\\hline
\hline
&&$m_{t^\prime} - m_{b^\prime} = 50$~GeV&&\\\hline
7 &$Basic$                       &443, 113, 73&99, 25, 17&15, 4, 3\\
&$Basic + H_T > 350$~GeV         &368, 97, 61&95, 24, 17&15, 4, 
3\\\hline
10 &$Basic$                      &1467, 353, 244&367, 98, 60&66, 18, 11\\
&$Basic + H_T > 350$~GeV         &1236, 309, 210&351, 94, 58&65, 18, 
11\\\hline
14 &$Basic$                      &3954, 997, 626&1067, 
289, 186&222, 61, 39\\
&$Basic + H_T > 350$~GeV         &3327, 871, 540&1024, 279, 178&222, 60, 
38\\\hline
\hline
\end{tabular}
\caption{
This table shows the number of events with the same cuts as in 
Table~\ref{tpbp} with the addition of the jet isolation cut that 
all the jets are separated with $\Delta R_{jj} > 0.4$.
}
\label{tpbpr}
\end{table}

\section{Standard Model Backgrounds}\label{backgrounds}

The leading SM background for the aforementioned processes is due to top 
pair production, where at least one top decays semileptonically.  To 
this end, in our analysis, we also include process where a top pair is 
produced with up to 3 jets, i.e. $tt + n j$; with $n \leq 3$. The other 
subdominant backgrounds that can affect our signal are due to tri-gauge 
boson production processes: $VVV$; $V=W^\pm, Z$. It is to be noted that 
there can be other sources as well such as $ttVV$ and $VVVV$ but these 
will be quite small as compared to the aforementioned ones, therefore, 
we ignored such background in our analysis. We list all the relevant 
processes and their respective production cross-section in 
Table~\ref{smbkg}.

\begin{table}[h]
\centering
\begin{tabular}{|c|c|c|c|c|c|c|}\hline\hline
SM Background & $t\bar{t} + 0 j $ & $t\bar{t} + 1 j$ & 
$t\bar{t} + 2j$ &  $t\bar{t} + 3j$ & $t\bar{t} V$ & $VVV$\\\hline
~7~TeV &~89.7&~20.7&~19.4&~10.2&0.24&0.06\\\hline
10~TeV &233.6&~64.2&~60.2&~23.3&0.57&0.12\\\hline
14~TeV &535.1&443.2&247.3&108.3&1.2&0.21\\\hline
\hline
\end{tabular}
\caption{Cross-section (in pb) for various SM background processes at 
the LHC with $\sqrt{s} =$ 7, 10 and 14~TeV with basis cuts.} 
\label{smbkg} 
\end{table}

Clearly the jets or lepton arising due to the background processes will 
be relatively softer than our signal. Therefore, with the demand of 
higher transverse momentum jets and leptons or alternatively a higher 
scalar sum of transverse momentum of visible final state particles and 
the missing transverse energy, $H_T$ will reduce the background 
considerably as shown in Tables~\ref{big-table-1a} 
and~\ref{big-table-2a} 
for each case.

\section{Summary and Conclusions}\label{summary}

Extending the Standard Model by adding a fourth fermion generation may 
be helpful in understanding how electro-weak symmetry breaking works and 
possibly also in elucidating the mechanism of baryogenesis in the early 
universe. The LHC will pair produce fourth generation quarks at a 
relatively high rate if their mass is in the range 400--600~GeV. Decay 
channels which result in final states with one or two leptons provide 
signals which have only modest Standard Model backgrounds.

In this study we considered the case where fourth generation quarks 
decay predominantly to quarks from the first three generations assuming 
that the mass splitting between the fourth generation quarks is small. 
This is 
likely the decay mechanism of the lightest fourth generation quark and 
also could be the decay mode for both of the heavy quarks. For such a 
quark we investigated the signals and backgrounds in three different 
channels: (1) Single lepton plus jets and missing momentum; (2) Opposite 
sign dilepton pair plus jets and missing momentum and (3) Same sign 
dilepton pair plus jets and missing momentum. In all cases, we first 
consider a basic set of cuts which eliminates a large portion of the 
background but generally leaves the signal intact. To further isolate 
the signal we use the kinematics of the event to reconstruct the heavy 
quark mass. 

In the case of a single lepton plus jets and missing momentum, we can 
use the fact that the missing momentum arises form a single neutrino 
which is associated with the lepton in the decay of a W-boson. This 
constraint over determines the kinematics once the role of the jets is 
correctly assigned. In particular one can extract independently a mass 
for the heavy quark both from the hadronic and leptonic sides of the 
event. We find that the combinatorial background from the many possible 
jet assignments can be greatly reduced if you accept only the jet 
assignment which minimizes the discrepancy between the two mass 
determinations. Other cuts which are helpful in reducing the 
combinatorial background are a cut on $H_T$ and a cut on $\Delta m_W$ 
which tends to force jet pairs which arise from the decay of a single 
$W$ boson to be assigned to the same side of the event. 

If the single lepton event arises from the decay of a heavy quark to a 
light quark other than the top (i.e. which we have generally taken to be 
a $\tp$ in our discussion), there are a total of four jets in the final 
state so the combinatorial background is not particularly severe. If, 
however a $\bp$ quark decays to a $tW$ final state, then there are 
eight jets in total and the combinatorial background may pose a problem 
for quarks in the heavier mass range. Thus we have considered what mass 
resolution is required in order to satisfactorily implement the minimum 
$\Delta m_Q$ selection of the correct jet assignment.

In the case of an opposite sign dilepton pair, if the heavy quark 
does not decay to a top quark, there are only 2 jets so there is no 
large combinatorial background; however, the kinematics of such events 
is underdetermined so the quark mass cannot precisely be extracted on an 
event by event basis. But if we make the approximation that the two 
heavy 
quarks are at rest in their center of mass frame, we can deduce roughly 
the heavy quark mass which evidently separates the signal from the 
background well enough to provide a prominent signal and determine the 
heavy quark mass.

In the case where a $\bp$ decays to $tW$ there are three cases which 
could apply depending on whether 0, 1 or 2 of the leptons arise from top 
decay. For each lepton which arises from top decay there is an 
additional constraint on the momentum of the associated neutrino; 
therefore, in these scenarios the kinematics is either under determined, 
critically determined or over determined respectively. For a given event 
we do not initially know which of these scenarios applies, however, if 
we iterate over the three scenarios and all jet assignments, most of the 
wrong assignments lead to unphysical neutrino momenta so that retaining 
the cases which can be reconstructed we produce a satisfactory signal 
and reconstruction of the heavy quark mass.

The same sign dilepton signal only arises in the scenario where $\bp\to 
tW$ and results in kinematics where the neutrino momenta are critically 
determined. This channel has the advantage that the Standard Model 
background is negligible. As in the opposite sign case, we can resolve 
the combinatorial background by taking only cases where there are 
physical solutions for the neutrino momenta and thus obtain the heavy 
quark mass. Thus the mass of the heavy quark may be determined by the 
kinematic reconstruction or through the characteristics of the invariant 
mass distribution of the dilepton pair.

\section*{Acknowledgements}

We would like to thank Michael Begel, Shaouly Bar-Shalom, Thomas 
Gadfort, Michael Wilson and Daniel Whiteson for discussions. In addition 
S.~K.~G. would like to thank Nils Krumnack and Marzia Rosati for their 
useful help with the package ROOT~\cite{Brun:1997pa}. The work of D.~A., 
S.~K.~G. and A.~S. are supported in part by US DOE grant Nos. 
DE-FG02-94ER40817 (ISU) and DE-AC02-98CH10886 (BNL).


\begin{thebibliography}{99}
\bibitem{sqrtsNote}
For clarity,
we show the invariant mass distributions for 
$\sqrt{s}=14$~TeV only. 
The invariant mass distributions for 
$\sqrt{s}=7-14$~TeV generally have the same shape but are rescaled by 
the total cross section.
\end{thebibliography}
\end{document}